\documentclass[sigplan,nonacm,final]{acmart-mjolnir}

\acmConference[LICS20]{ACM/IEEE Symposium on Logic in Computer Science}{June 2020}{Beijing, China}
\startPage{1}

\setcopyright{none}

\usepackage{amsmath}

\newcommand{\lo}[1]{}
\newcommand{\cm}[1]{}

\makeatletter
\newcommand\ztag[1]{%
\gdef\tmp{%
\def\theequation{\footnotesize #1}}%
\aftergroup\aftergroup\aftergroup\aftergroup\aftergroup\aftergroup
\aftergroup\aftergroup\aftergroup\aftergroup\aftergroup\aftergroup
\aftergroup\aftergroup\aftergroup\aftergroup\aftergroup\aftergroup
\aftergroup\aftergroup\aftergroup\aftergroup\aftergroup\aftergroup
\aftergroup\aftergroup\aftergroup\aftergroup\aftergroup\aftergroup
\aftergroup
\tmp}

\newcommand{\Real}{\mathbb{R}}

\newcommand{\Function}[5]{
  \begin{align*}
    #1: & & #2 & \;\longrightarrow\; #3 \\
        & & #4 & \;\longmapsto\;     #5
  \end{align*}
}

\newcommand{\st}{\ |\ }

\newcommand{\Id}{\mathsf{Id}}
\newcommand{\set}[1]{\{#1\}}
\newcommand{\anglebr}[1]{\langle #1\rangle}
\newcommand{\sqbr}[1]{\llbracket #1 \rrbracket}
\let\oldint\int
\renewcommand\int{\displaystyle\oldint}
\let\oldoint\oint
\renewcommand\oint{\displaystyle\oldoint}
\let\oldfrac\frac
\renewcommand\frac{\displaystyle\oldfrac}
\let\oldbigcap\bigcap
\renewcommand\bigcap{\displaystyle\oldbigcap}
\let\oldbigcup\bigcup
\renewcommand\bigcup{\displaystyle\oldbigcup}
\let\oldsum\sum
\renewcommand\sum{\displaystyle\oldsum}
\newcommand{\cat}[1]{\mathbf{#1}}
\newcommand{\catev}{\mathsf{ev}}
\newcommand{\Arrow}{\Rightarrow}

\newcommand{\deno}[1]{\sqbr{#1}} 

\newcommand{\PCFReal}{\mathsf{R}}
\newcommand{\PCF}[1]{\underline{#1}}

\newcommand{\Let}[2]{\mathsf{let}\ #1\ \mathsf{in}\ #2}

\newcommand{\curry}[1]{\mathsf{cur}(#1)}
\newcommand{\uncurry}[1]{\mathsf{uncur}(#1)}

\renewcommand\terms{\Lambda}

\newcommand{\variables}{\mathcal{V}}

\newcommand{\freevar}[1]{\mathsf{FV}(#1)}

\newcommand{\diff}[2]{\mathsf{D}{#1}\cdot{#2}}
\newcommand{\jacob}{\mathcal{J}}
\newcommand{\jac}[2]{\mathcal{J}\PCF{#1}\cdot{#2}}
\newcommand{\pbmap}[2]{(\oneform \, {#1})\cdot{#2}}
\newcommand{\pbof}[1]{\oneform \, \PCF{#1}}
\newcommand{\pbat}[2]{{#1}\,{#2}}

\newcommand{\dualmap}[2]{\dual{#1}\cdot{#2}}
\newcommand{\proj}[2]{{#1}_{\pi {#2}}}

\newcommand{\pbterms}{\mathbb{P}}

\newcommand{\letterms}{\mathbb{L}}
\newcommand{\elemterms}{\mathbb{E}}
\newcommand{\simterms}{\mathbb{S}}
\newcommand{\values}{\mathbb{V}}

\newcommand{\red}{\longrightarrow}
\newcommand{\Ared}{\red_A}
\newcommand{\Pred}{\red}
\newcommand{\Predtext}[1]{\xrightarrow{#1}}
\newcommand{\Predtextm}[2]{\xrightarrow[#2]{#1}}

\newcommand{\redplus}{\red^*}
\newcommand{\Aredplus}{\Ared^*}
\newcommand{\Predplus}{\Pred^*}
\newcommand{\Predplustext}[1]{{\xrightarrow{#1}}^*}

\newcommand{\config}[2]{\anglebr{{#1}\,|\,{#2}}}
\newcommand{\dsubup}[3]{{\tfrac{{\partial} {#1}}{{\partial} {#3}}}\cdot {#2}}
\newcommand{\dsub}[3]{{\tfrac{{\partial}}{{\partial} {#3}}{#1}}\cdot {#2}}
\newcommand{\mathdsubup}[2]{{\tfrac{{\partial} {#1}}{{\partial} {#2}}}}
\newcommand{\mathdsubupat}[3]{
  {\tfrac{{\partial} {#1}}{{\partial} {#2}}\Big|_{#3}}
}

\newcommand{\colvec}[2][.8]{%
  \scalebox{#1}{%
    \renewcommand{\arraystretch}{.8}%
    $\begin{bmatrix}#2\end{bmatrix}$%
  }
}
\newcommand{\mat}[2][.9]{%
  \scalebox{#1}{%
    \renewcommand{\arraystretch}{.8}%
    $\begin{pmatrix*}[l]#2\end{pmatrix*}$%
  }
}
\newcommand{\letarray}[3][.9]{%
  \scalebox{#1}{%
    \renewcommand{\arraystretch}{.8}%
    $\begin{array}{rl}
      \mathsf{let}\ #2
      \ \mathsf{in}\ #3
    \end{array}$%
  }
}
\newcommand{\idmat}{\mathbf{I}}

\newcommand{\oneform}{\Omega}

\newcommand{\lang}[1]{\mathcal{L}_{#1}}
\newcommand{\trans}[1]{{#1}_t}

\newcommand{\smooth}[2]{C^{\infty}(#1,#2)}
\newcommand{\easilydiff}{\mathcal{F}}

\newcommand{\lin}[1]{\mathsf{lin}{(#1)}}

\newcommand{\dual}[1]{{#1}^*}

\newcommand{\ConS}{\cat{Con}^{\infty}}

\usepackage[utf8]{inputenc} 
\usepackage[T1]{fontenc} 

\usepackage[normalem]{ulem}
\usepackage{amsthm}
\usepackage{amssymb}
\usepackage{amsmath}
\usepackage{stmaryrd}
\usepackage{mathtools}
\usepackage{dsfont}
\usepackage{IEEEtrantools}
\usepackage{relsize} 
\usepackage{bussproofs} 
\usepackage{algorithm} 
\usepackage{listings} 
\usepackage{qtree} 
\usepackage{physics} 
\allowdisplaybreaks

\usepackage[toc,page]{appendix} 
\usepackage{centernot}
\PassOptionsToPackage{hyphens}{url}
\usepackage{hyperref} 
\usepackage{tabularx,tabulary,multirow,booktabs} 
\usepackage[all]{xy} 
\usepackage{tikz}
\usepackage{fancyhdr} 
\usepackage[style=british]{csquotes} 
\usepackage{enumerate}
\makeatletter
\preto{\@tabularx}{\parskip=5pt}
\makeatother
\usepackage{environ} 
\NewEnviron{killcontents}{}
\usepackage[caption=false]{subfig}

\renewcommand{\arraystretch}{0.2} 

\newcommand{\ie}{i.e.\ }
\usepackage{setspace}
\definecolor{oxfordblue}{RGB}{0, 33, 71}
\definecolor{teal}{RGB}{0, 128, 128}
\definecolor{midnightblue}{RGB}{25, 25, 112}

\usepackage{thm-restate}

\begin{document}

\setlength{\pdfpageheight}{\paperheight}
\setlength{\pdfpagewidth}{\paperwidth}





\title[Differential-form Pullback Programming Language and Reverse-mode AD]{A Differential-form Pullback Programming Language for Higher-order Reverse-mode Automatic Differentiation}%

\author{Carol Mak}
\author{Luke Ong}


\begin{abstract}

  \lo{An abstract (especially LICS abstract) should be concise, and summarise the technical contributions. It should not contain  motivational remarks such as the preceding text (now commented out).}

  Building on the observation that
  reverse-mode automatic differentiation (AD) ---
  a generalisation of backpropagation ---
  can naturally be expressed as pullbacks of differential 1-forms,
  we design a simple
  higher-order programming language
  with a first-class differential operator, and
  present a reduction strategy which
  exactly simulates reverse-mode AD.
  We justify our reduction strategy by interpreting
  our language
  in any differential $\lambda$-category that satisfies
  the Hahn-Banach Separation Theorem, and
  show that the reduction strategy
  precisely captures reverse-mode AD
  in a truly higher-order setting.

\end{abstract}

\maketitle





\addtolength{\jot}{-.4em}

\lo{- To hide comments and changed-highlighting, use {\tt final} option in {\tt documentclass}.

- To submit, choose line 4 in {\tt main.tex} (for double-blind review).}

\lo{Please spell check.}

\sloppy

\section{Introduction}

  \emph{Automatic differentiation} (AD) \cite{Wengert64} is widely considered the most efficient and accurate algorithm for computing derivatives, thanks largely to the chain rule.
  There are two modes of AD:
  \begin{itemize}
    \item \emph{Forward-mode AD}
  evaluates the chain rule from inputs to outputs;
  it has time complexity that scales with the number of inputs, and constant space complexity.
  \item \emph{Reverse-mode AD} --- a generalisation of backpropagation --- evaluates the chain rule (in dual form) from outputs to inputs;
  it has time complexity that scales with the number of outputs, and space complexity that scales with the number of intermediate variables.
  \end{itemize}

  In machine learning applications such as neural networks, the number of input parameters is usually considerably larger than the number of outputs.
  For this reason, reverse-mode AD has been the preferred method of differentiation, especially in deep learning applications.
  (See \citet{DBLP:journals/jmlr/BaydinPRS17} for an excellent survey of AD.)

  The only downside of reverse-mode AD is its rather involved definition,
  which has led to a variety of complicated implementations in
  neural networks.
  On the one hand,
  TensorFlow \cite{DBLP:conf/osdi/AbadiBCCDDDGIIK16} and
  Theano \cite{DBLP:journals/corr/Al-RfouAAa16}
  employ the \emph{define-and-run} approach
  where the model is constructed as a computational graph
  before execution.
  On the other hand,
  PyTorch \cite{DBLP:journals/corr/abs-1912-01703} and
  Autograd \cite{MaclaurinDA15}
  employ the \emph{define-by-run} approach
  where the computational graph is constructed dynamically during the execution.

  \emph{Can we replace the traditional graphical representation of reverse-mode AD by a simple yet expressive framework?}
  Indeed, there have been calls from the neural network community for the development of \emph{differentiable programming} \cite{Olah15,LeCun,Dalrymple},
  based on a higher-order functional language with a built-in differential operator that returns the derivative of a given program via reverse-mode AD.
  Such a language would free the programmer from implementational details of differentiation.
  Programmers would be able to concentrate on the construction of machine learning models, and train them by calling the built-in differential operator on the cost function of their models.

  The goal of this work is to present
  a simple higher-order programming language with
  an explicit differential operator,
  such that its reduction semantics is exactly reverse-mode AD, in a truly higher-order manner.

  The syntax of our language is inspied by \citet{DBLP:journals/tcs/EhrhardR03}'s
  differential $\lambda$-calculus, which is
  an extension of simply-typed $\lambda$-calculus
  with a differential operator that
  mimics standard symbolic differentiation
  (but not reverse-mode AD).
  Their definition of differentiation
  via a linear substitution
  provides a good foundation for our language.

  The reduction strategy of our language uses differential $\lambda$-category
  \cite{DBLP:journals/entcs/BucciarelliEM10}
  (the model of differential $\lambda$-calculus)
  as a guide.
  Differential $\lambda$-category is a Cartesian closed differential category \cite{blute2009cartesian},
  and hence enjoys the fundamental properties of derivatives, and behaves well with exponentials (curry).



  \paragraph{Contributions.}

  Our starting point (Section~\ref{sec:reverse-mode AD}) is the observation that the computation of reverse-mode AD can naturally be expressed as a transformation of \emph{pullbacks of differential 1-forms}.
  We argue that this viewpoint is essential for understanding reverse-mode AD in a functional setting.
  Standard reverse-mode AD (as presented in
  \cite{Bauer1974,DBLP:journals/jmlr/BaydinPRS17}) is only defined in Euclidean spaces.

      We present (in Section \ref{sec:dppl})
      a simple higher-order programming language,
      extending the simply-typed $\lambda$-calculus \cite{Church65} with
      an explicit differential operator called the \emph{pullback},
      $\pbmap{\lambda x.\pbterms}{\simterms}$, which
      serves as a reverse-mode AD simulator.

      Using differential $\lambda$-category
      \cite{DBLP:journals/entcs/BucciarelliEM10}
      as a guide,
      we design a reduction strategy for our language so that
      the reduction of the application,
      $\big(\pbmap{\lambda x.\pbterms}{(\lambda x.\dual{e_p})}\big)\,\simterms$,
      mimics reverse-mode AD in computing the
      $p$-th row of the Jacobian matrix (derivative) of the function
      $\lambda x.\pbterms$ at the point $\simterms$, where $e_p$ is the column vector
  with $1$ at the $p$-th position
  and $0$ everywhere else.
      Moreover, we show how our reduction semantics can be adapted to a
      continuation passing style evaluation (Section \ref{subsec:cps}).

      Owing to the higher-order nature of our language,
      standard differential calculus is not enough to model our language
      and hence cannot justify our reductions.
      Our final contribution
      (in Section \ref{sec:model}) is
      to show that any differential $\lambda$-category
      \cite{DBLP:journals/entcs/BucciarelliEM10}
      that satisfies the Hahn-Banach Separation Theorem
      is a model of our language
      (Theorem \ref{thm: correctness of reductions}).
      Our reduction semantics is faithful to reverse-mode AD, in that it is exactly reverse-mode AD when restricted to first-order; moreover we can
  perform reverse-mode AD on \emph{any}  higher-order abstraction,
  which may contain higher-order terms, duals, pullbacks, and free variables as subterms (Corollary \ref{cor:Reverse-mode AD is correct}).

  Finally, we discuss related works in Section \ref{sec:comparison} and
  conclusion and future directions in Section \ref{sec:conclusion}.

  Throughout this paper, we will point to the attached Appendix for additional content.
  All proofs are in Appendix \ref{appendix:proofs}, unless stated otherwise.

\section{Reverse-mode Automatic Differentiation}
  We introduce forward- and reverse-mode automatic differentiation (AD),
  highlighting their respective benefits in practice.
  Then we explain how reverse-mode AD can naturally be expressed as the pullback of differential 1-forms.
  {(The examples used to illustrate the above methods are collated in Figure~\ref{fig: AD example})}.

  \lo{Main changes to description of reverse-mode.

  Following the pattern of the description of forward-mode

  - first present computation of whole matrix

  - then present column-by-column computation

  - explain why it is efficient when n >> m}


  \subsection{Forward- and Reverse-mode AD}
  \label{subsec:technique or rAD}
  Recall that the Jacobian matrix
  of a smooth real-valued function $f:\Real^n \to \Real^m$
  at $x_0 \in \Real^n$ is
  $$
    \jacob(f)(x_0)
     :=
    \colvec{
      \mathdsubupat{f_1}{z_1}{x_0} & \mathdsubupat{f_1}{z_2}{x_0} & \dots  &
      \mathdsubupat{f_1}{z_n}{x_0} \\
      \mathdsubupat{f_2}{z_1}{x_0} & \mathdsubupat{f_2}{z_2}{x_0} & \dots  &
      \mathdsubupat{f_2}{z_n}{x_0} \\
      \vdots & \vdots & \ddots & \vdots \\
      \mathdsubupat{f_m}{z_1}{x_0} & \mathdsubupat{f_m}{z_2}{x_0} & \dots  &
      \mathdsubupat{f_m}{z_n}{x_0}
    }
  $$
  where $f_j := \pi_j \circ f : \Real^n \to \Real$.
  We call the function
  $\jacob:\smooth{\Real^n}{\Real^m} \to \smooth{\Real^n}{L(\Real^n,\Real^m)}$
  the \emph{Jacobian};
  \footnote{
    $\smooth{A}{B}$ is the set of all smooth functions from $A$ to $B$, and
    $L(A,B)$ is the set of all linear functions from $A$ to $B$,
    for Euclidean spaces $A$ and $B$.
  }
  $\jacob(f)$ the Jacobian of $f$;
  $\jacob(f)(x)$ the Jacobian of $f$ at $x$;
  $\jacob(f)(x)(v)$ the Jacobian of $f$ at $x$ along $v \in \Real^n$ and
  $\lambda x.\jacob(f)(x)(v)$ the Jacobian of $f$ along $v$.


  \paragraph{Symbolic Differentiation}
  {Numerical derivatives are standardly computed using \emph{symbolic differentiation}}:
  first compute
  $\mathdsubup{f_j}{z_i}$ for all $i,j$
  using rules (e.g.\ product and chain rules),
  then substitute $x_0$ for $z$
  to obtain $\jacob(f)(x_0)$.

  For example, to compute the Jacobian of
  $f: \anglebr{x,y} \mapsto \big((x + 1)(2x + y^2)\big)^2$
  at $\anglebr{1,3}$
  by symbolic differentiation,
  first compute
  $\mathdsubup{f}{x} = 2(x+1)(2x+y^2)(2x+y^2+2(x +1))$
  and
  $\mathdsubup{f}{y} = 2(x+1)(2x+y^2)(2y(x+1))$.
  Then, substitute $1$ for $x$ and $3$ for $y$ to obtain
  $
    \jacob(f)(\anglebr{1,3}) =
    \colvec{660 & 528}.
  $

  Symbolic differentiation is accurate but inefficient.
  Notice that the term $(x+1)$ appears twice in $\mathdsubup{f}{x}$,
  and $(1+1)$ is evaluated twice in $\mathdsubupat{f}{x}{1}$ (because
  for $h: \anglebr{x,y} \mapsto (x + 1)(2x + y^2)$,
  both $h(\anglebr{x,y})$ and $\mathdsubup{h}{x}$ contain the term $(x+1)$, and the product rule tells us to calculate them separately).
  This duplication is a cause of the so-called \emph{expression swell problem},
  resulting in exponential time-complexity.


  \paragraph{Automatic Differentiation}
  \emph{Automatic differentiation} (AD)
  avoids this problem by a simple divide-and-conquer approach:
  first arrange $f$ as a composite of elementary\footnote{in the sense of being easily differentiable} functions, $g_1,\dots, g_k$
  (\ie~$f = g_k \circ \dots \circ g_1$),
  then compute the Jacobian of each of these elementary functions,
  and finally combine them via the chain rule to yield the desired Jacobian of $f$.

  \paragraph{Forward-mode AD}
  Recall the chain rule:
  \[
    \jacob(f)(x_0)
    =
    \jacob(g_k)(x_{k-1}) \times
    \dots \times
    \jacob(g_2)(x_1) \times
    \jacob(g_1)(x_0)
  \]
  for $f = g_k \circ \dots \circ g_1$,
  where
  $x_i := g_i(x_{i-1})$.
  \emph{Forward-mode} AD computes the Jacobian matrix $\jacob(f)(x_0)$
  by calculating
  $\alpha_i := \jacob(g_i)(x_{i-1}) \times \alpha_{i-1}$
  and $x_i := g_i(x_{i-1})$,
  with $\alpha_0 := \idmat$ (identity matrix) and $x_0$.
  Then, $\alpha_k = \jacob(f)(x_0)$ is the Jacobian of $f$ at $x_0$.
  This computation can neatly be presented as an iteration of
  the {$\config{\cdot}{\cdot}$-reduction},
  $
    \config{x}{\alpha}
    \xrightarrow{g}
    \config{g(x)}{\jacob(g)(x) \times \alpha},
  $
  for $g = g_1, \dots, g_k$,
  starting from the pair $\config{x_0}{\idmat}$.
  Besides being easy to implement,
  {forward-mode AD computes the new pair from the current pair $\config{x}{\alpha}$, requiring no additional memory}.

  To compute the Jacobian of
  $f: \anglebr{x,y} \mapsto \big((x + 1)(2x + y^2)\big)^2$
  at $\anglebr{1,3}$ by
  forward-mode AD, first decompose $f$ into elementary functions as
  $
    \Real^2
      \xrightarrow{\ g\ }
    \Real^2
      \xrightarrow{\ *\ }
    \Real
      \xrightarrow{(-)^2}
    \Real,
  $
  where $g(\anglebr{x,y}) := \anglebr{x+1,2x+y^2}$.
  Then, starting from $\config{\anglebr{3,1}}{\idmat}$,
  {iterate the $\config{\cdot}{\cdot}$-reduction}
  \begin{align*}
    &
    \config{\anglebr{1,3}}{
      \colvec{
        1 & 0 \\
        0 & 1
      }
    }
    \xrightarrow{g}
    \config{\anglebr{\overset{1+1}{2},\overset{2*1+3^2}{11}}}{
      \colvec{
        1 & 0 \\
        2 & 6
      }
    }
    \xrightarrow{*}
    \config{\overset{2*11}{22}}{
      \colvec{
        15 & 12
      }
    } \\
    &
    \xrightarrow{(-)^2}
    \config{\overset{22^2}{484}}{
      \colvec{
        660 & 528
      }
    }
  \end{align*}
  yielding
  $\colvec{660 & 528}$
  as the Jacobian of $f$ at $\anglebr{1,3}$.
  Notice that $(1 + 1)$ is only evaluated once,
  even though its result is used in various calculations.

  {In practice, because storing the intermediate matrices $\alpha_i$ can be expensive},
  the matrix $\jacob(f)(x_0)$ is computed \emph{column-by-column},
  by simply changing the starting pair from $\config{x_0}{\idmat}$ to $\config{x_0}{e_p}$,
  where
  $e_p \in \Real^n$ is the column vector
  with $1$ at the $p$-th position
  and $0$ everywhere else.
  Then, the computation becomes a reduction of
  a vector-vector pair, and
  $\alpha_k = \jacob(f)(x_0) \times e_p$ is the
  $p$-th column of the Jacobian matrix $\jacob(f)(x_0)$.
  Since $\jacob(f)(x_0)$ is a $m$-by-$n$ matrix,
  $n$ {runs} are required to
  compute the whole Jacobian matrix.

  For example, if we start from
  $\config{\anglebr{1,3}}{\colvec{1\\0}}$,
  the reduction
  $$
    \config{\anglebr{1,3}}{\colvec{1\\0}}
    \xrightarrow{g}
    \config{\anglebr{2,11}}{
      \colvec{
        1 \\
        2
      }
    }
    \xrightarrow{*}
    \config{22}{
      \colvec{
        15
      }
    }
    \xrightarrow{(-)^2}
    \config{484}{
      \colvec{
        660
      }
    }
  $$
  gives us the first column of the Jacobian matrix $\jacob(f)(\anglebr{1,3})$.



  \paragraph{Reverse-mode AD}
  {By contrast, \emph{reverse-mode} AD computes the dual of the Jacobian matrix, $\dual{(\jacob(f)(x_0))}$, using the chain rule in \emph{dual (transpose) form}}
  \[
    \dual{(\jacob(f)(x_0))}
      =
    \dual{(\jacob(g_1)(x_0))} \times
    \dots \times
    \dual{(\jacob(g_k)(x_{k-1}))}
  \]
  as follows: first compute $x_i := g_i(x_{i-1})$ for $i = 1,\dots,k-1$ (Forward {Phase});
  then compute
  $\beta_i := (\jacob(g_i)(x_{i-1}))^* \times \beta_{i+1}$
  for $i = k,\dots,1$ with {$\beta_{k+1} := \idmat$}
  (Reverse Phase).

  For example, the reverse-mode AD computation on $f$ is as follows.
  $$
  \arraycolsep=1.4pt\def\arraystretch{1}
  \begin{array}{lccccccc}
    \text{Forward Phase: }
    &
    \anglebr{1,3}
    &
    \xrightarrow{g}
    &
    \anglebr{2,11}
    &
    \xrightarrow{*}
    &
    22
    &
    \xrightarrow{(-)^2}
    &
    484
    \\
    \text{Reverse Phase: }
    &
    {
      \colvec{
        660 \\
        528
      }
    }
    &
      \xleftarrow{g}
    &
    {
      \colvec{
        484 \\
        88
      }
    }
    &
      \xleftarrow{*}
    &
    {
      \colvec{
        44
      }
    }
    &
      \xleftarrow{(-)^2}
    &
    \idmat
  \end{array}
  $$

  {In practice, like forward-mode AD, the matrix $(\jacob(f)(x_0))^*$ is computed column-by-column, by simply setting $\beta_{k+1} := \pi_p$, where $\pi_p\in L(\Real^m,\Real)$ is the $p$-th projection.}
  Thus, {a run (comprising Forward and Reverse Phase)} computes $\dual{(\jacob(f)(x_0))}(\pi_p)$,
  the $p$-th row of the Jacobian of $f$ at $x_0$.
  It follows that $m$ runs are required to compute the $m$-by-$n$ Jacobian matrix.

  In many machine learning (e.g.~deep learning) problems,
  the functions $f : \Real^n \to \Real^m$ we need to differentiate have many more inputs than outputs, in the sense that $n \gg m$.
  Whenever this is the case, reverse-mode AD is more efficient than forward-mode.

  \begin{remark}
  {Unlike forward-mode AD, we cannot interleave the iteration of $x_i$ and
  the computation of $\beta_i$.
  In fact, according to \citet{Hoffmann2016},
  nobody knows how to do reverse-mode AD using pairs $\config{\cdot}{\cdot}$, as employed by forward-mode AD to great effect.
  In other words, reverse-mode AD does not seem presentable as an \emph{in-place} algorithm.}
  \end{remark}

  \subsection{Geometric Perspective of Reverse-mode AD}
  \label{subsec:explain pb}

  Reverse-mode AD can naturally be expressed using
  pullbacks and differential 1-forms,
  as alluded to by \citet{betancourt2018geometric}
  and discussed in \cite{PearlmutterLAFI19}.

  Let $E := \Real^n$ and $F := \Real^m$.
  A \emph{differential 1-form} of $E$ is
  a smooth map $\omega \in \smooth{E}{L(E,\Real)}$.
  Denote the set of all differential 1-forms of $E$ as $\oneform E$.
  E.g.~$\lambda x.\pi_p \in \oneform \, {\Real^m}$.
  (Henceforth, by \emph{1-form}, we mean differential 1-form.)
  The \emph{pullback} of a 1-form $\omega \in \oneform F$
  along a smooth map $f:E \to F$ is
  a 1-form $\oneform(f)(\omega) \in \oneform \, E$ where
  \Function
    {\oneform(f)(\omega)}{E}{L(E,\Real)}
    {x}{\dual{(\jacob(f)(x))} (\omega(f x))}

  Notice the result of an iteration of reverse-mode AD
  $\dual{(\jacob(f)(x_0))}(\pi_p)$
  can be expressed as $\oneform(f)(\lambda x.\pi_p)(x_0)$,
  which can be expanded to
  $\big(\oneform(g_1) \circ \dots \circ \oneform(g_k)\big)(\lambda x.\pi_p)(x_0)$.
  Hence, reverse-mode AD can be expressed as:
  {first iterate the reduction of 1-forms,
  $\omega \xrightarrow{g} \oneform(g)(\omega)$,
  for $g = g_k, \dots, g_1$,
  starting from the 1-form $\lambda x.\pi_p$;
  then compute $\omega_0(x_0)$, which yields the $p$-th row of $\jacob(f)(x_0)$.}

  Returning to our example,
  \begin{align*}
    &\oneform(f)(\lambda x.\colvec{1})(\anglebr{1,3}) \\
    & = \big(\oneform(g) \circ \oneform(*) \circ \oneform((-)^2)\big)
    (\lambda x.\dual{\colvec{1}})(\anglebr{1,3}) \\
    & =
    \dual{(\jacob(g)(\anglebr{1,3}))}\big(\oneform(*) \circ \oneform((-)^2)\big)
    (\lambda x.\dual{\colvec{1}})(\anglebr{2,11}) \\
    & =
    \dual{(\jacob(g)(\anglebr{1,3}))}\dual{(\jacob(*)(\anglebr{2,11}))}
    \big(\oneform((-)^2)\big)
    (\lambda x.\dual{\colvec{1}})(22) \\
    & =
    \dual{(\jacob(g)(\anglebr{1,3}))}\dual{(\jacob(*)(\anglebr{2,11}))}
    \dual{(\jacob((-)^2)(22))}
    \big((\lambda x.\dual{\colvec{1}})(484)\big) \\
    & =
    \dual{(\jacob(g)(\anglebr{1,3}))}\dual{(\jacob(*)(\anglebr{2,11}))}
    \dual{\colvec{44}} \\
    & =
    \dual{(\jacob(g)(\anglebr{1,3}))}\dual{\colvec{484\\88}} \\
    & =
    \dual{\colvec{660\\528}}
  \end{align*}
  which is the Jacobian $\jacob(f)(\anglebr{1,3})$.


  The pullback-of-1-forms perspective gives us a way to perform reverse-mode AD beyond Euclidean spaces
  (for example on the function $\mathsf{sum}:List(\Real)\to \Real$, which returns the sum of the elements of a list);
  and it shapes our language and reduction presented in Section \ref{sec:dppl}.
  (Example \ref{eg:running-sum-term}
  {shows} how $\mathsf{sum}$ can be defined in our language and
  Appendix \ref{appendix:sum example}
  {shows} how reverse-mode AD can be performed on $\mathsf{sum}$.)

  \begin{remark}
  Pullbacks can be generalised to arbitrary $p$-forms, using essentially the same approach.
  However the pullbacks of general $p$-forms no longer resemble reverse-mode AD as it is commonly understood.
  \end{remark}



\label{sec:reverse-mode AD}

\section{A Differential-form Pullback Programming Language}
\label{sec:dppl}

  \subsection{Syntax}
  \label{subsec:syntax}


  \renewcommand\vec[1]{\mathbf{#1}}

  \begin{figure}[t]
    \begin{align*}
      \text{Simple terms} & &
      \simterms
      & ::=
          x
      \mid \lambda x.\simterms
      \mid \simterms\,\pbterms
      \mid \pi_i({\simterms})
      \mid \anglebr{\simterms,\simterms}
      \mid \PCF{r}
      \mid \PCF{f}(\pbterms)
      \\
      & & & \st
          \jac{f}{\simterms}
      \mid \dualmap{(\lambda x.\simterms)}{\simterms}
      \mid \pbmap{(\lambda x.\pbterms)}{\simterms}
      \mid \dual{\PCF{\vec{r}}}
      \\
      \text{Pullback terms} & &
      \pbterms
      & ::=
          0
      \mid \simterms
      \mid \simterms + \pbterms
    \end{align*}
    \caption{
      Grammar of
      \emph{simple terms} $\simterms$ and
      \emph{pullback terms} $\pbterms$.
      Assume a collection $\variables$ of variables (typically $x,y,z,\omega$),
      and a collection $\easilydiff$ (typically $f, g, h$) of easily-differentiable real-valued functions,
      in the sense that the Jacobian of $f$,
      $\jacob(f)$, can be called by the language,
      $r$ and $\vec{r}$ range over $\Real$ and $\Real^n$ respectively.
    }
    \label{fig: simple and pullback terms}
  \end{figure}

  \begin{figure*}[t]
    \def\arraystretch{2}
    \begin{center}
      \small
      \begin{tabular}{c}
        $
          \sigma,\tau
          \ ::=\
              \PCFReal
          \mid \sigma_1\times\sigma_2
          \mid \sigma_1\Arrow\sigma_2
          \mid \dual{\sigma}
        $
        \\
        \AxiomC{\vphantom{$\langle$}}
        \UnaryInfC{$\Gamma \vdash 0:\sigma$}
        \DisplayProof
        $\hskip 1em$
        \AxiomC{$\Gamma \vdash \simterms :\sigma$}
        \AxiomC{$\Gamma \vdash \pbterms :\sigma$}
        \BinaryInfC{$\Gamma \vdash \simterms + \pbterms:\sigma$}
        \DisplayProof
        $\hskip 1em$
        \AxiomC{\vphantom{$\langle$}}
        \UnaryInfC{$\Gamma\cup\set{x:\sigma} \vdash x:\sigma$}
        \DisplayProof
        $\hskip 1em$
        \AxiomC{$\Gamma\cup\{x:\sigma\} \vdash \simterms:\tau$}
        \UnaryInfC{$\Gamma \vdash \lambda x.\simterms:\sigma\Arrow\tau$}
        \DisplayProof
        $\hskip 1em$
        \AxiomC{$\Gamma \vdash \simterms:\sigma\Arrow\tau$}
        \AxiomC{$\Gamma \vdash \pbterms:\sigma$}
        \BinaryInfC{$\Gamma \vdash \simterms\,\pbterms:\tau$}
        \DisplayProof
        $\hskip 1em$
        \\
        \AxiomC{$\Gamma \vdash \simterms :\sigma_1 \times \sigma_2$}
        \UnaryInfC{$\Gamma \vdash \pi_i (\simterms):\sigma_i$}
        \DisplayProof
        $\hskip 1em$
        \AxiomC{$\Gamma \vdash \simterms_1 :\sigma_1$}
        \AxiomC{$\Gamma \vdash \simterms_2 :\sigma_2$}
        \BinaryInfC{$\Gamma \vdash \anglebr{\simterms_1,\simterms_2}:\sigma_1\times\sigma_2$}
        \DisplayProof
        $\hskip 1em$
        \AxiomC{$r \in \Real$}
        \UnaryInfC{$\Gamma \vdash \PCF{r}:\PCFReal$}
        \DisplayProof
        $\hskip 1em$
        \AxiomC{$\Gamma \vdash \pbterms:\PCFReal^n$}
        \UnaryInfC{$\Gamma \vdash \PCF{f}(\pbterms):\PCFReal^m$}
        \DisplayProof
        $\hskip 1em$
        \AxiomC{$\Gamma \vdash \simterms:\PCFReal^n$}
        \UnaryInfC{$\Gamma \vdash \jac{f}{\simterms}:\PCFReal^n \Arrow\PCFReal^m$}
        \DisplayProof
        $\hskip 1em$
        \AxiomC{$\vec{r} \in \Real^n$}
        \UnaryInfC{$\Gamma \vdash \dual{\PCF{\vec{r}}}:\dual{\PCFReal^n}$}
        \DisplayProof
        $\hskip 1em$
        \\
        \AxiomC{$\Gamma\cup\{x:\sigma\} \vdash \simterms_1:\tau$}
        \AxiomC{$\Gamma \vdash \simterms_2:\dual{\tau}$}
        \AxiomC{$x \in \lin{\simterms_1}$}
        \TrinaryInfC{$\Gamma \vdash \dualmap{(\lambda x.\simterms_1)}{\simterms_2}:\dual{\sigma}$}
        \DisplayProof
        $\hskip 1em$
        \AxiomC{$\Gamma \cup \set{x:\sigma} \vdash \pbterms: \tau$}
        \AxiomC{$\Gamma \vdash \simterms: \oneform{\tau}$}
        \BinaryInfC{$\Gamma \vdash \pbmap{\lambda x.\pbterms}{\simterms}:\oneform{\sigma}$}
        \DisplayProof
        $\hskip 1em$
      \end{tabular}
    \end{center}
    \caption{
      The types and typing rules for DPPL.
      $\oneform{\sigma}\equiv\sigma\Arrow\dual{\sigma}$ and
      $f:\Real^n \to \Real^m$ is easily differentiable, \ie $f \in \easilydiff$.
    }
    \label{fig: typing rules}
  \end{figure*}

  Figure~\ref{fig: simple and pullback terms}
  presents the grammar of simple terms $\simterms$ and pullback terms $\pbterms$,
  and Figure~\ref{fig: typing rules}
  presents the type system.
  While the definition of simple terms $\simterms$
  is relatively standard
  (except for the new constructs which will be discussed later),
  the definition of pullback terms $\pbterms$
  as sums of simple terms is not.

  \subsubsection{Sum and Linearity}

  The idea of sum is important
  since it specifies the ``linear positions''
  in a simple term,
  just as it specifies the algebraic notion of linearity in Mathematics.
  For example, $x(y+z)$ is a term but $(x+y)z$ is not.
  This is because $(x+y)z$ is the same as $xz + yz$,
  but $x(y+z)$ cannot.
  Hence in $\simterms\,\pbterms$,
  $\simterms$ is in a linear position but not $\pbterms$.
  Similarly, in Mathematics
  $(f_1+f_2)(x_1) = f_1(x_1) + f_2(x_1)$
  but
  in general
  $f_1(x_1 + x_2) \not= f_1(x_1) + f_1(x_2)$
  for smooth functions $f_1,f_2$ and $x_1,x_2$.
  Hence, the function $f$ in an application $f(x)$ is in a linear position
  while the argument $x$ is not.

  Formally we define the set $\lin{\simterms}$ of \emph{linear variables} in a simple term $\simterms$ by
  $y \in \lin{\simterms}$
  if, and only if,
  $y$ is in a linear position in $\simterms$.
  \begin{align*}
    \lin{x} & :=
      \set{x} \\
    \lin{\lambda x.\simterms} & :=
      \lin{\simterms}\setminus \set{x} \\
    \lin{\simterms\,\pbterms} & :=
      \lin{\simterms} \setminus \freevar{\pbterms} \\
    \lin{\pi_i{(\simterms)}} & :=
      \lin{\simterms} \\
    \lin{\anglebr{\simterms_1,\simterms_2}} & :=
      \lin{\simterms_1} \cap \lin{\simterms_2}\\
    \lin{\jac{f}{\simterms}} & :=
      \lin{\simterms} \\
    \lin{\dualmap{(\lambda x.\simterms_1)}{\simterms_2}} & :=
      \big(\lin{\simterms_1}\setminus\freevar{\simterms_2}\big) \cup
      \big(\lin{\simterms_2}\setminus\freevar{\simterms_1}\big)  \\
    \lin{\simterms} & :=
      \varnothing \qquad\qquad\qquad\qquad \text{otherwise.}
  \end{align*}
  For example, $\lin{x \, z \, (y \, z)} = \set{x}$.

  \subsubsection{Dual Type, Jacobian, Dual Map and Pullback}


  Any term of the \emph{dual type} $\dual{\sigma}$
  is considered a linear functional of $\sigma$.
  For example, $\dual{\PCF{e_p}}$ has the dual type $\dual{\PCFReal^n}$.
  Then the term $\dual{\PCF{e_p}}$ mimics the linear functional
  $\pi_p \in L(\Real^n,\Real)$.

  The \emph{Jacobian} $\jac{f}{\simterms}$
  is considered as the Jacobian of $f$ along $\simterms$,
  which is a smooth function.
  For example,
  let $f:\Real^m \to \Real^n$ be ``easily differentiable'',
  then
  $\jac{f}{v}$
  mimics the Jacobian along $v$, \ie
  the function $\lambda x.\jacob(f)(x)(v)$.

  The \emph{dual map} $\dualmap{(\lambda x.\simterms_1)}{\simterms_2}$
  is considered the dual of
  the linear functional $\simterms_2$
  along the function $\lambda x.\simterms_1$,
  where $x \in \lin{\simterms_1}$.
  For example,
  let $\vec{r} \in \Real^m$.
  The dual map
  $\dualmap{(\lambda v.(\jac{f}{v})\,\vec{r})}{\dual{\PCF{e_p}}}$ mimics
  $(\jacob(f)(\vec{r}))^*(\pi_p) \in L(\Real^m,\Real)$, which is
  the dual of $\pi_p$
  along the Jacobian $\jacob(f)(\vec{r})$.

  The \emph{pullback} $\pbmap{\lambda x.\pbterms}{\simterms}$ is considered the pullback of the 1-form $\simterms$
  along the function $\lambda x.\pbterms$.
  For example,
  $\pbmap{\lambda x.\PCF{f}(x)}{(\lambda x.\dual{\PCF{e_p}})}$
  mimics $\Omega(f)(\lambda x.\pi_p) \in \Omega(\Real^m)$, which is
  the pullback of
  the 1-form $\lambda x.\pi_p$
  along $f$.

  Hence, to perform reverse-mode AD on a term
  $\lambda x.\pbterms$ at $\pbterms'$
  with respect to $\omega$,
  we consider the term
  $\big(\pbmap{\lambda x.\pbterms}{\omega}\big)\pbterms'$.

  \subsubsection{Notations}

  We use syntactic sugars to ease writing.
  For $n \geq 1$ and $z$ a fresh variable.
  \begin{align*}
    \PCFReal^{n+1} & \equiv \PCFReal^n \times \PCFReal &
    \Omega\sigma & \equiv \sigma\Arrow\dual{\sigma} \\
    \PCF{\colvec{r_1\\\vdots\\r_n}} & \equiv \anglebr{\PCF{r_1},\dots,\PCF{r_n}} &
    \anglebr{\pbterms_1,\pbterms_2,\pbterms_3} & \equiv
    \anglebr{\anglebr{\pbterms_1,\pbterms_2},\pbterms_3} \\
    \proj{\simterms}{i} & \equiv \pi_i(\simterms) &
    \Let{x=t}{s} & \equiv (\lambda x.s)\,t \\
    \pbof{\vec{r}} & \equiv \lambda x.\dual{\PCF{\vec{r}}} &
    \lambda \anglebr{x,y}.\simterms & \equiv
    \lambda z.\simterms[\proj{z}{1}/x][\proj{z}{2}/y]
  \end{align*}

  Capture-free substitution is applied recursively, e.g.\
  $
    \big(\dualmap{(\lambda x.\simterms_1)}{\simterms_2}\big) [{\pbterms'}/z]
    \equiv \dualmap{(\lambda x.\simterms_1[{\pbterms'}/z])}{(\simterms_2[{\pbterms'}/z])}
  $ and
  $
    \big(\pbmap{\lambda x.\pbterms}{\simterms}\big) [{\pbterms'}/z]
    \equiv \pbmap{(\lambda x.\pbterms[{\pbterms'}/z])}{(\simterms[{\pbterms'}/z])}.
  $
  We treat $0$ as the unit of our sum terms, \ie
  $0 \equiv 0 + 0$, $\simterms \equiv 0 + \simterms$ and
  $\simterms \equiv \simterms + 0$; and
  consider $+$ as a associative and commutative operator.
  We also define
  $\simterms[\simterms_1 + \simterms_2/y]
  \equiv \simterms[\simterms_1/y] + \simterms[\simterms_2/y]$
  if and only if
  $y \in \lin{\simterms}$.
  For example,
  $(\simterms_1 + \simterms_2)\,\pbterms
  \equiv \simterms_1\,\pbterms + \simterms_2\,\pbterms$.


  We finish this subsection with some examples
  that can be expressed in this language.

  \begin{example}
    \label{eg:running-ex-term}
    Consider the running example
    in computing the Jacobian of
    $f: \anglebr{x,y} \mapsto \big((x + 1)(2x + y^2)\big)^2$
    at $\anglebr{1,3}$.
    Assume $g(\anglebr{x,y}) := \anglebr{x+1,2x+y^2}$,
    $\mathsf{mult}$ and $\mathsf{pow2}$
    are in the set of easily differentiable functions, \ie
    $g, \mathsf{mult}, \mathsf{pow2} \in \easilydiff$.
    The function $f$ can be presented by the term
    $
      \set{\anglebr{x,y}:\PCFReal^2} \vdash
      \PCF{\mathsf{pow2}}(\PCF{\mathsf{mult}}(\PCF{g}(\anglebr{x,y}))): \PCFReal.
    $
    More interestingly,
    the Jacobian of $f$ at $\anglebr{1,3}$, \ie $\jacob(f)(\anglebr{1,3})$,
    can be presented by the term
    $$
      \vdash
      \pbat{\big(
        \pbmap{\lambda \anglebr{x,y}.
          \PCF{\mathsf{pow2}}(\PCF{\mathsf{mult}}(\PCF{g}(\anglebr{x,y})))}
        {(\pbof{\colvec{1}})}
      \big)}{\anglebr{\PCF{1},\PCF{3}}}
      : \dual{\PCFReal^2}.
    $$
    This is the application of
    the pullback
    $\Omega(f)(\lambda x.\dual{\colvec{1}})$
    to the point $\anglebr{1,3}$,
    which we saw in Subsection~\ref{subsec:explain pb} is
    the Jacobian of $f$ at $\anglebr{1,3}$.
  \end{example}

  \begin{example}
    \label{eg:running-sum-term}
    Consider the function that takes a list of real numbers and
    returns the sum of the elements of a list.
    Using the standard Church encoding of List, \ie
    $List(X) \equiv (X \to D \to D) \to (D \to D)$, and
    $[x_1,x_2,\dots, x_n] \equiv
    \lambda fd.f\,x_n\big(\dots (f\,x_2\,(f\,x_1\,d))\big)$
    for some dummy type $D$,
    $\mathsf{sum}:List(\PCFReal)\to \PCFReal$ is defined to be
    $\lambda l. l\,(\lambda xy.x+y)\,\PCF{0}$.
    Hence the Jacobian of $\mathsf{sum}$ at a list
    $[\PCF{7},\PCF{-1}]$ can be expressed as
    $$
      \set{\omega:\oneform(List(\PCFReal))} \vdash
      \big(\pbmap{(\mathsf{sum})}{\omega}\big)\,[\PCF{7},\PCF{-1}]:
      \dual{\PCFReal}.
    $$
  \end{example}

  Now the question is how we could perform reverse-mode AD on this term.
  Recall
  the result of a reverse-mode AD on a function $f:\Real^n\to\Real^m$ at
  $x \in \Real^n$, \ie
  the $p$-th row of the Jacobian matrix of $f$ at $x$,
  can be expressed as
  $\oneform(f)(\lambda x.\pi_p)(x)$,
  which is
  $(\jacob(f)(x))^*((\lambda x.\pi_p) (fx)) = (\jacob(f)(x))^*\times \pi_p$.

  In the rest of this Section,
  we consider how the term
  $\pbat{(\pbmap{\lambda y.\pbterms'}{\omega})}{\pbterms}$,
  which mimics
  $\oneform(f)(\omega)(x)$,
  can be reduced.
  To avoid expression swell, we first perform A-reduction:
  $\pbterms' \Aredplus \letterms$
  which decompose a term into a series of ``smaller'' terms,
  as explained in Subsection~\ref{subsec:A-reduction}.
  Then, we reduce
  $\pbat{(\pbmap{\lambda y.\letterms}{\omega})}{\pbterms}$
  by induction on $\letterms$,
  as explained in Subsection~\ref{subsec:pb reduction}.
  Lastly,
  we complete our reduction strategy in Subsection~\ref{subsec:combine reduction}.

  We use the term in Example \ref{eg:running-ex-term} as a running example
  in our reduction strategy to illustrate that
  this reduction is faithful to reverse-mode AD
  (in that it is exactly reverse-mode AD when restricted to first-order).
  The reduction of the term in Example \ref{eg:running-sum-term}
  is given in Appendix \ref{appendix:sum example}.
  It illustrates how reverse-mode AD can be performed on a higher-order function.


  \subsection{Divide: Administrative Reduction}
  \label{subsec:A-reduction}

  We use the administrative reduction (A-reduction) of \citet{DBLP:conf/lfp/SabryF92}
  to decompose a pullback term $\pbterms$ into
  a let series $\letterms$ of elementary terms, \ie
  \[
    \pbterms
    \Aredplus
    \Let{
      x_1 = \elemterms;\
      \dots;\
      x_n = \elemterms
    }{x_n},
  \]
  where
  \emph{elementary terms} $\elemterms$
  and \emph{let series} $\letterms$ are defined as
  \begin{align*}
    \elemterms & ::=
        0
    \mid z_1 + z_2
    \mid z
    \mid \lambda x.\letterms
    \mid z_1\,z_2
    \mid {z}_i
    \mid \anglebr{z_1,z_2}
    \mid \PCF{r}
    \mid \PCF{f}(z)
    \\
    & \st
        \jac{f}{z}
    \mid \dualmap{(\lambda x.\letterms)}{z}
    \mid \pbmap{\lambda x.\letterms}{z}
    \mid \dual{\PCF{\vec{r}}}
    \\
    \letterms & ::=
      \Let{z = \elemterms}{\letterms} \st
      \Let{z = \elemterms}{z}.
  \end{align*}
  Note that elementary terms $\elemterms$ should be ``fine enough''
  to avoid expression swell.
  The complete set of A-reductions on $\pbterms$ can be found in
  Appendix \ref{appendix:administrative reduction}.
  We write $\Aredplus$ for the
  reflexive and transitive closure of $\Ared$.


  \begin{example}
    \label{eg:running-ex-A-red}
    We decompose the term
    considered in Example \ref{eg:running-ex-term},
    $\PCF{\mathsf{pow2}}(\PCF{\mathsf{mult}}(\PCF{g}(\anglebr{x,y})))$,
    via administrative reduction.
    $$
      \PCF{\mathsf{pow2}}(\PCF{\mathsf{mult}}(\PCF{g}(\anglebr{x,y})))
      \Aredplus
      \letarray{
        z_1 & \mkern-14mu = \anglebr{x,y}; \\
        z_2 & \mkern-14mu = \PCF{g}(z_1); \\
        z_3 & \mkern-14mu = \PCF{\mathsf{mult}}(z_2); \\
        z_4 & \mkern-14mu = \PCF{\mathsf{pow2}}(z_3)
      }{z_4.}
    $$
    This is reminiscent {of} the {decomposition} of $f$ into
    $
      \Real^2
        \xrightarrow{\ g\ }
      \Real^2
        \xrightarrow{\ *\ }
      \Real
        \xrightarrow{(-)^2}
      \Real
    $
    before performing AD.
  \end{example}


  \subsection{Conquer: Pullback Reduction}
  \label{subsec:pb reduction}

  \begin{figure*}
    \begin{align*}
      \text{Let Series:} & &
      \pbmap{(\lambda y.\Let{x=\elemterms}{x})}{\omega}
        & \Predtext{\hypertarget{eq:7}{(7)}}
        \pbmap{\lambda y.\elemterms}{\omega}
      \\
      & &
      \pbmap{(\lambda y.\Let{x=\elemterms}{\letterms})}{\omega}
        & \Predtext{\hypertarget{eq:8}{(8)}}
        \pbmap
          {\lambda y. \anglebr{y,\elemterms}}
          {\big(\pbmap{\lambda \anglebr{y,x}.\letterms}{\omega}\big)}
      \\
      \text{Constant Functions:} & &
      \pbat{\big(\pbmap{\lambda y.\elemterms}{\omega}\big)}{\values}
        & \Predtext{\hypertarget{eq:9}{(9)}} 0
        \quad \text{for } y \not\in \freevar{\elemterms}.
      \\
      \text{Linear Functions:} & &
      \pbat{\big(\pbmap{\lambda y.z+\proj{y}{j}}{\omega}\big)}{\values}
        & \Predtext{\hypertarget{eq:10a}{(10a)}}
        \dualmap
          {(\lambda v.\proj{v}{j})}
          {
            \big(\pbat{\omega}{(z + \proj{\values}{j})}\big)
          } \\
      & &
      \pbat{\big(\pbmap{\lambda y.\proj{y}{i}+\proj{y}{j}}{\omega}\big)}{\values}
        & \Predtext{\hypertarget{eq:10b}{(10b)}}
        \dualmap
          {(\lambda v.\proj{v}{i}+\proj{v}{j})}
          {
            \big(\pbat{\omega}{(\proj{\values}{i} + \proj{\values}{j})}\big)
          } \\
      & &
      \pbat{\big(\pbmap{\lambda y.y}{\omega}\big)}{\values}
        & \Predtext{\hypertarget{eq:11}{(11)}}
        \dualmap{(\lambda v.v)}{(\pbat{\omega}{\values})} \\
      & &
      \pbat{\big(\pbmap{\lambda y.\proj{y}{i}}{\omega}\big)}{\values}
        & \Predtext{\hypertarget{eq:12}{(12)}}
        \dualmap{(\lambda v.\proj{v}{i})}{(\pbat{\omega}{\proj{\values}{i}})} \\
      & &
      \pbat{\big(
        \pbmap{\lambda y. \anglebr{\proj{y}{i},z}}{\omega}
      \big)}{\values}
        & \Predtext{\hypertarget{eq:13a}{(13a)}}
        \dualmap
          {(\lambda v.\anglebr{\proj{v}{i},0})}
          {
            (\pbat{\omega}{\anglebr{\proj{\values}{i},z})}
          } \\
      & &
      \pbat{\big(
        \pbmap{\lambda y. \anglebr{z,\proj{y}{j}}}{\omega}
      \big)}{\values}
        & \Predtext{\hypertarget{eq:13b}{(13b)}}
        \dualmap
          {(\lambda v.\anglebr{0,\proj{v}{j}})}
          {
            (\pbat{\omega}{\anglebr{z,\proj{\values}{j}})}
          } \\
      & &
      \pbat{\big(
        \pbmap{\lambda y. \anglebr{\proj{y}{i},\proj{y}{j}}}{\omega}
      \big)}{\values}
        & \Predtext{\hypertarget{eq:13c}{(13c)}}
        \dualmap
          {(\lambda v.\anglebr{\proj{v}{i},\proj{v}{j}})}
          {
            (\pbat{\omega}{\anglebr{\proj{\values}{i},\proj{\values}{j}})}
          } \\
      & &
      \pbat{\big(
        \pbmap{\lambda y. \jac{f}{\proj{y}{i}}}{\omega}
      \big)}{\values}
        & \Predtext{\hypertarget{eq:14}{(14)}}
        \dualmap{(\lambda v.
          \jac{f}{\proj{v}{i}}
        )}{\big(\pbat{\omega}{(\jac{f}{\proj{\values}{i}})}\big)}
      \\
      \text{Function Symbols:} &&
      \pbat{\big(\pbmap{\lambda y. \PCF{f}(\proj{y}{i})}{\omega}\big)}{\values}
        & \Predtext{\hypertarget{eq:15}{(15)}}
        \dualmap
          {(\lambda v.(\jac{f}{\proj{v}{i}})\proj{\values}{i})}
          {\big(\pbat{\omega}{(\PCF{f}{(\proj{\values}{i})})}\big)}
      \\
      \text{Dual Maps:} &&
        \pbat{\Big(\pbmap{\lambda y.
          \dualmap{(\lambda x.\letterms)}{\proj{y}{i}}
        }{\omega}\Big)}{\values}
        & \Predtext{\hypertarget{eq:16a}{(16a)}}
          \dualmap{(
            \lambda v.
              \dualmap{(\lambda x.\letterms)}{\proj{v}{i}}
          )}{\big(
            \pbat{\omega}{(
              \dualmap{(\lambda x.\letterms)}{\proj{\values}{i}}
            )}
          \big)}
          \qquad \text{if }y \not\in \freevar{\lambda x.\letterms} \\
      &&
      \omit\rlap{
      \AxiomC{$
        \pbat{\big(\pbmap{\lambda y.\letterms}{\omega'}\big)}{\values}
        \Predplus
        \dualmap{(\lambda v.\simterms)}{\pbat{\omega'}{\values'}}
      $}
      \AxiomC{$y \not\in \freevar{z}$}
      \BinaryInfC{$
        \pbat{\big(
          \pbmap{\lambda y.\dualmap{(\lambda x.\letterms)}{z}}{\omega}
        \big)}{\values}
        \Predtext{\hypertarget{eq:16b}{(16b)}}
        \dualmap{\big(
          \lambda v. \dualmap{(\lambda x.\simterms)}{z}
        \big)}{
          \pbat{\omega}{\big(
            \dualmap{(\lambda x.\letterms[\values/y])}{z}
          \big)}
        }
      $}
      \DisplayProof
      } \\
      &&
      \omit\rlap{
      \AxiomC{$
        \pbat{\big(\pbmap{\lambda y.\letterms}{\omega'}\big)}{\values}
        \Predplus
        \dualmap{(\lambda v.\simterms)}{\pbat{\omega'}{\values'}}
      $}
      \UnaryInfC{$\begin{array}{l}
        \pbat{\big(\pbmap{\lambda y.
          \dualmap{(\lambda x.\letterms)}{\proj{y}{i}}
        }{\omega}\big)}{\values}
        \Predtext{\hypertarget{eq:16c}{(16c)}} \\[0.5mm]
        \qquad\quad \dualmap{\big(
          \lambda v.
            \dualmap{(\lambda x.\letterms[\values/y])}{\proj{v}{i}} +
            \dualmap{(\lambda x.\simterms)}{\values_{\pi i}}
        \big)}{\big(
          \pbat{\omega}{\big(
            \dualmap{(\lambda x.\letterms[\values/y])}{\values_{\pi i}}
          \big)}
        \big)}
      \end{array}$}
      \DisplayProof
      } \\
      \text{Pullback Terms:} &&
      \omit\rlap{
        \AxiomC{$
          \pbat{\big(\pbmap{\lambda x.\letterms}{z})}{a}
          \Predplus
          \dualmap{(\lambda v.\simterms)}{(\pbat{z}{\letterms[a/x]})}
        $}
        \UnaryInfC{$
          \pbat{\big(\pbmap{
            \lambda y. \pbmap{\lambda x.\letterms}{z}
          }{\omega}\big)}{\values}
          \Predtext{\hypertarget{eq:17}{(17)}}
          \pbat{\big(\pbmap{\lambda y. \lambda a.
            \dualmap
              {(\lambda v.\simterms)}
              {(\pbat{z}{\letterms[a/x]})}
          }{\omega}\big)}{\values}
        $}
        \DisplayProof
      } \\
      \text{Abstraction:} &&
      \omit\rlap{
        \AxiomC{$
          \pbat{\big(\pbmap{\lambda y.\letterms}{\omega'}\big)}{\values}
          \Predplus
          \dualmap{(\lambda v.\simterms)}{(\pbat{\omega'}{\letterms[\values/y]})}
        $}
        \AxiomC{$x \not\in \freevar{\values}$}
        \BinaryInfC{$
          \pbat{\big(\pbmap{\lambda y.\lambda x.\letterms}{\omega}\big)}{\values}
          \Predtext{\hypertarget{eq:18}{(18)}}
          \dualmap{(
            \lambda v.\lambda x.\simterms
          )}{\big(
            \pbat{\omega}{(\lambda x.\letterms[\values/y])}
          \big)}
        $}
        \DisplayProof
      }
      \\
      \text{Application:} &&
      \pbat
        {\big(\pbmap{\lambda y.\proj{y}{i}\,z}{\omega}\big)}
        {\values}
      & \Predtext{\hypertarget{eq:19a}{(19a)}}
        \dualmap
          {(\lambda v.\proj{v}{i}\,z)}
          {(\pbat{\omega}{(\proj{\values}{i}\,z)})}
      \\
      &&
      \omit\rlap{
        \AxiomC{$
          \pbat{\big(\pbmap{\lambda z.\pbterms'}{\omega'}\big)}{\proj{\values}{j}}
          \Pred
          \dualmap{(\lambda v'.\simterms')}{\pbat{\omega'}{(
            \pbterms'[\proj{\values}{j}/z]}
          )}
        $}
        \AxiomC{$\proj{\values}{i} \equiv \lambda z.\pbterms'$}
        \BinaryInfC{$
          \pbat{\big(
            \pbmap{\lambda y.\proj{y}{i}\,\proj{y}{j}}{\omega}
          \big)}{\values}
          \Predtext{\hypertarget{eq:19b}{(19b)}}
          \dualmap{(\lambda v.
            \proj{v}{i}\,\proj{\values}{j} +
            \simterms'[\proj{v}{j}/v']
          )}{\big(
            \pbat{\omega}{(\proj{\values}{i}\,\proj{\values}{j})}
          \big)}
        $}
        \DisplayProof
      } \\
      &&
      \omit\rlap{
        \AxiomC{$
          \pbat{\big(\pbmap{\lambda z.\values'}{\omega'}\big)}{\proj{\values}{j}}
          \Pred 0
        $}
        \AxiomC{$\proj{\values}{i} \equiv \lambda z.\values'$}
        \BinaryInfC{$
          \pbat{\big(
            \pbmap{\lambda y.\proj{y}{i}\,\proj{y}{j}}{\omega}
          \big)}{\values}
          \Predtext{\hypertarget{eq:19c}{(19c)}}
          \dualmap{(\lambda v.
            \proj{v}{i}\,\proj{\values}{j}
          )}{\big(
            \pbat{\omega}{(\proj{\values}{i}\,\proj{\values}{j})}
          \big)}
        $}
        \DisplayProof
      }
      \\
      \text{Pair:} &&
      \omit\rlap{
        \AxiomC{$
          \pbat{\big(\pbmap{\lambda y.\elemterms}{\omega'}\big)}{\values}
          \Pred
          \dualmap{(\lambda v.\simterms)}
          {(\pbat{\omega'}{(\elemterms[\values/y])})}
        $}
        \AxiomC{$y \in \freevar{\elemterms}$}
        \BinaryInfC{$
          \pbat
            {\big(\pbmap{\lambda y.\anglebr{y,\elemterms}}{\omega}\big)}
            {\values}
          \Predtext{\hypertarget{eq:20a}{(20a)}}
          \dualmap
            {(\lambda v.\anglebr{v, \simterms})}
            {(\pbat{\omega}{\anglebr{\values,\elemterms[\values/y]}})}
        $}
        \DisplayProof
      } \\
      &&
      \pbat
        {\big(\pbmap{\lambda y.\anglebr{y,\elemterms}}{\omega}\big)}
        {\values}
      & \Predtext{\hypertarget{eq:20b}{(20b)}}
      \dualmap
        {(\lambda v.\anglebr{v, 0})}
        {(\pbat{\omega}{\anglebr{\values,\elemterms }})}
      \qquad \text{for }y \not\in \freevar{\elemterms}
    \end{align*}
    \caption{Pullback Reductions}
    \label{fig:pullback reductions}
  \end{figure*}



  \subsubsection{Let Series}


  After decomposing
  $\pbterms'$ to a let series $\letterms$ of elementary terms
  via A-reductions
  in
  $\pbmap{\lambda y.\pbterms'}{\omega}$,
  we reduce
  $\pbmap{\lambda y.\letterms}{\omega}$
  by induction on $\letterms$
  as shown in Figure \ref{fig:pullback reductions} (Let series).
  Reduction $\hyperlink{eq:7}{7}$ is the base case and
  reduction $\hyperlink{eq:8}{8}$ expresses
  the contra-variant property of pullbacks.

  \begin{example}
    \label{eg:running-ex-omega-split}
    Take $\pbmap{\lambda \anglebr{x,y}.
      \PCF{\mathsf{pow2}}(\PCF{\mathsf{mult}}(\PCF{g}(\anglebr{x,y})))}
    {(\pbof{\colvec{1}})}$
    discussed in Example \ref{eg:running-ex-term}, as
    when applied to the point $\anglebr{1,3}$ is
    the Jacobian $\jacob(f)(\anglebr{1,3})$
    where $f(\anglebr{x,y}) := \big((x + 1)(2x + y^2)\big)^2$.
    In Example \ref{eg:running-ex-A-red}, we showed that
    $\PCF{\mathsf{pow2}}(\PCF{\mathsf{mult}}(\PCF{g}(\anglebr{x,y})))$
    is A-reduced to a let series $\letterms$.
    Now via reduction $\hyperlink{eq:7}{7}$ and $\hyperlink{eq:8}{8}$,
    $\pbmap{\lambda \anglebr{x,y}.
      \letterms
    }{\omega}$
    is reduced to a series of pullback along elementary terms.
    \begin{align*}
      & \pbmap{\lambda \anglebr{x,y}.
        \letarray{
          z_1 & \mkern-14mu = \anglebr{x,y}; \\
          z_2 & \mkern-14mu = \PCF{g}(z_1); \\
          z_3 & \mkern-14mu = \PCF{\mathsf{mult}}(z_2); \\
          z_4 & \mkern-14mu = \PCF{\mathsf{pow2}}(z_3)
        }{z_4}
      }{\omega} \\
      & \Predplus
        \mat{
          \pbmap{\lambda \anglebr{x,y}.\anglebr{\anglebr{x,y},\anglebr{x,y}}}{} \\
          \pbmap{\lambda \anglebr{\anglebr{x,y},z_1}.
            \anglebr{\anglebr{x,y},z_1,\PCF{g}(z_1)}}{} \\
          \pbmap{\lambda \anglebr{\anglebr{x,y},z_1,z_2}.
            \anglebr{\anglebr{x,y},z_1,z_2,\PCF{\mathsf{mult}}(z_2)}}{} \\
          \pbmap{\lambda \anglebr{\anglebr{x,y},z_1,z_2,z_3}.
            \PCF{\mathsf{pow2}}(z_3)
          }{\omega}
        }
    \end{align*}
  \end{example}

  Via A-reductions and
  reductions $\hyperlink{eq:7}{7}$ and $\hyperlink{eq:8}{8}$,
  $\pbmap{\lambda y.\pbterms'}{\omega}$ is reduced to
  a series of pullback along elementary terms
  $\pbmap{\lambda y.\elemterms_1}{(\dots (\pbmap{\lambda y.\elemterms_n}{\omega}))}$.
  Now, we define the reduction of
  pullback along elementary terms when applied to a value\footnote{
    A value is a normal form of the reduction strategy.
    Its definition will be made precise in the next subsection.
  } $\values$,
  \ie $\pbat{(\pbmap{\lambda y.\elemterms}{\omega})}{\values}$.

  Recall the pullback
  of a 1-form $\omega \in \Omega(F)$
  along a smooth function $f:E \to F$
  is defined to be
  $$
    \oneform(f)(\omega): \quad
    x \longmapsto (\jacob(f)(x))^*(\omega (f \, x)).
  $$
  Hence,
  we have the following pullback reduction
  $$
    \pbat{\big(\pbmap{\lambda y.\elemterms}{\omega}\big)}{\values}
    \Pred
    \dualmap
    {(\lambda v.\simterms)}
    {(\pbat{\omega}{(\elemterms[\values/y])})}
  $$
  of the application
  $\pbat{\big(\pbmap{\lambda y.\elemterms}{\omega}\big)}{\values}$
  which mimics the pullback
  of a variable $\omega$
  along an abstraction $\lambda y.\elemterms$
  at a term $\values$.
  But how should one define the simple term $\simterms$ in
  $\dualmap
  {(\lambda v.\simterms)}
  {(\pbat{\omega}{(\elemterms[\values/y])})}$
  so that
  $\lambda v.\simterms$ mimics the Jacobian of $f$ at $x$, \ie $\jacob(f)(x)$?
  We 
  do so by induction on the elementary terms
  $\elemterms$,
  shown
  in Figure \ref{fig:pullback reductions}
  Reductions $\hyperlink{eq:9}{9}$-$\hyperlink{eq:20}{20}$.

  \begin{remark}
    For readers familiar with
    differential $\lambda$-calculus \cite{DBLP:journals/tcs/EhrhardR03},
    $\simterms$ is the result of
    substituting a linear occurrence of $y$ by $v$,
    and then
    substituting all free occurrences of $y$ by $\values$
    in the term $\elemterms$.
    Our approach is different from differential $\lambda$-calculus in that we define a reduction strategy instead of a substitution.
    A comprehensive comparison between our language and differential $\lambda$-calculus is given in Section~\ref{sec:comparison}.
  \end{remark}

  \subsubsection{Constant Functions}

  If $y$ is not a free variable in $\elemterms$,
  $\lambda y.\elemterms$ is mimicking a constant function.
  The Jacobian of a constant function is $0$,
  hence we reduce
  $\pbat{\big(\pbmap{\lambda y.\elemterms}{\omega}\big)}{\values}$
  to
  $\dualmap
  {(\lambda v.0)}
  {(\pbat{\omega}{(\elemterms[\values/y])})}$, which
  is the sugar for $0$
  as shown in Figure~\ref{fig:pullback reductions} (Constant Functions)
  Reduction $\hyperlink{eq:9}{9}$.
  The redexes
  $\pbat{\big(\pbmap{\lambda y.0}{\omega}\big)}{\values}$,
  $\pbat{\big(\pbmap{\lambda y.\PCF{r}}{\omega}\big)}{\values}$ and
  $\pbat{\big(\pbmap{\lambda y.\dual{\vec{r}}}{\omega}\big)}{\values}$
  all reduce to $0$.

  Henceforth, we assume $y \in \freevar{\elemterms}$.

  \subsubsection{Linear Functions}
  \label{subsubsec:linear functions}


  We consider the redexes where
  $y \in \lin{\elemterms}$.
  Then $\lambda y.\elemterms$ is mimicking a linear function,
  whose Jacobian is itself.
  Hence
  $\pbat{\big(\pbmap{\lambda y.\elemterms}{\omega}\big)}{\values}$
  is reduced to
  $\dualmap
  {(\lambda v.\simterms)}
  {(\pbat{\omega}{(\elemterms[\values/y])})}$
  where
  $\simterms$ is the result of substituting $y$ by $v$ in $\elemterms$.
  Figure \ref{fig:pullback reductions} (Linear Functions)
  Reductions $\hyperlink{eq:10a}{10}$-$\hyperlink{eq:14}{14}$
  shows how they are reduced.

  \subsubsection{Smooth Functions}

  Now consider the redexes
  where $y$ might not be a linear variable in $\elemterms$.
  All reductions are shown in Figure \ref{fig:pullback reductions}.

  \paragraph{Function Symbols}

  Let $f$ be ``easily differentiable''.
  Then, $\lambda y.\PCF{f}(\proj{y}{i})$ is mimicking $f \circ \pi_i$,
  whose Jacobian at $x$ is
  $\jacob(f)(\pi_i(x)) \circ \pi_i$.
  Hence
  the Jacobian of $\lambda y.\PCF{f}(\proj{y}{i})$
  is
  $\lambda v.(\jac{f}{\proj{v}{i}})\proj{\values}{i}$ and
  $\pbat{\big(\pbmap{\lambda y. \PCF{f}(\proj{y}{i})}{\omega}\big)}{\values}$
  is reduced to
  $\dualmap
  {(\lambda v.(\jac{f}{\proj{v}{i}})\proj{\values}{i})}
  {(\pbat{\omega}{(\PCF{f}(\proj{\values}{i}))})}$
  as shown in
  Reduction $\hyperlink{eq:15}{15}$.

  \paragraph{Dual Maps}

  Consider the Jacobian of $\lambda y.\dualmap{(\lambda x.\letterms)}{z}$
  at $\values$.
  It is easy to see that the result varies depending on
  where the variable $y$ is located in
  the dual map $\dualmap{(\lambda x.\letterms)}{z}$.
  We consider three cases.

  First, if $y \not\in \freevar{\lambda x.\letterms}$,
  we must have $z \equiv \proj{y}{i}$.
  Then $y$ is a linear variable in
  $\dualmap{(\lambda x.\letterms)}{\proj{y}{i}}$ and so
  the Jacobian of
  $\lambda y.\dualmap{(\lambda x.\letterms)}{\proj{y}{i}}$
  at $\values$ is
  $\lambda v.\dualmap{(\lambda x.\letterms)}{\proj{v}{i}}$.
  Hence, we have Reduction $\hyperlink{eq:16a}{16a}$.

  Second, say
  $y \not\in \freevar{z}$.
  Since dual and abstraction are both linear operations,
  and $y$ is only free in $\letterms$,
  the Jacobian of
  $\lambda y.\dualmap{(\lambda x.\letterms)}{z}$ at $\values$.
  should be
  $\lambda v.\dualmap{(\lambda x.\simterms')}{z}$ where
  $\lambda v.\simterms'$ is the Jacobian of $\lambda y.\letterms$ at $\values$.
  To find the Jacobian of $\lambda y.\letterms$ at $\values$,
  we reduce
  $\pbat{\big(\pbmap{\lambda y.\letterms}{\omega}\big)}{\values}$ to
  $\dualmap
  {(\lambda v.\simterms')}
  {(\pbat{\omega}{\letterms[\values/y]})}$.
  Then $\lambda v.\simterms'$
  is the Jacobian of $\lambda y.\letterms$ at $\values$.
  The reduction is given in
  Reduction $\hyperlink{eq:16b}{16b}$.
  Note that this reduction avoids expression swell,
  as we are reducing the let series $\letterms$
  in $\lambda y.\dualmap{(\lambda x.\letterms)}{z}$
  using our pullback reductions,
  which does not suffer from expression swell.

  Finally, for $y \in \freevar{\lambda x.\letterms} \cap \freevar{z}$,
  the Jacobian of
  $\lambda y.\dualmap{(\lambda x.\letterms)}{z}$ at $\values$
  is the ``sum'' of the results we have for the two cases above, \ie
  $\lambda v.
    \dualmap{(\lambda x.\letterms)}{\proj{v}{i}} +
    \dualmap{(\lambda x.\simterms)}{\proj{y}{i}},
  $
  where the remaining free occurrences of $y$
  are substituted by $\values$,
  since the Jacobian of a bilinear function
  $l:X_1 \times X_2 \to Y$ is
  $
    \jacob(l)(\anglebr{x_1,x_2}(\anglebr{v_1,v_2}) =
    l\anglebr{x_1,v_2} + l\anglebr{v_1,x_2}
  $.
  Hence, we have Reduction $\hyperlink{eq:16c}{16c}$.

  \paragraph{Pullback Terms}

  Consider
  $\pbat{\big(\pbmap{
    \lambda y. \pbmap{\lambda x.\letterms}{z}
  }{\omega}\big)}{\values}$.
  Instead of reducing it
  to some
  $\dualmap
  {(\lambda v.\simterms)}
  {(\pbat{\omega}{(
    \pbmap{\lambda y. \pbmap{\lambda x.\letterms}{z}}{\omega}
  )[\values/y]})}$
  like the others,
  here we simply reduce
  $\pbat{\big(\pbmap{\lambda x.\letterms}{z})}{a}$ to
  $\dualmap{(\lambda v.\simterms)}{(\pbat{z}{\letterms[a/x]})}$,
  where $a$ is a fresh variable and
  $z \not\equiv x$,
  and replace
  $\pbmap{\lambda x.\letterms}{z}$
  by
  $\lambda a.\dualmap{(\lambda v.\simterms)}{(\pbat{z}{\letterms[a/x]})}$ in
  $\pbat{\big(\pbmap{
    \lambda y. \pbmap{\lambda x.\letterms}{z}
  }{\omega}\big)}{\values}$ as shown in
  Reduction $\hyperlink{eq:17}{17}$.

  \paragraph{Abstraction}

  Consider the Jacobian of $\lambda y.\lambda x.\letterms$
  at $\values$.

  We follow the treatment of exponentials in
  differential $\lambda$-category \cite{DBLP:journals/entcs/BucciarelliEM10}
  where the (D-curry) rule states that
  for all $f:Y \times X \to A$,
  $
    D[\curry{f}]
     = \curry{D[f] \circ \anglebr{\pi_1\times 0_X, \pi_2 \times \Id_X}}
  $, which means $\jacob(\curry{f})(y)$
  is equal to
  $$
    \lambda v.\jacob(\curry{f})(y)(v)
    =
    \lambda v x.\jacob(f\anglebr{-,x})(y)(v).
  $$
  According to this (D-curry) rule,
  the Jacobian of $\lambda y.\lambda x.\letterms$ at $\values$
  should be
  $\lambda v.\lambda x.\simterms$
  where $\lambda v.\simterms$ is
  the Jacobian of $\lambda y.\letterms$ at $\values$.
  Hence
  similar to the dual map case,
  we first reduce
  $\pbat{\big(\pbmap{\lambda y.\letterms}{\omega}\big)}{\values}$
  to
  $\dualmap{(\lambda v.\simterms)}{(\pbat{\omega}{\letterms[\values/y]})}$
  and obtain
  the Jacobian of $\lambda y.\letterms$ at $\values$, \ie $\lambda v.\simterms$
  and then reduce
  $\pbat{\big(\pbmap{\lambda y.\lambda x.\letterms}{\omega}\big)}{\values}$
  to
  $\dualmap{(
    \lambda v.\lambda x.\simterms
  )}{\big(
    \pbat{\omega}{(\lambda x.\letterms[\values/y])}
  \big)}$
  as shown in
  Reduction $\hyperlink{eq:18}{18}$.

  \paragraph{Application}

  Consider the Jacobian of $\lambda y.z_1\,z_2$
  at $\values$.
  Note that $z_1$ and $z_2$
  may or may not contain $y$ as a free variable.
  Hence, there are two cases.

  First, we consider $\lambda y.\proj{y}{i}\,z$ where $z$ is fresh.
  Since $y\in \lin{\proj{y}{i}\,z}$,
  $\lambda y.\proj{y}{i}\,z$ mimics a linear function, and hence
  its Jacobian at $\values$ is $\lambda v.\proj{v}{i}\,z$.
  So
  $\pbat{\big(\pbmap{\lambda y.\proj{y}{i}\,z}{\omega}\big)}{\values}$
  is reduced to
  $\dualmap
  {(\lambda v.\proj{v}{i}\,z)}
  {(\pbat{\omega}{(\proj{\values}{i}\,z)})}$
  as shown in
  Reduction $\hyperlink{eq:19a}{19a}$.

  Second, we consider the Jacobian of
  $\lambda y.\proj{y}{i}\,\proj{y}{j}$ at $\values$.
  Now $y$ is not a linear variable in $\proj{y}{i}\,\proj{y}{j}$,
  since it occurs in the argument $\proj{y}{j}$.
  As proved in Lemma 4.4 of \cite{DBLP:journals/mscs/Manzonetto12},
  every differential $\lambda$-category satisfies the (D-eval) rule,
  $
    D[\catev \circ \anglebr{h,g}]
    =
    \catev \circ \anglebr{D[h],g\circ \pi_2} +
    D[\uncurry{h}] \circ
      \anglebr{\anglebr{0,D[g]},\anglebr{\pi_2,g\circ \pi_2}}
  $
  which means
  $\jacob(\catev \circ \anglebr{h,g})(x)(v)$
  is equal to
  $$
    \big(\jacob(h)(x)(v)\big)(g(x))+
    \jacob(h(x))(g(x))(\jacob(g)(x)(v))
  $$
  for all $h:C\to (A\Arrow B)$ and $g:C \to A$.
  Hence, the Jacobian of
  $\catev \circ \anglebr{\pi_i,\pi_j}$
  at $x$ along $v$, \ie
  $\jacob(\catev \circ \anglebr{\pi_i,\pi_j})(x)(v)$, is
  $$
    \pi_i(v)(\pi_j(x)) +
    \jacob(\pi_i(x))(\pi_j(x))(\pi_j(v)).
  $$
  So the Jacobian of $\lambda y.\proj{y}{i}\,\proj{y}{j}$
  at $\values$ is
  $\lambda v.
    \proj{v}{i}\,\proj{\values}{j} +
    \simterms'[\proj{v}{j}/v']$
  where $\lambda v'.\simterms'$ is the Jacobian of
  $\proj{\values}{i}$ at $\proj{\values}{j}$.
  Hence assuming
  $\proj{\values}{i} \equiv \lambda z.\pbterms'$,
  we first reduce
  $\pbat{\big(\pbmap{\lambda z.\pbterms'}{\omega}\big)}{\proj{\values}{j}}$
  to
  $\dualmap{(\lambda v'.\simterms')}{\pbat{\omega}{(
    \pbterms'[\proj{\values}{j}/z]}
  )}$
  and obtain $\lambda v'.\simterms'$
  as the Jacobian of
  $\lambda z.\pbterms'$ at $\proj{\values}{j}$.
  Then, we reduce
  $\pbat{\big(
    \pbmap{\lambda y.\proj{y}{i}\,\proj{y}{j}}{\omega}
  \big)}{\values}$
  to
  $\dualmap{(\lambda v.
    \proj{v}{i}\,\proj{\values}{j} +
    \simterms'[\proj{v}{j}/v']
  )}{\big(
    \pbat{\omega}{(\proj{\values}{i}\,\proj{\values}{j})}
  \big)}$ as shown in
  Reduction $\hyperlink{eq:19b}{19b}$.

  If
  $\pbat{\big(\pbmap{\lambda z.\values'}{\omega}\big)}{\proj{\values}{j}}$
  reduces to $0$, which means
  $\lambda z.\values' \equiv \proj{\values}{i}$ is a constant function,
  the Jacobian of $\lambda y.\proj{y}{i}\,\proj{y}{j}$ at $\values$
  is just
  $\lambda v.\proj{v}{i}\,\proj{\values}{j}$
  and we have
  Reduction $\hyperlink{eq:19c}{19c}$.

  \begin{remark}
    Doing induction on elementary terms defined in
    Subsection \ref{subsec:A-reduction},
    we can see that there are a few elementary terms $\elemterms$ where
    $\pbat{\big(\pbmap{\lambda y.\elemterms}{\omega}\big)}{\values}$
    is \emph{not} a redex, namely
    \begin{enumerate}[{value} 1:]
      \item $\pbat{\big(\pbmap{\lambda y.z\,\proj{y}{i}}{\omega}\big)}{\values}$
        where $z$ is a free variable,
      \item $\pbat{\big(\pbmap{\lambda y.\proj{y}{i}\,\proj{y}{j}}{\omega})}{\values}$ where
        $\values_{\pi i} \not\equiv \lambda z.\pbterms'$.
    \end{enumerate}

    Having these terms as values makes sense intuitively, since
    they have ``inappropriate'' values in positions.
    Values 1 has a free variable $z$ in a function position.
    Value 2 substitutes $\proj{y}{i}$ by $\proj{\values}{i}$
    which is a non-abstraction, to a function position.
  \end{remark}

  \paragraph{Pair}

  Last but not least, we consider the Jacobian of
  $\lambda y.\anglebr{y,\elemterms}$ at $\values$.
  It is easy to see that Jacobian is
  $\lambda v.\anglebr{v,\simterms}$
  where $\lambda v.\simterms$ is the Jacobian of $\lambda y.\elemterms$,
  as shown in
  Reduction $\hypertarget{eq:20a}{20a}$
  and Reduction $\hypertarget{eq:20b}{20b}$.

  \begin{example}
    \label{eg:running-ex-pb-red}
    Take
    our running example.
    In Examples \ref{eg:running-ex-A-red} and \ref{eg:running-ex-omega-split}
    we showed that via A-reductions and
    Reductions $\hyperlink{eq:7}{7}$ and $\hyperlink{eq:8}{8}$,
    $\pbmap{\lambda \anglebr{x,y}.
      \PCF{\mathsf{pow2}}(\PCF{\mathsf{mult}}(\PCF{g}(\anglebr{x,y})))}
    {\omega}$
    is reduced to
    $$
      \mat{
        \pbmap{\lambda \anglebr{x,y}.\anglebr{\anglebr{x,y},\anglebr{x,y}}}{} \\
        \pbmap{\lambda \anglebr{\anglebr{x,y},z_1}.
          \anglebr{\anglebr{x,y},z_1,\PCF{g}(z_1)}}{} \\
        \pbmap{\lambda \anglebr{\anglebr{x,y},z_1,z_2}.
          \anglebr{\anglebr{x,y},z_1,z_2,\PCF{\mathsf{mult}}(z_2)}}{} \\
        \pbmap{\lambda \anglebr{\anglebr{x,y},z_1,z_2,z_3}.
          \PCF{\mathsf{pow2}}(z_3)
        }{\omega}
      }
    $$

    We show how it can be reduced when applied to $\anglebr{\PCF{1},\PCF{3}}$.
    \begin{align*}
      &
      \mat{
        \pbmap{\lambda \anglebr{x,y}.\anglebr{\anglebr{x,y},\anglebr{x,y}}}{} \\
        \pbmap{\lambda \anglebr{\anglebr{x,y},z_1}.
          \anglebr{\anglebr{x,y},z_1,\PCF{g}(z_1)}}{} \\
        \pbmap{\lambda \anglebr{\anglebr{x,y},z_1,z_2}.
          \anglebr{\anglebr{x,y},z_1,z_2,\PCF{\mathsf{mult}}(z_2)}}{} \\
        \pbmap{\lambda \anglebr{\anglebr{x,y},z_1,z_2,z_3}.
          \PCF{\mathsf{pow2}}(z_3)
        }{\omega}
      }\,\PCF{\colvec{1\\3}} \\
      & \Predtextm{\hyperlink{eq:20.1}{20.1}}{\hyperlink{eq:11}{11}}
      \mat{
        \dualmap{(\lambda \anglebr{v_1,v_2}.
          \anglebr{\anglebr{v_1,v_2},\anglebr{v_1,v_2}}
        )}{} \\
        \mat{
          \pbmap{\lambda \anglebr{\anglebr{x,y},z_1}.
            \anglebr{\anglebr{x,y},z_1,\PCF{g}(z_1)}}{} \\
          \pbmap{\lambda \anglebr{\anglebr{x,y},z_1,z_2}.
            \anglebr{\anglebr{x,y},z_1,z_2,\PCF{\mathsf{mult}}(z_2)}}{} \\
          \pbmap{\lambda \anglebr{\anglebr{x,y},z_1,z_2,z_3}.
            \PCF{\mathsf{pow2}}(z_3)
          }{\omega}
        }\,
        \anglebr{\PCF{\colvec{1\\3}},\PCF{\colvec{1\\3}}}
      }
      \\
      & \Predtextm{\hyperlink{eq:20.1}{20.1}}{\hyperlink{eq:14}{14},
      \hyperlink{eq:3}{3}}
      \mat{
        \dualmap{(\lambda \anglebr{v_1,v_2}.
          \anglebr{\anglebr{v_1,v_2},\anglebr{v_1,v_2}}
        )}{} \\
        \dualmap{(\lambda \anglebr{\anglebr{v_1,v_2},v_3}.
          \anglebr{
            \anglebr{v_1,v_2},v_3,
            (\jac{g}{v_3})\anglebr{\PCF{1},\PCF{3}}}
        )}{}
        \\
        \Big(
          \mat{
            \pbmap{\lambda \anglebr{\anglebr{x,y},z_1,z_2}.
              \anglebr{\anglebr{x,y},z_1,z_2,\PCF{\mathsf{mult}}(z_2)}}{}
            \\
            {(
              \pbmap{\lambda \anglebr{\anglebr{x,y},z_1,z_2,z_3}.\PCF{\mathsf{pow2}}(z_3)
              }{\omega}
            )}
          }
          \anglebr{
            \PCF{\colvec{1\\3}},
            \PCF{\colvec{1\\3}},
            \PCF{\colvec{2\\11}}}
        \Big)
      } \tag{$\star$} \\
      & \Predtextm{\hyperlink{eq:20.1}{20.1}}{\hyperlink{eq:14}{14},
      \hyperlink{eq:3}{3}}
      \mat{
        \dualmap{(\lambda \anglebr{v_1,v_2}.
          \anglebr{\anglebr{v_1,v_2},\anglebr{v_1,v_2}}
        )}{} \\
        \dualmap{(\lambda \anglebr{\anglebr{v_1,v_2},v_3}.
          \anglebr{
            \anglebr{v_1,v_2},v_3,
            (\jac{g}{v_3})\anglebr{\PCF{1},\PCF{3}}}
        )}{} \\
        \dualmap{(\lambda \anglebr{\anglebr{v_1,v_2},v_3,v_4}.
          \anglebr{
            \anglebr{v_1,v_2},v_3,v_4,
            (\jac{\mathsf{mult}}{v_4})\anglebr{\PCF{2},\PCF{11}}}
        )}{} \\
        \big(
          \mat{
            \pbmap{\lambda \anglebr{\anglebr{x,y},z_1,z_2,z_3}.\PCF{\mathsf{pow2}}(z_3)
            }{\omega}
          }
          \anglebr{
            \PCF{\colvec{1\\3}},
            \PCF{\colvec{1\\3}},
            \PCF{\colvec{2\\11}},
            \PCF{22}
          }
        \big)
      } \\
      & \Predtextm{\hyperlink{eq:20.1}{20.1}}{\hyperlink{eq:14}{14},
      \hyperlink{eq:3}{3}}
      \mat{
        \dualmap{(\lambda \anglebr{v_1,v_2}.
          \anglebr{\anglebr{v_1,v_2},\anglebr{v_1,v_2}}
        )}{} \\
        \dualmap{(\lambda \anglebr{\anglebr{v_1,v_2},v_3}.
          \anglebr{
            \anglebr{v_1,v_2},v_3,
            (\jac{g}{v_3})\anglebr{\PCF{1},\PCF{3}}}
        )}{} \\
        \dualmap{(\lambda \anglebr{\anglebr{v_1,v_2},v_3,v_4}.
          \anglebr{
            \anglebr{v_1,v_2},v_3,v_4,
            (\jac{\mathsf{mult}}{v_4})\anglebr{\PCF{2},\PCF{11}}}
        )}{} \\
        \dualmap{(\lambda \anglebr{\anglebr{v_1,v_2},v_3,v_4,v_5}.
            (\jac{\mathsf{pow2}}{v_5})\PCF{22}
        )}{(\pbat{\omega}{
          \PCF{484}
        })}
      }
    \end{align*}
    Notice how
    this is reminiscent of the forward phase of reverse-mode AD
    performed on $f: \anglebr{x,y} \mapsto \big((x + 1)(2x + y^2)\big)^2$
    at $\anglebr{1,3}$ considered in Subsection \ref{subsec:technique or rAD}.

    Moreover,
    we used the reduction
    $\PCF{f}(\PCF{r})\xrightarrow{\hyperlink{eq:3}{3}}\PCF{f(r)}$
    couples of times in the argument position of an application.
    This is to avoid expression swell.
    Note $1+1$ is only evaluated once in $(\star)$
    even when the result is used in various computations.
    Hence, we must have a \emph{call-by-value} reduction strategy
    as presented below.
  \end{example}

  \subsection{Combine}
  \label{subsec:combine reduction}

  Reductions in
  Subsections \ref{subsec:A-reduction} and \ref{subsec:pb reduction}
  are the most interesting development of the paper.
  However, they alone are not enough to
  complete a reduction strategy.
  In this subsection, we define contexts and redexes so that
  any non-value term can be reduced.

  The definition of \emph{context} $C$ is the standard
  call-by-value context, extended with duals and pullbacks.
  Notice that the context $\pbmap{\lambda y.C_A}{\simterms}$
  contains a A-context defined in Subsection \ref{subsec:A-reduction}.
  This follows from the idea of reverse-mode AD to
  decompose a term into elementary terms before
  differentiating them.
  \begin{align*}
    C & ::=
        []
    \mid C + \pbterms
    \mid \values + C_A
    \mid C\,\pbterms
    \mid {\values}\,C
    \mid \pi_i({C})
    \mid \anglebr{C,\simterms}
    \mid \anglebr{{\values},C}
    \\
    & \st
          \PCF{f}(C)
      \mid \jac{f}{C}
      \st
          \dualmap{(\lambda x.\simterms)}{C}
      \mid \dualmap{(\lambda x.C)}{\values}
      \\
    & \st
          \pbmap{\lambda y.C_A}{\simterms}
      \mid \pbmap{\lambda y.\elemterms}{C}
      \mid \pbmap{\lambda y.\anglebr{y,\elemterms}}{C}
  \end{align*}

  Our \emph{redex} $r$ extend the standard call-by-value redex with four sets of terms.

  \begin{align*}
    r & ::=
        (\lambda x.\simterms)\,{\values}
    \mid \pi_i{(\anglebr{\values_1,\values_2})}
    \mid \PCF{f}(\PCF{r})
    \mid (\jac{f}{\PCF{\vec{r}}})\,\PCF{\vec{r'}}
    \\
  & \st
        \dualmap{(\lambda v.(\jac{f}{v})\,\PCF{\vec{r}})}{\dual{\PCF{\vec{r'}}}}
    \mid \dualmap{(\lambda v_1.\values_1)}{\big(
          \dualmap{(\lambda v_2.{\values_2})}{{\values_3}}
        \big)}
    \\
   & \st
        \pbmap{\lambda y.\letterms}{\simterms}
    \mid \pbat{\big(\pbmap{\lambda y.\elemterms}{\values_1}\big)}{\values_2}
    \mid \pbat{\big(\pbmap{\lambda y.\anglebr{y,\elemterms}}{\values_1}\big)}{\values_2}
  \end{align*}
  where either
  ${\values_2} \not\equiv (\jac{f}{v_2})\,\PCF{\vec{r}}$ or
  ${\values_3} \not\equiv \dual{\PCF{\vec{r'}}}$.
  A \emph{value} $\values$ is a pullback term $\pbterms$
  that cannot be reduced further, \ie
  a term in normal form.

  The following standard lemma,
  which is proved by induction on $\pbterms$,
  tells us that
  there is at most one redex to reduce.
  \begin{lemma}
    \label{lemma: Context/redex/value are well-defined}
    Every term $\pbterms$ can be expressed as
    either $C[r]$ for some unique context $C$ and redex $r$
    or a value $\values$.
  \end{lemma}

  Let's look at the reductions of redexes.
  ($\hyperlink{eq:1}{1}$-$\hyperlink{eq:4}{4}$) are the standard call-by-value reductions.
  ($\hyperlink{eq:5}{5}$) reduces the dual along a linear map $l$ and
  ($\hyperlink{eq:6}{6}$) is the contra-variant property of dual maps.
  \begin{align*}
    (\lambda x.\simterms)\,{\values}
      & \xrightarrow{\hypertarget{eq:1}{(1)}} \simterms[\values/x] &
    \pi_i{(\anglebr{\values_1,\values_2})}
      & \xrightarrow{\hypertarget{eq:2}{(2)}} \values_i
    \\
    \PCF{f}(\PCF{r})
      & \xrightarrow{\hypertarget{eq:3}{(3)}} \PCF{f(r)} &
    (\jac{f}{\PCF{\vec{r}}})\,\PCF{\vec{r'}}
      & \xrightarrow{\hypertarget{eq:4}{(4)}}
      \PCF{\jacob{(f)}{(\vec{r'})}(\vec{r})}
    \\
    \omit\rlap{$
    \dualmap{(\lambda v.(\jac{f}{v})\,\PCF{\vec{r}})}{\dual{\PCF{\vec{r'}}}}
      \xrightarrow{\hypertarget{eq:5}{(5)}}
      \dual{\PCF{(\dual{(\jacob(f)(\vec{r}))}(\vec{r'}))}}
    $} \\
    \omit\rlap{$
    \dualmap{(\lambda v_1.\values_1)}{\big(
        \dualmap{(\lambda v_2.{\values_2})}{{\values_3}}
      \big)}
      \xrightarrow{\hypertarget{eq:6}{(6)}}
      \dualmap{(\lambda v_1.{\values_2}[\values_1/v_2])}{{\values_3}}
    $}
  \end{align*}
  where either
  ${\values_2} \not\equiv (\jac{f}{v_2})\,\PCF{\vec{r}}$ or
  ${\values_3} \not\equiv \dual{\PCF{\vec{r'}}}$.

  We say $C[r] \red C[\values]$ if $r \red \values$
  for all reductions
  except for those with a proof tree, \ie
  Reductions
  $\hyperlink{eq:16b}{16b}$,
  $\hyperlink{eq:16c}{16c}$,
  $\hyperlink{eq:17}{17}$,
  $\hyperlink{eq:18}{18}$,
  $\hyperlink{eq:19b}{19b}$,
  $\hyperlink{eq:19c}{19c}$ and
  $\hyperlink{eq:20a}{20a}$,
  where we have\\
  \AxiomC{$
    r \redplus \values
  $}
  \UnaryInfC{$
    C[r'[\values_1/\omega][\values_2/\values]] \red
    C[\values'[\values_1/\omega][\values_2/\values]]
  $}
  \DisplayProof
  if
  \AxiomC{$
    r \redplus \values
  $}
  \UnaryInfC{$
    r' \red \values'
  $}
  \DisplayProof

  \begin{example}
    \label{eg:running-ex-result}
    Consider our running example
    $\pbterms \equiv \pbat{\big(
      \pbmap{\lambda \anglebr{x,y}.
        \PCF{\mathsf{pow2}}(\PCF{\mathsf{mult}}(\PCF{g}(\anglebr{x,y})))}
      {\pbof{\colvec{1}}}
    \big)}{\anglebr{\PCF{1},\PCF{3}}}$
    which represents the Jacobian of
    $f: \anglebr{x,y} \mapsto \big((x + 1)(2x + y^2)\big)^2$
    at $\anglebr{1,3}$,
    as shown in Example \ref{eg:running-ex-term}.
    Replacing $\omega$ by
    $\pbof{\colvec{1}} \equiv \lambda x.\dual{\colvec{1}}$
    in Examples \ref{eg:running-ex-A-red},
    \ref{eg:running-ex-omega-split} and
    \ref{eg:running-ex-pb-red},
    $\pbterms$ is reduced to
    $$
      \mat{
        \dualmap{(\lambda \anglebr{v_1,v_2}.
          \anglebr{\anglebr{v_1,v_2},\anglebr{v_1,v_2}}
        )}{} \\
        \dualmap{(\lambda \anglebr{\anglebr{v_1,v_2},v_3}.
          \anglebr{
            \anglebr{v_1,v_2},v_3,
            (\jac{g}{v_3})\anglebr{\PCF{1},\PCF{3}}}
        )}{} \\
        \dualmap{(\lambda \anglebr{\anglebr{v_1,v_2},v_3,v_4}.
          \anglebr{
            \anglebr{v_1,v_2},v_3,v_4,
            (\jac{\mathsf{mult}}{v_4})\anglebr{\PCF{2},\PCF{11}}}
        )}{} \\
        \dualmap{(\lambda \anglebr{\anglebr{v_1,v_2},v_3,v_4,v_5}.
            (\jac{\mathsf{pow2}}{v_5})\PCF{22}
        )}{(\pbat{\omega}{
          \PCF{484}
        })}
      }.
    $$
    Via reduction $\hyperlink{eq:5}{5}$ and $\beta$ reduction,
    $\pbterms$ is reduced to
    \begin{align*}
      & \mat{
        \dualmap{(\lambda \anglebr{v_1,v_2}.
          \anglebr{\anglebr{v_1,v_2},\anglebr{v_1,v_2}}
        )}{} \\
        \dualmap{(\lambda \anglebr{\anglebr{v_1,v_2},v_3}.
          \anglebr{
            \anglebr{v_1,v_2},v_3,
            (\jac{g}{v_3})\anglebr{\PCF{1},\PCF{3}}}
        )}{} \\
        \dualmap{(\lambda \anglebr{\anglebr{v_1,v_2},v_3,v_4}.
          \anglebr{
            \anglebr{v_1,v_2},v_3,v_4,
            (\jac{\mathsf{mult}}{v_4})\anglebr{\PCF{2},\PCF{11}}}
        )}{} \\
        \dualmap{(\lambda \anglebr{\anglebr{v_1,v_2},v_3,v_4,v_5}.
            (\jac{\mathsf{pow2}}{v_5})\PCF{22}
        )}{\PCF{\colvec{1}}}
      } \\
      & \red
      \mat{
        \dualmap{(\lambda \anglebr{v_1,v_2}.
          \anglebr{\anglebr{v_1,v_2},\anglebr{v_1,v_2}}
        )}{} \\
        \dualmap{(\lambda \anglebr{\anglebr{v_1,v_2},v_3}.
          \anglebr{
            \anglebr{v_1,v_2},v_3,
            (\jac{g}{v_3})\anglebr{\PCF{1},\PCF{3}}}
        )}{} \\
        \dualmap{(\lambda \anglebr{\anglebr{v_1,v_2},v_3,v_4}.
          \anglebr{
            \anglebr{v_1,v_2},v_3,v_4,
            (\jac{\mathsf{mult}}{v_4})\anglebr{\PCF{2},\PCF{11}}}
        )}{}
      }
      \dual{\PCF{
        \colvec{0\\0\\0\\0\\0\\0\\44}
      }}
      \\
      & \red
      \mat{
        \dualmap{(\lambda \anglebr{v_1,v_2}.
          \anglebr{\anglebr{v_1,v_2},\anglebr{v_1,v_2}}
        )}{} \\
        \dualmap{(\lambda \anglebr{\anglebr{v_1,v_2},v_3}.
          \anglebr{
            \anglebr{v_1,v_2},v_3,
            (\jac{g}{v_3})\anglebr{\PCF{1},\PCF{3}}}
        )}{}
      }
      \dual{\PCF{
        \colvec{0\\0\\0\\0\\484\\88}
      }}
      \\
      & \red
      \mat{
        \dualmap{(\lambda \anglebr{v_1,v_2}.
          \anglebr{\anglebr{v_1,v_2},\anglebr{v_1,v_2}}
        )}{}
      }
      \dual{\PCF{
        \colvec{0\\0\\660\\528}
      }}
      \\
      & \red
      \dual{\PCF{
        \colvec{660\\528}
      }}
    \end{align*}
    Notice how this mimics the reverse phase of reverse-mode AD
    on $f: \anglebr{x,y} \mapsto \big((x + 1)(2x + y^2)\big)^2$ at $\anglebr{1,3}$
    considered in Subsection \ref{subsec:technique or rAD}.
  \end{example}

  Examples \ref{eg:running-ex-A-red},
  \ref{eg:running-ex-omega-split} and
  \ref{eg:running-ex-pb-red} demonstrates that
  our reduction strategy is faithful to reverse-mode AD
  (in that it is exactly reverse-mode AD when restricted to first-order).

  \subsection{Continuation-Passing Style}
  \label{subsec:cps}

  Differential 1-forms
  $\oneform E := \smooth{E}{L(E,\Real)}$
  is similar to the continuation of $E$ with the ``answer'' $\Real$.
  We can indeed write our reduction in a
  continuation passing style (CPS) manner.
  Let $\config{\pbterms}{\simterms}_y \equiv
  \pbmap{\lambda y.\pbterms}{\simterms}$,
  then
  we can treat $\config{\pbterms}{\simterms}_y$ as
  a configuration of
  an element
  $\Gamma \cup \set{y:\sigma} \vdash \pbterms:\tau$
  and a ``continuation''
  $\Gamma \vdash \simterms:\oneform \tau$.
  The rules for the redexes
  $\config{\letterms}{\simterms}_y$,
  $\pbat{\config{\elemterms}{\values_1}_y}{\values_2}$ and
  $\pbat{\config{\anglebr{y,\elemterms}}{\values_1}_y}{\values_2}$
  can be directly converted from
  Reductions $\hyperlink{eq:7}{7}$-$\hyperlink{eq:20}{20}$.
  For example, Reduction $\hyperlink{eq:8}{8}$ can be written as
  ${\config{\Let{x = \elemterms}{\letterms}}{\omega}_y}$
  $\red
  \config{\anglebr{y,\elemterms}}{
        \config{\letterms}{\omega}_{\anglebr{y,x}}
      }_y.
  $

  We prefer to write our language without the explicit mention of CPS
  since this paper focuses on the syntactic notion of reverse-mode AD
  using pullbacks and 1-forms.
  Also, 1-form of the type $\sigma$
  is more precisely described as
  an element of the function type
  $\oneform\sigma \equiv \sigma \Arrow \dual{\sigma}$,
  than of the continuation of $\sigma$, \ie
  $\sigma \Arrow (\sigma\Arrow\PCFReal)$.

\section{Model}
\label{sec:model}
  We show that any differential $\lambda$-category satisfying
  the Hahn-Banach Separation Theorem
  can soundly model our language.

  \subsection{Differential Lambda-Category}

  \emph{Cartesian differential category}
  \cite{blute2009cartesian}
  aims to axiomatise fundamental properties of derivative.
  Indeed, any model of synthetic differential geometry
  has an associated Cartesian differential category.
  \cite{DBLP:journals/acs/CockettC14}

  \paragraph{Cartesian differential category}
  A category $\cat{C}$
  is a \emph{Cartesian differential category} if
  \begin{itemize}
    \item
      every homset $\cat{C}(A, B)$ is enriched with
      a commutative monoid $(\cat{C}(A, B), +_{AB}, 0_{AB})$ and
      the additive structure is preserved by composition on the left. \ie
      $(g+h)\circ f = g \circ f + h \circ f$ and $0\circ f = 0.$
    \item
      it has products and
      projections and pairings of additive maps are additive.
      A morphism $f$ is \emph{additive} if
      it preserves the additive structure of the homset on the right. \ie
      $f \circ (g+h) = f \circ g + f \circ h$ and $f\circ 0 = 0.$
  \end{itemize}
  and
  it has an operator
  $D[-]:\cat{C}(A, B)\rightarrow\cat{C}(A\times A, B)$
  that satisfies the following axioms:
  \begin{itemize}
    \item[\lbrack CD1\rbrack]
      $D$ is linear:
      $D[f+g]=D[f]+D[g]$ and $D[0]=0$
    \item[\lbrack CD2\rbrack]
      $D$ is additive in its first coordinate:
      $D[f]\circ \anglebr{h+k, v} =
        D[f]\circ\anglebr{h,v} + D[f]\circ\anglebr{k, v}$,
      $D[f]\circ\anglebr{0, v} = 0$
    \item[\lbrack CD3\rbrack]
      $D$ behaves with projections:
      $D[\mathsf{Id}] = \pi_1$,
      $D[\pi_1] = \pi_1\circ\pi_1$ and
      $D[\pi_2] = \pi_2 \circ \pi_1$
    \item[\lbrack CD4\rbrack]
      $D$ behaves with pairings:
      $D[\langle f, g\rangle]=\langle D[f], D[g]\rangle$
    \item[\lbrack CD5\rbrack]
      Chain rule:
      $D[g \circ f]=D[g]\circ \langle D[f], f\circ \pi_2\rangle$
    \item[\lbrack CD6\rbrack]
      $D[f]$ is linear in its first component: \\
      $D[D[f]]\circ\langle\langle g, 0\rangle, \langle h, k\rangle\rangle = D[f]\circ \langle g, k\rangle $
    \item[\lbrack CD7\rbrack]
      Independence of order of partial differentiation: \\
      $D[D[f]]\circ\langle\langle 0, h\rangle, \langle g, k\rangle\rangle = D[D[f]]\circ\langle\langle 0, g\rangle, \langle h, k\rangle\rangle $
  \end{itemize}
  We call $D$ the \emph{Cartesian differential operator} of $\cat{C}$.

  \begin{example}
    The category $\cat{FVect}$ of
    finite dimensional vector spaces and differentiable functions
    is a Cartesian differential category,
    with the Cartesian differential operator
    $D[f]\anglebr{\vec{v}, \vec{x}} = \jacob(f)(\vec{x})(\vec{v})$,
  \end{example}

  Cartesian differential operator
  does not necessarily behave well with exponentials.
  Hence, \citet{DBLP:journals/entcs/BucciarelliEM10}
  added the (D-curry) rule
  and introduced differential $\lambda$-category.

  \paragraph{Differential $\lambda$-category}
  A Cartesian differential category is a \emph{differential $\lambda$-category} if
  \begin{itemize}
    \item
      it is Cartesian closed,
    \item
      $\lambda(-)$ preserves the additive structure, \ie
      $\lambda(f + g) = \lambda(f)+\lambda(g)$ and $\lambda(0) = 0$,
    \item
      $D[-]$ satisfies the (D-curry) rule:
      for any $f:A_1\times A_2 \rightarrow B$,
      $D[\lambda(f)] = \lambda(D[f]\circ \langle \pi_1\times 0_{A_2}, \pi_2\times \mathsf{Id}_{A_2}\rangle)$
  \end{itemize}

  \paragraph{Linearity}
  A morphism $f$ in a differential $\lambda$-category is \emph{linear} if
  $D[f] = f \circ \pi_1$.

  \begin{example}
    The category $\ConS$
    of convenient vector space and smooth maps,
    considered by \cite{DBLP:journals/corr/abs-1006-3140},
    is a differential $\lambda$-category with
    the Cartesian differential operator
    $
      D[f]\anglebr{v,x} :=
      \lim_{t \to 0} (f(x+tv)-f(x))/t
    $,
    as shown in Lemma \ref{lemma: cons is a differential lambda-cat}.
  \end{example}

  \subsection{Hahn-Banach Separation Theorem}

  We say a differential $\lambda$-category $\cat{C}$
  satisfies \emph{Hahn-Banach Separation Theorem} if
  $\Real$ is an object in $\cat{C}$ and
  for any object $A$ in $\cat{C}$ and distinct elements $x,y$ in $A$,
  there exists a linear morphism $l:A \to \Real$ that separates $x$ and $y$,
  \ie $l(x) \not= l(y)$.

  \begin{example}
    The category $\ConS$
    of convenient vector space and smooth maps
    satisfies the Hahn-Banach Separation Theorem,
    as shown in Proposition \ref{prop:ConS is separated}.
  \end{example}

  \subsection{Interpretation}

  Let $\cat{C}$ be a differential $\lambda$-category that
  satisfies Hahn-Banach Separation Theorem.
  Since $\cat{C}$ is Cartesian closed,
  the interpretations for the $\lambda$-calculus terms
  are standard, and hence omitted.
  The full set of interpretations can be found
  in Appendix \ref{appendix:interpretation}.
  \begin{align*}
    \deno{\PCFReal} & := \Real &
    \deno{\sigma_1 \times \sigma_2} & := \deno{\sigma_1} \times \deno{\sigma_2} \\
    \deno{\dual{\sigma}} & := L(\deno{\sigma},\Real) &
    \deno{\sigma_1 \Arrow \sigma_2} & := \cat{C}(\deno{\sigma_1},\deno{\sigma_2})
  \end{align*}
  where $L(\deno{\sigma},\Real) :=
  \set{f \in \cat{C}(\deno{\sigma,\Real}) \mid D[f] = f \circ \pi_1 }$
  is the set of all linear morphisms from $\deno{\sigma}$ to $\Real$.
  $$
    \begin{array}{rcl}
      \deno{0} \gamma &\mkern-14mu := \mkern-14mu &
      0
      \\[4pt]
      \deno{ \simterms + \pbterms} \gamma &\mkern-14mu := \mkern-14mu&
      \deno{\simterms} \gamma +
      \deno{\pbterms} \gamma
    \end{array}
    \quad
    \deno{ \dual{\PCF{\colvec{r_1\\\vdots\\r_n}}}} \gamma :=
    \colvec{v_1\\\vdots\\v_n} \mapsto \sum_{i=1}^n r_i\,v_i
  $$
  \vspace{-.5cm}
  \begin{align*}
    \deno{
      \dualmap{(\lambda x.\simterms_1)}{\simterms_2}
    } \gamma & :=
    \lambda v.
      \deno{\simterms_2} \gamma
      \big(\deno{\simterms_1}\anglebr{\gamma,v}\big)
    \\
    \deno{
      \pbmap{\lambda x.\pbterms}{\simterms}
    } \gamma & :=
    \lambda xv.
      \deno{\simterms}\gamma
      (\deno{\pbterms}\anglebr{\gamma,x})
      (D[\curry{\deno{\pbterms}}\gamma]\anglebr{v,x})
  \end{align*}

  \subsection{Correctness}

  We verify our definitions of linearity and substitution
  in Lemma \ref{lemma:linearity of syntax} and
  Lemma \ref{lemma:betasublemma} respectively.

  \begin{restatable}[Linearity]{lemma}{linearity} \label{lemma:linearity of syntax}
    Let $\Gamma_1 \cup \set{x:\sigma_1} \vdash \pbterms_1:\tau$
    and
    $\Gamma_2 \vdash \pbterms_2:\dual{\sigma}$.
    Let $\gamma_1 \in \deno{\Gamma_1}$
    and $\gamma_2 \in \deno{\Gamma_2}$. Then,
    \begin{enumerate}[1.]
      \item
        if $x \in \lin{\pbterms_1}$, then
        $\curry{\deno{\pbterms_1}}\gamma_1$ is linear, \ie
        $D[\curry{\deno{\pbterms_1}}\gamma_1]
        = (\curry{\deno{\pbterms_1}}\gamma_1) \circ \pi_1$,
      \item
        $\deno{\pbterms_2}\gamma$ is linear, \ie
        $D[\deno{\pbterms_2}\gamma] = (\deno{\pbterms_2}\gamma) \circ \pi_1$.
    \end{enumerate}
  \end{restatable}


  \begin{restatable}[Substitution]{lemma}{betasublemma}
    \label{lemma:betasublemma}
    $
      \deno{\Gamma \vdash \simterms[\pbterms/x]:\tau}
      =
      \deno{\Gamma \cup \set{x:\sigma} \vdash \simterms:\tau}
      \circ
      \anglebr{
        \Id_{\deno{\Gamma}},
        \deno{\Gamma \vdash \pbterms:\sigma}
      }
    $
  \end{restatable}

  Any differential $\lambda$-category
  satisfying Hahn-Banach Separation Theorem
  is a sound model of our language.
  Note that the Hahn-Banach Separation Theorem is crucial in the proof.

  \begin{restatable}[Correctness of Reductions]{theorem}{adequacy}
    \label{thm: correctness of reductions}
    Let $\Gamma \vdash \pbterms:\sigma$.
    \begin{enumerate}
      \item $\pbterms \red_A \pbterms'$ implies $\deno{\pbterms} = \deno{\pbterms'}$.
      \item $\pbterms \red \pbterms'$ implies $\deno{\pbterms} = \deno{\pbterms'}$.
    \end{enumerate}
  \end{restatable}

  \begin{proof}
    The full proof can be found in Appendix \ref{appendix:proofs}.
    \begin{enumerate}[1.]
      \item[2.] Case analysis on reductions of pullback terms.
        Consider Reduction $\hyperlink{eq:16.2}{16.2}$.

        Let $\gamma \in \deno{\Gamma}$.
        By IH, and $\proj{\values}{i} \equiv \lambda z.\pbterms'$, we have
        $
          \deno{\pbat{\big(
            \pbmap{\lambda z.\pbterms'}{\omega}
          \big)}{\proj{\values}{j}}}
          =
          \deno{\dualmap{(\lambda v'.\simterms')}{
            \pbat{\omega}{(\pbterms'[\proj{\values}{j}/z])}}
          }
        $
        which means for any 1-form $\phi$ and $v$,
        \begin{align*}
          & \phi\,
          (\deno{\pbterms'} \anglebr{\gamma,\deno{\proj{\values}{j}} \gamma})\,
          \big(
            D[\curry{\deno{\pbterms'}}\gamma]
            \anglebr{v,\deno{\proj{\values}{j}}\gamma}
          \big) \\
          & =
          \phi\,
          (\deno{\pbterms'} \anglebr{\gamma,\deno{\proj{\values}{j}} \gamma})\,
          (\deno{\simterms'}\anglebr{\gamma,v}).
        \end{align*}
        Let $l$ be a linear morphism to $\Real$,
        then $\lambda x.l$ is a 1-form and hence
        we have
        $
          l\big(
            D[\curry{\deno{\pbterms'}}\gamma]
            \anglebr{v,\deno{\proj{\values}{j}}\gamma}
          \big) =
          l(\deno{\simterms'}\anglebr{\gamma,v}).
        $
        By the contra-positive of the
        Hahn-Banach Separation Theorem,
        it implies
        $
          D[\curry{\deno{\pbterms'}}\gamma]
          \anglebr{v,\deno{\proj{\values}{j}}\gamma}
          =
          \deno{\simterms'}\anglebr{\gamma,v}.
        $

        Note that by (D-eval) in \cite{DBLP:journals/mscs/Manzonetto12},
        $
          D[\catev \circ \anglebr{\pi_i,\pi_j}]\anglebr{v,x}
          =
          \pi_i(v)(\pi_j(x)) +
          D[\pi_i(x)]\anglebr{\pi_j(v),\pi_j(x)}.
        $
        Hence we have
        \begin{align*}
          & \deno{
            \pbat{\big(
              \pbmap{\lambda y.\proj{y}{i} \proj{y}{j}}{\omega}
            \big)}{\values}} \gamma \\
          & =
            \lambda v.
            \deno{\omega} \gamma
            \big(\deno{\proj{y}{i} \proj{y}{j}}
              \anglebr{\gamma, \deno{\values}\gamma}
            \big)
            \big(D[\catev\circ\anglebr{\pi_i,\pi_j}]
              \anglebr{v,\deno{\values}\gamma)}
            \big)
          \\
          & =
            \lambda v.
            \deno{\omega} \gamma
            \big(
              \deno{\proj{\values}{i}\,\proj{\values}{j}}\gamma
            \big)
          \\
          & \qquad\qquad\qquad
            \big(
              \proj{v}{i}(\deno{\proj{\values}{j}}\gamma) +
              D[\deno{\proj{\values}{i}}\gamma]
              \anglebr{\proj{v}{j},\deno{\proj{\values}{j}}\gamma}
            \big)
          \\
          & =
            \lambda v.
            \deno{\omega} \gamma
            \big(
              \deno{\proj{\values}{i}\,\proj{\values}{j}}\gamma
            \big)
          \\
          & \qquad\qquad\qquad
            \big(
              \proj{v}{i}(\deno{\proj{\values}{j}}\gamma) +
              D[\curry{\deno{\pbterms'}}\gamma]
              \anglebr{\proj{v}{j},\deno{\proj{\values}{j}}\gamma}
            \big)
          \\
          & =
            \lambda v.
            \deno{\omega} \gamma
            \big(
              \deno{\proj{\values}{i}\,\proj{\values}{j}}\gamma
            \big)
            \big(
              \proj{v}{i}(\deno{\proj{\values}{j}}\gamma) +
              \deno{\simterms'}\anglebr{\gamma,\deno{\proj{\values}{j}}\gamma}
            \big)
          \\
          & =
            \deno{
              \dualmap{(\lambda v.
                \proj{v}{i}\,{\values}_{\pi j} +
                \simterms'[\proj{\values}{j}/v]
              )}{
                \pbat{\omega}{({\proj{\values}{i}}\,{\proj{\values}{j}})}
              }
            }\gamma
        \end{align*}
    \end{enumerate}
  \end{proof}

  A simple corollary of Theorem \ref{thm: correctness of reductions} is that
  types are invariant under reductions.

  \begin{corollary}{(Subject Reduction)}
    \label{cor:subject reduction}
    For any pullback terms $\pbterms$ and $\pbterms'$
    where $\pbterms \red \pbterms'$.
    If $\Gamma \vdash \pbterms:\sigma$,
    then $\Gamma \vdash \pbterms':\sigma$.
  \end{corollary}

  \subsection{Reverse-mode AD}

  Recall performing reverse-mode AD on a real-valued function
  $f:\Real^n \to \Real^m$ at a point $x_0 \in \Real^n$ computes
  a row of the Jacobian matrix $\jacob(f)(x_0)$, \ie
  $\dual{(\jacob(f)(x_0))}(\pi_p)$.

  The following corollary tells us that
  our reduction is faithful to reverse-mode AD
  (in that it is exactly reverse-mode AD when restricted to first-order)
  and
  we can
  perform reverse-mode AD on \emph{any} abstraction
  which might contain higher-order terms, duals, pullbacks
  and free variables.


  \begin{corollary}
    \label{cor:Reverse-mode AD is correct}
    Let $\Gamma \cup \set{y:\sigma} \vdash \pbterms_1:\tau$,
    $\Gamma \vdash \pbterms_2:\sigma$,
    $\gamma \in \deno{\Gamma}$.
    \begin{enumerate}[1.]
      \item Let $\sigma \equiv \PCFReal^n$,
        $\tau \equiv \PCFReal^m$.
        If
        $
          \pbat{\big(
            \pbmap{\lambda y.\pbterms_1}{\pbof{e_p}}
          \big)}{\pbterms_2}
          \redplus
          \values,
        $
        then
        the $p$-th row of the Jacobian matrix of
        $\deno{\pbterms_1}\anglebr{\gamma,-}$
        at $\deno{\pbterms_2}\gamma$ is
        $\dual{(\deno{\values}\gamma)}$.
      \item
        Let $l$ be a linear morphism from $\deno{\tau}$ to $\Real$.
        If
        $
          \pbat{\big(
            \pbmap{\lambda y.\pbterms_1}{\omega}
          \big)}{\pbterms_2}
          \redplus
          \dualmap{(\lambda v.\pbterms'_1)}{\pbat{\omega}{\pbterms'_2}}
        $
        for some fresh variable $\omega$,
        then
        the derivative of $l \circ (\deno{\pbterms_1}\anglebr{\gamma,-})$
        at $\deno{\pbterms_2}\gamma$ along some $v \in \deno{\sigma}$
        is
        $
          l\,(\deno{\pbterms'_1}\anglebr{\gamma,\lambda x.l,v})
        $ \ie
        $
          D[l \circ (\deno{\pbterms_1}\anglebr{\gamma,-})]
          \anglebr{v,\deno{\pbterms_2}\gamma}
          =
          l\,(\deno{\pbterms'_1}\anglebr{\gamma,\lambda x.l,v})
        $
    \end{enumerate}
  \end{corollary}

  \begin{example}
    In Example \ref{eg:running-ex-result},
    we showed that
    $\pbat{\big(
       \pbmap{\lambda \anglebr{x,y}.
         \PCF{\mathsf{pow2}}(\PCF{\mathsf{mult}}(\PCF{g}(\anglebr{x,y})))}
       {\pbof{\colvec{1}}}
     \big)}{\anglebr{\PCF{1},\PCF{3}}}
     \redplus
     \dual{\PCF{
      \colvec{660\\528}
    }}$
    Note that $\colvec{660&528}$ is exactly the
    Jacobian matrix of
    $f: \anglebr{x,y} \mapsto \big((x + 1)(2x + y^2)\big)^2$
    at $\anglebr{1,3}$.
  \end{example}

\section{Related Work}
\label{sec:comparison}
  We discuss recent works on calculi / languages that provide differentiation capabilities.

  \subsection{Differential Lambda-Calculus}

  The standard bearer is none other than
  differential $\lambda$-calculus \cite{DBLP:journals/tcs/EhrhardR03},
  which has inspired the design of our language.

  The implementation induced by
  differential $\lambda$-calculus is
  a form of symbolic differentiation,
  which suffers from expression swell.
  For this reason, \citet{DBLP:journals/entcs/Manzyuk12}
  introduced the \emph{perturbative $\lambda$-calculus},
  a $\lambda$-calculus with a forward-mode AD operator.
  {Our language is complementary to these calculi, in that it implements higher-order reverse-mode AD; moreover, it is call-by-value,
  which is crucial for reverse-mode AD to avoid expression swell, as illustrated in Example \ref{eg:running-ex-pb-red}.
  }

  {What is the relationship between our language and differential $\lambda$-calculus?
  We can give a precise answer via a compositional translation $\trans{(-)}$ to a differential $\lambda$-calculus extended by real numbers, function symbols, pairs and projections, defined as follows:}
  \begin{align*}
    s, t ::= & \
      x \mid
      \lambda x.s \mid
      s\,T \mid
      \diff{s}{t} \mid
      \pi_i(s) \mid
      \anglebr{s,t} \mid
      \PCF{r} \mid
      \PCF{f}(T) \mid
      \diff{\PCF{f}}{t} \\
    S, T ::= & \
      0 \mid
      s \mid
      s+T \qquad \text{where }r \in \Real, f \in \easilydiff
  \end{align*}
  The major cases of the definition of $\trans{(-)}$ are;
  $$
    \begin{array}{rcl}
      \trans{(\dual{\sigma})}
      & \mkern-14mu := \mkern-14mu &
      \trans{\sigma} \Arrow \PCFReal \\[4pt]
      \trans{(\jac{f}(\simterms))}
      & \mkern-14mu := \mkern-14mu &
      \diff{\PCF{f}}{\trans{\simterms}}
    \end{array}
    \quad
    \trans{\dual{\PCF{\colvec{r_1\\\vdots\\r_n}}}}
    := \lambda v.\sum_{i=1}^n \PCF{f_i}(\pi_i(v))
  $$
  \vspace{-.5cm}
  \begin{align*}
    \trans{\big(\dualmap{(\lambda y.\simterms_1)}{\simterms_2}\big)}
      & := \lambda v.\trans{(\simterms_2)}\big((\lambda y.\trans{(\simterms_1)})v\big) \\
    \trans{(\pbmap{\lambda y.\pbterms}{\simterms})}
      & := \lambda xv.
          \trans{\simterms}
          \big((\lambda y.\trans{\pbterms})x\big)
          \big((\diff{(\lambda y.\trans{\pbterms})}{v})x\big)
  \end{align*}
  for $f_i := r_i \times -$. (The definitions are provided in full in Appendix \ref{appendix:extended diff lam cal}.)


  Because differential $\lambda$-calculus does not have linear function type,
  $\trans{(\simterms_1)}$ is no longer in a linear position
  in $\trans{\big(\dualmap{(\lambda x.\simterms_1)}{\simterms_2}\big)}$.
  Though the translation does not preserve linearity,
  it does preserve reductions and interpretations
  (Lemma \ref{lemma:translation respects reduction and model}).

  \begin{restatable}{lemma}{TransRedModel}
    \label{lemma:translation respects reduction and model}
    Let $\pbterms$ be a term.
    \begin{enumerate}[1.]
      \item
        If $\pbterms \red \pbterms'$, then
        there exists a reduct $s$ of $\trans{\pbterms'}$ such that
        $\trans{\pbterms} \redplus s$ in $\lang{D}$.
      \item
        $\deno{\pbterms} = \deno{\trans{\pbterms}}$ in $\cat{C}$.
    \end{enumerate}
  \end{restatable}


  A corollary of Lemma \ref{lemma:translation respects reduction and model} (1) is that our reduction strategy is strongly normalizing.

  \begin{restatable}[Strong Normalization]{corollary}{StrongNorm}
    \label{cor:strong normalization}
    {
    Any reduction sequence from any term is finite, and ends in a value.}
  \end{restatable}

  \subsection{Differentiable Programming Languages}

  Encouraged by calls \cite{Olah15,LeCun,Dalrymple} from the machine learning community,
  the development of \emph{reverse-mode AD programming language} has been an active research problem.
  Following \citet{DBLP:journals/toplas/PearlmutterS08},
  these languages usually treat reverse-mode AD
  as a \emph{meta}-operator on programs.

  \paragraph{First-order}
  \citet{DBLP:journals/pacmpl/Elliott18} gives a categorical presentation
  of reverse-mode AD.
  Using a functor over Cartesian categories,
  he presents a neat implementation of reverse-mode AD.

  As is well-known, conditional does not behave well with smoothness \cite{BECK1994119};
  nor does loops and recursion.
  \citet{DBLP:journals/pacmpl/AbadiP20}
  address this problem via a
  first-order language with conditionals,  recursively defined functions,
  and a construct for reverse-mode AD.
  Using real analysis, they prove the coincidence of operational and denotational semantics.

  To our knowledge, these treatments of reverse-mode AD are restricted to first-order functions.

  \paragraph{Towards higher-order}
  The first work that extends reverse-mode AD to higher orders is by \citet{DBLP:journals/toplas/PearlmutterS08};
  they use a non-compositional program transformation to implement reverse-mode AD.

  Inspired by \citet{DBLP:conf/nips/WangDWER18,DBLP:journals/pacmpl/WangZDWER19}, \citet{DBLP:journals/corr/abs-1909-13768} study a simply-typed $\lambda$-calculus augmented with a notion of linear negation type.
  Though our dual type may resemble their linear negation, they are actually quite different.
  In fact, our work can be viewed as providing a positive answer to the last paragraph of \cite[Sec.~7]{DBLP:journals/corr/abs-1909-13768}, where the authors address the relation between their work and differential lambda-calculus.
  They describe a ``na\"{i}ve'' approach of expressing reverse-mode AD in differential lambda-calculus in the sense that it suffers from ``expression swell'', which our approach does not (see Example~\ref{eg:running-ex-pb-red}).
  Moreover, Brunel et al.~use a program transformation to perform reverse-mode AD, whereas we use a first-class differential operator. Brunel et al. [1] prove correctness for performing reverse-mode AD on real-valued functions (Theorem 5.6, Corollary 5.7 in [1]), whereas we allow \emph{any} (higher-order) abstraction to be the argument of the pullback term and proved that the result of the reduction of such a pullback term is exactly the derivative of the abstraction (Corollary~\ref{cor:Reverse-mode AD is correct}).

  Building on \citet{DBLP:journals/pacmpl/Elliott18}'s categorical presentation of reverse-mode AD,
  and \citet{DBLP:journals/toplas/PearlmutterS08}'s idea of differentiating higher-order functions,
  \citet{Abadi} developed an implementation of a simply-typed differentiable programming language.

  {However, all these treatments are not \emph{purely} higher-order, in the sense that their differential operator can only compute the derivative of an ``end to end'' first-order program (which may be constructed using higher-order functions), but not the derivative of a higher-order function.}

  {As far as we know,
  our work gives the first implementation of reverse-mode AD in a higher-order programming language that directly computes the derivative of higher-order functions using reverse-mode AD (Corollary \ref{cor:Reverse-mode AD is correct} (2)).}

\section{Conclusion and Future Directions}
\label{sec:conclusion}




  After outlining the
  mathematical foundation of reverse-mode AD
  as the pullback of differential 1-forms (Section \ref{sec:reverse-mode AD}),
  we presented
  a simple higher-order programming language with
  an explicit differential operator,
  $\pbmap{(\lambda x.\pbterms)}{\simterms}$, (Subsection \ref{subsec:syntax})
  and
  a call-by-value reduction strategy
  to divide (A-reductions in Subsection \ref{subsec:A-reduction}),
  conquer (pullback reductions in Subsection \ref{subsec:pb reduction})
  and combine (Subsection \ref{subsec:combine reduction})
  the term
  $\big(\pbmap{(\lambda x.\pbterms)}{\omega}\big)\simterms$,
  such that its reduction exactly mimics reverse-mode AD.
  Examples are given to illustrate that our reduction
  is faithful to reverse-mode AD.
  Moreover, we show how our reduction can be adapted to a CPS evaluation
  (Subsection \ref{subsec:cps}).

  We showed (in Section \ref{sec:model}) that
  any differential $\lambda$-category that satisfies the
  Hahn-Banach Separation Theorem is a sound model of our language
  (Theorem \ref{thm: correctness of reductions}) and
  how our reduction precisely captures the notion of reverse-mode AD,
  in both first-order and higher-order settings
  (Corollary \ref{cor:Reverse-mode AD is correct}).

  \paragraph{Future Directions.}

  An interesting direction is to extend our language with probability, which can serve as a compiler intermediate representation for ``deep'' probabilistic frameworks
  such as Edward \cite{DBLP:journals/corr/TranHSBMB17} and Pyro \cite{Pyro}.
  Inference algorithms
  that require the computation of gradients,
  such as Hamiltonian Monte Carlo and variational inference,
  which Edward and Pyro rely on,
  can be expressed in such a language
  and allows us to prove correctness.

\bibliographystyle{abbrvnat}

\bibliography{database.bib}

\newpage
\clearpage

\section*{Appendix}
\appendix

  \section{Examples}

  \subsection{Simple Example}
  \label{appendix:simple example}

  \begin{figure*}
    \begin{tabular}{lc}
      Na\"{i}ve Forward Mode: &
      $$
      \xymatrix@C=17pt{
        \config{\anglebr{1,3}}{
          \colvec{
            1 & 0 \\
            0 & 1
          }
        }
          \ar[r]^-{g}
        &
        \config{\anglebr{\overset{1+1}{2},\overset{2*1+3^2}{11}}}{
          \colvec{
            1 & 0 \\
            2 & 6
          }
        }
          \ar[r]^-{*}
        &
        \config{\overset{2*11}{22}}{
          \colvec{
            15 & 12
          }
        }
          \ar[r]^-{(-)^2}
        &
        \config{\overset{22^2}{484}}{
          \colvec{
            660 & 528
          }
        }
      }
      $$ \\
      Forward Mode: &
      $$
      \xymatrix@R=3pt{
        \config{\anglebr{1,3}}{\colvec{1\\0}}
          \ar[r]^-{g}
        &
        \config{\anglebr{2,11}}{
          \colvec{
            1 \\
            2
          }
        }
          \ar[r]^-{*}
        &
        \config{22}{
          \colvec{
            15
          }
        }
          \ar[r]^-{(-)^2}
        &
        \config{484}{
          \colvec{
            660
          }
        }
      }
      $$ \\
      Reverse Mode: &
      $$
      \xymatrix@R=3pt{
        \text{Forward Phase: }
        &
        \anglebr{1,3}
          \ar[r]^-{g}
        &
        \anglebr{2,11}
          \ar[r]^-{*}
        &
        22
          \ar[r]^-{(-)^2}
        &
        484
        \\
        \text{Reverse Phase: }
        &
        {
          \colvec{
            660 \\
            528
          }
        }
        &
        {
          \colvec{
            484 \\
            88
          }
        }
          \ar[l]_-{g}
        &
        {
          \colvec{
            44
          }
        }
          \ar[l]_-{*}
        &
        \colvec{1}
          \ar[l]_-{(-)^2}
      }
      $$ \\
      Pullback: & {$
      \begin{array}{rl}
        & \big(\oneform(g) \circ \oneform(*) \circ \oneform((-)^2)\big)
        (\lambda x.\colvec{1})(\anglebr{1,3}) \\
        & =
        \dual{(\jacob(g)(\anglebr{1,3}))}\big(\oneform(*) \circ \oneform((-)^2)\big)
        (\lambda x.\colvec{1})(\anglebr{2,11}) \\
        & =
        \dual{(\jacob(g)(\anglebr{1,3}))}\dual{(\jacob(*)(\anglebr{2,11}))}
        \big(\oneform((-)^2)\big)
        (\lambda x.\colvec{1})(22) \\
        & =
        \dual{(\jacob(g)(\anglebr{1,3}))}\dual{(\jacob(*)(\anglebr{2,11}))}
        \dual{(\jacob((-)^2)(22))}
        \big((\lambda x.\colvec{1})(484)\big) \\
        & =
        \dual{(\jacob(g)(\anglebr{1,3}))}\dual{(\jacob(*)(\anglebr{2,11}))}
        \colvec{44} \\
        & =
        \dual{(\jacob(g)(\anglebr{1,3}))}\colvec{484\\88} \\
        & =
        \colvec{660\\528}
      \end{array}
      $}
    \end{tabular}
    \caption{
      Different modes of automatic differentiation performed on the function
      $f: \anglebr{x,y} \mapsto \big((x + 1)(2x + y^2)\big)^2$
      at $\anglebr{1,3}$,
      after $f$ is decomposed into elementary functions:
      $
        \Real^2
          \xrightarrow{\ g\ }
        \Real^2
          \xrightarrow{\ *\ }
        \Real
          \xrightarrow{(-)^2}
        \Real,
      $
      where $g(\anglebr{x,y}) := \anglebr{x+1,2x+y^2}$.
    }
    \label{fig: AD example}
  \end{figure*}

  We focus on how to compute the derivative of
  $f: \anglebr{x,y} \mapsto \big((x + 1)(2x + y^2)\big)^2$
  at $\anglebr{1,3}$ by different modes of AD.

  First $f$ is decomposed into elementary functions as
  $
    \Real^2
      \xrightarrow{\ g\ }
    \Real^2
      \xrightarrow{\ *\ }
    \Real
      \xrightarrow{(-)^2}
    \Real,
  $
  where $g(\anglebr{x,y}) := \anglebr{x+1,2x+y^2}$.
  Then, Figure \ref{fig: AD example} summarize the iterations
  of different modes of AD.

  Now we show how Section \ref{sec:dppl} tells us how to perform
  reverse-mode AD on $f$.

  \paragraph{Term}
  Assuming $g, \mathsf{mult}, \mathsf{pow2} \in \easilydiff$,
  we can define the following term in the language.
  $$
    \vdash
    \pbat{\big(
      \pbmap{\lambda \anglebr{x,y}.
        \PCF{\mathsf{pow2}}(\PCF{\mathsf{mult}}(\PCF{g}(\anglebr{x,y})))}
      {(\pbof{\colvec{1}})}
    \big)}{\anglebr{\PCF{1},\PCF{3}}}
    : \dual{\PCFReal^2}
  $$
  This term is the application of
  the pullback
  $\Omega(f)(\lambda x.\dual{\colvec{1}})$
  to the point $\anglebr{1,3}$,
  which is exactly the Jacobian of $f$ at $\anglebr{1,3}$.

  \paragraph{Administrative Reduction}

  We decompose the term
  $\PCF{\mathsf{pow2}}(\PCF{\mathsf{mult}}(\PCF{g}(\anglebr{x,y})))$,
  via administrative reduction,
  into a let series of elementary terms.
  $$
    \PCF{\mathsf{pow2}}(\PCF{\mathsf{mult}}(\PCF{g}(\anglebr{x,y})))
    \Aredplus
    \letterms \equiv
    \letarray{
      z_1 & \mkern-14mu = \anglebr{x,y}; \\
      z_2 & \mkern-14mu = \PCF{g}(z_1); \\
      z_3 & \mkern-14mu = \PCF{\mathsf{mult}}(z_2); \\
      z_4 & \mkern-14mu = \PCF{\mathsf{pow2}}(z_3)
    }{z_4.}
  $$
  This is reminiscent {of} the {decomposition} of $f$ into
  $
    \Real^2
      \xrightarrow{\ g\ }
    \Real^2
      \xrightarrow{\ *\ }
    \Real
      \xrightarrow{(-)^2}
    \Real
  $
  before performing AD.

  \paragraph{Splitting the Omega}

  Now via reduction $\hyperlink{eq:7}{7}$ and $\hyperlink{eq:8}{8}$,
  $\pbmap{\lambda \anglebr{x,y}.
    \letterms
  }{\omega}$
  is reduced to a series of pullback along elementary terms.
  \begin{align*}
    & \pbmap{\lambda \anglebr{x,y}.
      \letarray{
        z_1 & \mkern-14mu = \anglebr{x,y}; \\
        z_2 & \mkern-14mu = \PCF{g}(z_1); \\
        z_3 & \mkern-14mu = \PCF{\mathsf{mult}}(z_2); \\
        z_4 & \mkern-14mu = \PCF{\mathsf{pow2}}(z_3)
      }{z_4}
    }{\omega} \\
    & \Predplus
      \mat{
        \pbmap{\lambda \anglebr{x,y}.\anglebr{\anglebr{x,y},\anglebr{x,y}}}{} \\
        \pbmap{\lambda \anglebr{\anglebr{x,y},z_1}.
          \anglebr{\anglebr{x,y},z_1,\PCF{g}(z_1)}}{} \\
        \pbmap{\lambda \anglebr{\anglebr{x,y},z_1,z_2}.
          \anglebr{\anglebr{x,y},z_1,z_2,\PCF{\mathsf{mult}}(z_2)}}{} \\
        \pbmap{\lambda \anglebr{\anglebr{x,y},z_1,z_2,z_3}.
          \PCF{\mathsf{pow2}}(z_3)
        }{\omega}
      }
  \end{align*}

  \paragraph{Pullback Reduction}

  We showed that via A-reductions and
  Reductions $\hyperlink{eq:7}{7}$ and $\hyperlink{eq:8}{8}$,
  $\pbmap{\lambda \anglebr{x,y}.
    \PCF{\mathsf{pow2}}(\PCF{\mathsf{mult}}(\PCF{g}(\anglebr{x,y})))}
  {\omega}$
  is reduced to
  $$
    \mat{
      \pbmap{\lambda \anglebr{x,y}.\anglebr{\anglebr{x,y},\anglebr{x,y}}}{} \\
      \pbmap{\lambda \anglebr{\anglebr{x,y},z_1}.
        \anglebr{\anglebr{x,y},z_1,\PCF{g}(z_1)}}{} \\
      \pbmap{\lambda \anglebr{\anglebr{x,y},z_1,z_2}.
        \anglebr{\anglebr{x,y},z_1,z_2,\PCF{\mathsf{mult}}(z_2)}}{} \\
      \pbmap{\lambda \anglebr{\anglebr{x,y},z_1,z_2,z_3}.
        \PCF{\mathsf{pow2}}(z_3)
      }{\omega}
    }
  $$

  We show how it can be reduced when applied to $\anglebr{\PCF{1},\PCF{3}}$.
  \begin{align*}
    &
    \mat{
      \pbmap{\lambda \anglebr{x,y}.\anglebr{\anglebr{x,y},\anglebr{x,y}}}{} \\
      \pbmap{\lambda \anglebr{\anglebr{x,y},z_1}.
        \anglebr{\anglebr{x,y},z_1,\PCF{g}(z_1)}}{} \\
      \pbmap{\lambda \anglebr{\anglebr{x,y},z_1,z_2}.
        \anglebr{\anglebr{x,y},z_1,z_2,\PCF{\mathsf{mult}}(z_2)}}{} \\
      \pbmap{\lambda \anglebr{\anglebr{x,y},z_1,z_2,z_3}.
        \PCF{\mathsf{pow2}}(z_3)
      }{\omega}
    }\,\PCF{\colvec{1\\3}} \\
    & \Predtextm{\hyperlink{eq:20.1}{20.1}}{\hyperlink{eq:11}{11}}
    \mat{
      \dualmap{(\lambda \anglebr{v_1,v_2}.
        \anglebr{\anglebr{v_1,v_2},\anglebr{v_1,v_2}}
      )}{} \\
      \mat{
        \pbmap{\lambda \anglebr{\anglebr{x,y},z_1}.
          \anglebr{\anglebr{x,y},z_1,\PCF{g}(z_1)}}{} \\
        \pbmap{\lambda \anglebr{\anglebr{x,y},z_1,z_2}.
          \anglebr{\anglebr{x,y},z_1,z_2,\PCF{\mathsf{mult}}(z_2)}}{} \\
        \pbmap{\lambda \anglebr{\anglebr{x,y},z_1,z_2,z_3}.
          \PCF{\mathsf{pow2}}(z_3)
        }{\omega}
      }\,
      \anglebr{\PCF{\colvec{1\\3}},\PCF{\colvec{1\\3}}}
    }
    \\
    & \Predtextm{\hyperlink{eq:20.1}{20.1}}{\hyperlink{eq:14}{14},
    \hyperlink{eq:3}{3}}
    \mat{
      \dualmap{(\lambda \anglebr{v_1,v_2}.
        \anglebr{\anglebr{v_1,v_2},\anglebr{v_1,v_2}}
      )}{} \\
      \dualmap{(\lambda \anglebr{\anglebr{v_1,v_2},v_3}.
        \anglebr{
          \anglebr{v_1,v_2},v_3,
          (\jac{g}{v_3})\anglebr{\PCF{1},\PCF{3}}}
      )}{}
      \\
      \Big(
        \mat{
          \pbmap{\lambda \anglebr{\anglebr{x,y},z_1,z_2}.
            \anglebr{\anglebr{x,y},z_1,z_2,\PCF{\mathsf{mult}}(z_2)}}{}
          \\
          {(
            \pbmap{\lambda \anglebr{\anglebr{x,y},z_1,z_2,z_3}.\PCF{\mathsf{pow2}}(z_3)
            }{\omega}
          )}
        }
        \anglebr{
          \PCF{\colvec{1\\3}},
          \PCF{\colvec{1\\3}},
          \PCF{\colvec{2\\11}}}
      \Big)
    } \tag{$\star$} \\
    & \Predtextm{\hyperlink{eq:20.1}{20.1}}{\hyperlink{eq:14}{14},
    \hyperlink{eq:3}{3}}
    \mat{
      \dualmap{(\lambda \anglebr{v_1,v_2}.
        \anglebr{\anglebr{v_1,v_2},\anglebr{v_1,v_2}}
      )}{} \\
      \dualmap{(\lambda \anglebr{\anglebr{v_1,v_2},v_3}.
        \anglebr{
          \anglebr{v_1,v_2},v_3,
          (\jac{g}{v_3})\anglebr{\PCF{1},\PCF{3}}}
      )}{} \\
      \dualmap{(\lambda \anglebr{\anglebr{v_1,v_2},v_3,v_4}.
        \anglebr{
          \anglebr{v_1,v_2},v_3,v_4,
          (\jac{\mathsf{mult}}{v_4})\anglebr{\PCF{2},\PCF{11}}}
      )}{} \\
      \big(
        \mat{
          \pbmap{\lambda \anglebr{\anglebr{x,y},z_1,z_2,z_3}.\PCF{\mathsf{pow2}}(z_3)
          }{\omega}
        }
        \anglebr{
          \PCF{\colvec{1\\3}},
          \PCF{\colvec{1\\3}},
          \PCF{\colvec{2\\11}},
          \PCF{22}
        }
      \big)
    } \\
    & \Predtextm{\hyperlink{eq:20.1}{20.1}}{\hyperlink{eq:14}{14},
    \hyperlink{eq:3}{3}}
    \mat{
      \dualmap{(\lambda \anglebr{v_1,v_2}.
        \anglebr{\anglebr{v_1,v_2},\anglebr{v_1,v_2}}
      )}{} \\
      \dualmap{(\lambda \anglebr{\anglebr{v_1,v_2},v_3}.
        \anglebr{
          \anglebr{v_1,v_2},v_3,
          (\jac{g}{v_3})\anglebr{\PCF{1},\PCF{3}}}
      )}{} \\
      \dualmap{(\lambda \anglebr{\anglebr{v_1,v_2},v_3,v_4}.
        \anglebr{
          \anglebr{v_1,v_2},v_3,v_4,
          (\jac{\mathsf{mult}}{v_4})\anglebr{\PCF{2},\PCF{11}}}
      )}{} \\
      \dualmap{(\lambda \anglebr{\anglebr{v_1,v_2},v_3,v_4,v_5}.
          (\jac{\mathsf{pow2}}{v_5})\PCF{22}
      )}{(\pbat{\omega}{
        \PCF{484}
      })}
    }
  \end{align*}
  Notice how
  this is reminiscent of the forward phase of reverse-mode AD
  performed on $f: \anglebr{x,y} \mapsto \big((x + 1)(2x + y^2)\big)^2$
  at $\anglebr{1,3}$ considered in Figure \ref{fig: AD example}.

  Moreover,
  we used the reduction
  $\PCF{f}(\PCF{r})\xrightarrow{\hyperlink{eq:3}{3}}\PCF{f(r)}$
  couples of times in the argument position of an application.
  This is to avoid expression swell.
  Note $1+1$ is only evaluated once in $(\star)$
  even when the result is used in various computations.

  \paragraph{Combine}

  Replacing $\omega$ by
  $\pbof{\colvec{1}} \equiv \lambda x.\dual{\colvec{1}}$,
  we have shown so far that
  $$
    \pbat{\big(
        \pbmap{\lambda \anglebr{x,y}.
          \PCF{\mathsf{pow2}}(\PCF{\mathsf{mult}}(\PCF{g}(\anglebr{x,y})))}
        {\pbof{\colvec{1}}}
      \big)}{\anglebr{\PCF{1},\PCF{3}}}
  $$
  is reduced to
  $$
    \mat{
      \dualmap{(\lambda \anglebr{v_1,v_2}.
        \anglebr{\anglebr{v_1,v_2},\anglebr{v_1,v_2}}
      )}{} \\
      \dualmap{(\lambda \anglebr{\anglebr{v_1,v_2},v_3}.
        \anglebr{
          \anglebr{v_1,v_2},v_3,
          (\jac{g}{v_3})\anglebr{\PCF{1},\PCF{3}}}
      )}{} \\
      \dualmap{(\lambda \anglebr{\anglebr{v_1,v_2},v_3,v_4}.
        \anglebr{
          \anglebr{v_1,v_2},v_3,v_4,
          (\jac{\mathsf{mult}}{v_4})\anglebr{\PCF{2},\PCF{11}}}
      )}{} \\
      \dualmap{(\lambda \anglebr{\anglebr{v_1,v_2},v_3,v_4,v_5}.
          (\jac{\mathsf{pow2}}{v_5})\PCF{22}
      )}{(\pbat{\omega}{
        \PCF{484}
      })}
    }.
  $$
  Now via reduction $\hyperlink{eq:5}{5}$ and $\beta$ reduction,
  we further reduce it to
  \begin{align*}
    & \mat{
      \dualmap{(\lambda \anglebr{v_1,v_2}.
        \anglebr{\anglebr{v_1,v_2},\anglebr{v_1,v_2}}
      )}{} \\
      \dualmap{(\lambda \anglebr{\anglebr{v_1,v_2},v_3}.
        \anglebr{
          \anglebr{v_1,v_2},v_3,
          (\jac{g}{v_3})\anglebr{\PCF{1},\PCF{3}}}
      )}{} \\
      \dualmap{(\lambda \anglebr{\anglebr{v_1,v_2},v_3,v_4}.
        \anglebr{
          \anglebr{v_1,v_2},v_3,v_4,
          (\jac{\mathsf{mult}}{v_4})\anglebr{\PCF{2},\PCF{11}}}
      )}{} \\
      \dualmap{(\lambda \anglebr{\anglebr{v_1,v_2},v_3,v_4,v_5}.
          (\jac{\mathsf{pow2}}{v_5})\PCF{22}
      )}{\PCF{\colvec{1}}}
    } \\
    & \red
    \mat{
      \dualmap{(\lambda \anglebr{v_1,v_2}.
        \anglebr{\anglebr{v_1,v_2},\anglebr{v_1,v_2}}
      )}{} \\
      \dualmap{(\lambda \anglebr{\anglebr{v_1,v_2},v_3}.
        \anglebr{
          \anglebr{v_1,v_2},v_3,
          (\jac{g}{v_3})\anglebr{\PCF{1},\PCF{3}}}
      )}{} \\
      \dualmap{(\lambda \anglebr{\anglebr{v_1,v_2},v_3,v_4}.
        \anglebr{
          \anglebr{v_1,v_2},v_3,v_4,
          (\jac{\mathsf{mult}}{v_4})\anglebr{\PCF{2},\PCF{11}}}
      )}{}
    }
    \dual{\PCF{
      \colvec{0\\0\\0\\0\\0\\0\\44}
    }}
    \\
    & \red
    \mat{
      \dualmap{(\lambda \anglebr{v_1,v_2}.
        \anglebr{\anglebr{v_1,v_2},\anglebr{v_1,v_2}}
      )}{} \\
      \dualmap{(\lambda \anglebr{\anglebr{v_1,v_2},v_3}.
        \anglebr{
          \anglebr{v_1,v_2},v_3,
          (\jac{g}{v_3})\anglebr{\PCF{1},\PCF{3}}}
      )}{}
    }
    \dual{\PCF{
      \colvec{0\\0\\0\\0\\484\\88}
    }}
    \\
    & \red
    \mat{
      \dualmap{(\lambda \anglebr{v_1,v_2}.
        \anglebr{\anglebr{v_1,v_2},\anglebr{v_1,v_2}}
      )}{}
    }
    \dual{\PCF{
      \colvec{0\\0\\660\\528}
    }}
    \\
    & \red
    \dual{\PCF{
      \colvec{660\\528}
    }}
  \end{align*}
  Notice how this mimics the reverse phase of reverse-mode AD
  on $f: \anglebr{x,y} \mapsto \big((x + 1)(2x + y^2)\big)^2$ at $\anglebr{1,3}$
  considered in Figure \ref{fig: AD example}.

  \subsection{Sum Example}
  \label{appendix:sum example}

  Consider the function that takes a list of real numbers and
  returns the sum of the elements of a list.
  We show how Section \ref{sec:dppl} tells us how to perform
  reverse-mode AD on such a higher-order function.

  \paragraph{Term}

  Using the standard Church encoding of List, \ie
  \begin{align*}
    List(X) & \equiv (X \to D \to D) \to (D \to D) \\
    [x_1,x_2,\dots, x_n] & \equiv
    \lambda fd.f\,x_n\big(\dots (f\,x_2\,(f\,x_1\,d))\big)
  \end{align*}
  for some dummy type $D$,
  $\mathsf{sum}:List(\PCFReal)\to \PCFReal$ can be expressed
  in our language described in Section \ref{sec:dppl} to be
  $\lambda l. l\,(\lambda xy.x+y)\,\PCF{0}$.
  Hence the derivative of $\mathsf{sum}$ at a list
  $[\PCF{7},\PCF{-1}]$ can be expressed as
  $$
    \set{\omega:\oneform(List(\PCFReal))} \vdash
    \big(\pbmap{(\mathsf{sum})}{\omega}\big)\,[\PCF{7},\PCF{-1}]:
    \dual{\PCFReal}.
  $$

  \paragraph{Administrative Reduction}

  We first decompose the body of the
  $\mathsf{sum}:List(\PCFReal)\to \PCFReal$ term,
  considered in Example \ref{eg:running-sum-term}, \ie
  $l\,(\lambda xy.x+y)\,\PCF{0}$
  via administrative reduction described in Subsection \ref{subsec:A-reduction}.
  \begin{align*}
    & l\,(\lambda xy.x+y)\,\PCF{0} \\
    & \Aredplus
      \big(
        (\Let{z_1'=l}{z_1'})\,
        (\lambda xy.\Let{z_2'=x+y}{z_2'})\,
      \big) \\
    & \qquad\qquad\qquad\qquad\qquad\qquad\qquad\qquad\quad
      (\Let{z_3'=0}{z_3'}) \\
    & \Aredplus
      \bigg(
        \letarray{
          z_1 & = l; \\
          z_2 & = \lambda xy.(\Let{z_2'=x+y}{z_2'}); \\
          z_3 & = z_1\,z_2
        }{z_3}
      \bigg)
      (\Let{z_3'=0}{z_3'}) \\
    & \Aredplus
        \letarray{
          z_1 & = l; \\
          z_2 & = \lambda xy.(\Let{z_2'=x+y}{z_2'}); \\
          z_3 & = z_1\,z_2; \\
          z_4 & = 0; \\
          z_5 & = z_3\,z_4
        }{z_5}
  \end{align*}

  \paragraph{Splitting the Omega}

  After the A-reductions where
  $l\,(\lambda xy.x+y)\,\PCF{0}$
  is A-reduced to a let series,
  we reduce
  $\pbmap{(\lambda l. l\,(\lambda xy.x+y)\,\PCF{0})}{\omega}$,
  via Reductions \hyperlink{eq:7}{7} and \hyperlink{eq:8}{8}.
  \begin{align*}
    &
    \pbmap{(\lambda l. l\,(\lambda xy.x+y)\,\PCF{0})}{\omega} \\
    & \Aredplus
    \pbmap{\lambda l.
      \letarray{
        z_1 & = l; \\
        z_2 & = \lambda xy.\Let{z_2'=x+y}{z_2'}; \\
        z_3 & = z_1\,z_2; \\
        z_4 & = 0; \\
        z_5 & = z_3\,z_4
      }{z_5}
    }{\omega} \\
    & \Predplus
      \mat{
        \pbmap{\lambda l.\anglebr{l,l}}{} \\
        \pbmap{\lambda \anglebr{l,z_1}.
          \anglebr{l,z_1,\lambda xy.\letterms}}{} \\
        \pbmap{\lambda \anglebr{l,z_1,z_2}.
          \anglebr{l,z_1,z_2,z_1\,z_2}}{} \\
        \pbmap{\lambda \anglebr{l,z_1,z_2,z_3}.
          \anglebr{l,z_1,z_2,z_3,0}}{} \\
        \pbmap{\lambda \anglebr{l,z_1,z_2,z_3,z_4}.z_3\,z_4
        }{\omega}
      }
  \end{align*}

  \paragraph{Pullback Reduction}

  \begin{figure*}
    \begin{align*}
      & \big(
        \pbmap{[\PCF{7},\PCF{-1}]}{\omega'}
      \big)(\lambda xy.\letterms) \\
      & \equiv
      \big(
        \pbmap{\lambda fd.f\,\PCF{-1}\,(f\,\PCF{7}\,d)}{\omega'}
      \big)(\lambda xy.\letterms) \\
      & \Aredplus
      \Bigg(\
        \pbmap{\lambda fd.
          \letarray{
            z_1 & = f \\
            z_2 & = \PCF{-1} \\
            z_3 & = z_1\,z_2 \\
            z_4 & = f \\
            z_5 & = \PCF{7} \\
            z_6 & = z_4\,z_5 \\
            z_7 & = d \\
            z_8 & = z_6\,z_7 \\
            z_9 & = z_3\,z_8
          }{z_9}
        }{\omega'}
      \Bigg)(\lambda xy.\letterms) \\
      & \Predplus
      \mat{
        \pbmap{\lambda f.\anglebr{f,f}}{} \\
        \pbmap{\lambda \anglebr{f,z_1}.
          \anglebr{f,z_1,\PCF{-1}}}{} \\
        \pbmap{\lambda \anglebr{f,z_1,z_2}.
          \anglebr{f,z_1,z_2,z_1\,z_2}}{} \\
        \pbmap{\lambda \anglebr{f,z_1,z_2,z_3}.
          \anglebr{f,z_1,z_2,z_3,f}}{} \\
        \pbmap{\lambda \anglebr{f,z_1,z_2,z_3,z_4}.
          \anglebr{f,z_1,z_2,z_3,z_4,\PCF{4}}}{} \\
        \pbmap{\lambda \anglebr{f,z_1,z_2,z_3,z_4,z_5}.
          \anglebr{f,z_1,z_2,z_3,z_4,z_5,z_4\,z_5}}{} \\
        \pbmap{\lambda \anglebr{f,z_1,z_2,z_3,z_4,z_5,z_6}.
          \anglebr{f,z_1,z_2,z_3,z_4,z_5,z_6,d}}{} \\
        \pbmap{\lambda \anglebr{f,z_1,z_2,z_3,z_4,z_5,z_6,z_7}.
          \anglebr{f,z_1,z_2,z_3,z_4,z_5,z_6,z_7,z_6\,z_7}}{} \\
        \pbmap{\lambda \anglebr{f,z_1,z_2,z_3,z_4,z_5,z_6,z_7,z_8}.
          \anglebr{f,z_1,z_2,z_3,z_4,z_5,z_6,z_7,z_8,z_3\,z_8}
        }{\omega'}
      }\,(\lambda xy.\letterms) \\
      & \Predplus
      \mat{
        \dualmap{(\lambda v.\anglebr{v,v})}{} \\
        \dualmap{(\lambda \anglebr{v,v_1}.\anglebr{v,v_1,0})}{} \\
        \dualmap{(\lambda \anglebr{v,v_1,v_2}.
          \anglebr{v,v_1,v_2,
            v_1\PCF{-1} +
            \lambda y.(\jac{{+}}{\anglebr{v_2,0}})\,\anglebr{\PCF{-1},y}
        })}{} \\
        \dualmap{(\lambda \anglebr{v,v_1,v_2,v_3}.
          \anglebr{v,v_1,v_2,v_3,v})}{} \\
        \dualmap{(\lambda \anglebr{v,v_1,v_2,v_3,v_4}.
          \anglebr{v,v_1,v_2,v_3,v_4,0})}{} \\
        \dualmap{(\lambda \anglebr{v,v_1,v_2,v_3,v_4,v_5}.
          \anglebr{v,v_1,v_2,v_3,v_4,v_5,
            v_4\PCF{7} +
            \lambda y.(\jac{{+}}{\anglebr{v_5,0}})\,\anglebr{\PCF{7},y}
        })}{} \\
        \dualmap{(\lambda \anglebr{v,v_1,v_2,v_3,v_4,v_5,v_6}.
          \anglebr{v,v_1,v_2,v_3,v_4,v_5,v_6,0})}{} \\
        \dualmap{(\lambda \anglebr{v,v_1,v_2,v_3,v_4,v_5,v_6,v_7}.
          \anglebr{v,v_1,v_2,v_3,v_4,v_5,v_6,v_7,
            v_6\,d +
            (\jac{+(\anglebr{7,-})}{v_7})\,d
        })}{} \\
        \dualmap{(\lambda \anglebr{v,v_1,v_2,v_3,v_4,v_5,v_6,v_7,v_8}.
          \anglebr{v,v_1,v_2,v_3,v_4,v_5,v_6,v_7,v_8,
            v_3\,(+(\anglebr{\PCF{7},d})) +
            (\jac{+(\anglebr{-1,-})}{v_8})\,(+(\anglebr{\PCF{7},d}))
        })}{\omega'\,A}
      } \\
      & \Predplustext{\ 6\ }
      \dualmap{(
        \lambda v.
          v\PCF{-1}(\PCF{+}(\anglebr{\PCF{7},d})) +
          (\jac{+(\anglebr{-1,-})}{(v\PCF{7}d)})\,(\PCF{+}\anglebr{\PCF{7},d})
      )}{\omega'\,A}
    \end{align*}
    where
    $A \equiv \anglebr{
      \lambda xy.\letterms,
      \lambda xy.\letterms,
      \PCF{-1},
      \lambda y.\PCF{+}(\anglebr{\PCF{-1},y}),
      \lambda xy.\letterms,
      \PCF{7},
      \lambda y.\PCF{+}(\anglebr{\PCF{7},y}),
      d,
      \PCF{+}(\anglebr{\PCF{7},d})
    }$
    \caption{
      Reduction of $\big(
        \pbmap{[\PCF{7},\PCF{-1}]}{\omega'}
      \big)(\lambda xy.\letterms)$
    }
    \label{fig:reduction of list}
  \end{figure*}

  First,
  Figure \ref{fig:reduction of list} shows that
  $
    \big(
      \pbmap{[\PCF{7},\PCF{-1}]}{\omega'}
    \big)(\lambda xy.\letterms)
  $
  is reduced to
  $
    \dualmap{(
      \lambda v.
        v\PCF{-1}(\PCF{+}(\anglebr{\PCF{7},d})) +
        (\jac{+(\anglebr{-1,-})}{(v\PCF{7}d)})\,(\PCF{+}\anglebr{\PCF{7},d})
    )}{\omega'\,A}
  $

  Then, we reduce
  $\big(\pbmap{(\mathsf{sum})}{\omega}\big)\,[\PCF{7},\PCF{-1}]$
  as follows.
  \begin{align*}
    &
    \mat{
      \pbmap{\lambda l.\anglebr{l,l}}{} \\
      \pbmap{\lambda \anglebr{l,z_1}.
        \anglebr{l,z_1,\lambda xy.\letterms}}{} \\
      \pbmap{\lambda \anglebr{l,z_1,z_2}.
        \anglebr{l,z_1,z_2,z_1\,z_2}}{} \\
      \pbmap{\lambda \anglebr{l,z_1,z_2,z_3}.
        \anglebr{l,z_1,z_2,z_3,0}}{} \\
      \pbmap{\lambda \anglebr{l,z_1,z_2,z_3,z_4}.z_3\,z_4
      }{\omega}
    }\,[\PCF{7},\PCF{-1}] \\
    & \Predplus
    \mat{
      \dualmap{(\lambda v.\anglebr{v,v})}{} \\
      \dualmap{(\lambda \anglebr{v,v_1}.
        \anglebr{v,v_1,0}
      )}{} \\
      (\lambda \anglebr{v,v_1,v_2}.
        \langle v,v_1,v_2,
          v_1(\lambda xy.\letterms)\ +
      \\
      \qquad
          v_2\PCF{-1}(\PCF{+}(\anglebr{\PCF{7},d})) +
          (\jac{+(\anglebr{-1,-})}{(v_2\PCF{7}d)})\,(\PCF{+}\anglebr{\PCF{7},d})
        \rangle
      )^*\cdot \\
      \dualmap{(\lambda \anglebr{v,v_1,v_2,v_3}.
        \anglebr{v,v_1,v_2,v_3,0}
      )}{} \\
      (\lambda \anglebr{v,v_1,v_2,v_3,v_4}.
        \langle v,v_1,v_2,v_3,v_4,
      \\
      \qquad
          v_3\PCF{0} +
          (\jac{+(\anglebr{-1,+(\anglebr{7,-})})}{v_4})\PCF{0}
        \rangle
      )^*\cdot{\omega B}
    } \\
    & \Predplus
    \dualmap{(\lambda v.v(\lambda xy.\letterms)\PCF{0})}{\omega B}
  \end{align*}
  where
  $B \equiv
  \anglebr{
    [\PCF{7},\PCF{-1}],
    [\PCF{7},\PCF{-1}],
    \lambda xy.\letterms,
    \lambda d.\PCF{+}(\anglebr{\PCF{-1},\PCF{+}(\anglebr{\PCF{7},d})}),
    \PCF{0},
    \PCF{6}
  }$.
  Hence,
  $\lambda v.v(\lambda xy.\letterms)\PCF{0}$
  is the derivative of
  $\mathsf{sum} \equiv \lambda l.l(\lambda xy.\letterms)\PCF{0}$
  at $[\PCF{7},\PCF{-1}]$.

  This sequence of reduction tells us how
  the derivative of
  $\mathsf{sum}$ at $[\PCF{7},\PCF{-1}]$
  can be computed using reverse-mode AD.

  \section{Administrative Reduction}
  \label{appendix:administrative reduction}

  Elementary terms $\elemterms$,
  let series $\letterms$,
  A-contexts $C_A$ and
  A-redexes $r_A$
  are defined as follows.
  \begin{align*}
    \elemterms & ::=
        0
    \mid z_1 + z_2
    \mid z
    \mid \lambda x.\letterms
    \mid z_1\,z_2
    \mid {z}_i
    \mid \anglebr{z_1,z_2}
    \mid \PCF{r}
    \\
    & \st
        \PCF{f}(z)
    \mid \jac{f}{z}
    \mid \dualmap{(\lambda x.\letterms)}{z}
    \mid \pbmap{\lambda x.\letterms}{z}
    \mid \dual{\PCF{\vec{r}}}
    \\
    \letterms & ::=
      \Let{z = \elemterms}{\letterms} \st
      \Let{z = \elemterms}{z}.
    \\
    C_A
    & ::=
        []
    \mid C_A + \pbterms
    \mid \letterms + C_A
    \mid \lambda z.C_A
    \mid C_A\,\pbterms
    \mid \letterms\,C_A
    \mid \pi_i{(C_A)}
    \\
    & \st
         \anglebr{C_A,\simterms}
    \mid \anglebr{\letterms,C_A}
    \mid \PCF{f}(C_A)
    \mid \jac{f}{C_A}
    \mid \dualmap{(\lambda x.C_A)}{\simterms}
    \\
    & \st
         \dualmap{(\lambda x.\letterms)}{C_A}
    \mid \pbmap{(\lambda x.C_A)}{\simterms}
    \mid \pbmap{(\lambda x.\letterms)}{C_A}
    \\
    r_A
    & ::=
        0
    \mid \letterms_1 + \letterms_2
    \mid x
    \mid \lambda z.\letterms
    \mid \letterms_1\,\letterms_2
    \mid \pi_i({\letterms})
    \mid \anglebr{\letterms_1,\letterms_2}
    \mid \PCF{r}
    \\
    & \st
         \PCF{f}(\letterms)
    \mid \jac{f}{\letterms}
    \mid \dualmap{(\lambda x.\letterms_1)}{\letterms_2}
    \mid \pbmap{\lambda x.\letterms_1}{\letterms_2}
    \mid \dual{\PCF{\vec{r}}}
  \end{align*}

  \begin{lemma}
    \label{lemma: A-context/redex/value are well-defined}
    Every pullback term $\pbterms$ can be expressed as
    either $C_A[r_A]$ for some unique A-context $C_A$ and A-redex $r_A$
    or a let series of elementary terms $\letterms$.
  \end{lemma}


  An A-redex $r_A$ is reduced to a let series $\letterms$ as follows.
  \begin{align*}
    0 & \red_A
      \Let{x_1 = 0}{x_1} \\
    \letterms_1 + \letterms_2 & \red_A
      \Let{
        x_1 = \letterms_1;\
        x_2 = \letterms_2;\
        x_3 = x_1 + x_2
      }{x_3} \\
    x & \red_A
      \Let{x_1 = x}{x_1} \\
    \lambda z.\letterms & \red_A
      \Let{x_1 = \lambda z.\letterms}{x_1} \\
    \letterms_1\,\letterms_2 & \red_A
      \Let{
        x_1 = \letterms_1;\
        x_2 = \letterms_2;\
        x_3 = x_1\,x_2
      }{x_3} \\
    \pi_i{(\letterms)} & \red_A
      \Let{x_1 = \letterms;\ x_2 = \pi_i{(x_1)}}{x_2} \\
    \anglebr{\letterms_1,\letterms_2} & \red_A
      \Let{
        x_1 = \letterms_1;\
        x_2 = \letterms_2;\
        x_3 = \anglebr{x_1,x_2}
      }{x_3} \\
    \PCF{r} & \red_A
      \Let{x_1 = \PCF{r}}{x_1} \\
    \PCF{f}(\letterms) & \red_A
      \Let{
        x_1 = \letterms;\
        x_2 = \PCF{f}(x_1)
      }{x_2} \\
    \jac{f}{\letterms} & \red_A
      \Let{
        x_1 = \letterms;\
        x_2 = \jac{f}{x_1}
      }{x_2} \\
    \dualmap{(\lambda x.\letterms_1)}{\letterms_2} & \red_A
      \Let{
        x_1 = \letterms_2;\
        x_2 = \dualmap{(\lambda x.\letterms_1)}{x_1}
      }{x_2} \\
    \pbmap{\lambda x.\letterms_1}{\letterms_2} & \red_A
      \Let{
        x_1 = \letterms_2;\
        x_2 = \pbmap{(\lambda x.\letterms_1)}{x_1}
      }{x_2} \\
    \dual{\PCF{\vec{r}}} & \red_A
      \Let{x_1 = \dual{\PCF{\vec{r}}}}{x_1}
  \end{align*}

  Any pullback term $\pbterms$ which can be expressed as
  $C_A[r_A]$ can be A-reduced to $C_A[\letterms]$ where
  $r_A \Ared \letterms$.

  \section{Interpretation}
  \label{appendix:interpretation}

  \begin{align*}
    \deno{\Gamma \vdash 0:\sigma} \gamma & =
    0
    \\
    \deno{\Gamma \vdash \simterms + \pbterms :\sigma} \gamma & =
    \deno{\simterms} \gamma +
    \deno{\pbterms} \gamma
    \\
    \deno{\Gamma\cup\set{x:\sigma} \vdash x:\sigma} \anglebr{\gamma,z} & =
    z
    \\
    \deno{\Gamma \vdash \lambda x.\simterms:\sigma_1\Arrow\sigma_2} \gamma & =
    \curry
      {\deno{\simterms}}\,
      \gamma
    \\
    \deno{\Gamma \vdash \simterms\,\pbterms:\sigma_2} \gamma & =
    (\deno{\simterms}\gamma)\,
    (\deno{\pbterms}\gamma)
    \\
    \deno{\Gamma \vdash \pi_i (\simterms):\sigma_i} \gamma & =
    \pi_i{(\deno{\simterms}\gamma)}
    \\
    \deno{\Gamma \vdash \anglebr{\simterms_1,\simterms_2}:\sigma_1\times\sigma_2} \gamma & =
    \anglebr{
      \deno{\simterms_1}\gamma,\
      \deno{\simterms_2}\gamma
    }
    \\
    \deno{\Gamma \vdash \PCF{r}:\PCFReal} \gamma & =
    r
    \\
    \deno{\Gamma \vdash \PCF{f}(\pbterms):\PCFReal^m} \gamma & =
    f(\deno{\pbterms}\gamma)
    \\
    \deno{\Gamma \vdash \jac{f}{\simterms}:\PCFReal^m\Arrow\PCFReal^m} \gamma & =
    \curry{D[f]}(\deno{\simterms}\gamma)
    \\
    \deno{\Gamma \vdash
      \dualmap{(\lambda x.\simterms_1)}{\simterms_2} :\dual{\sigma_1}
    } \gamma & =
    \lambda v.
      \deno{\simterms_2} \gamma
      \big(\deno{\simterms_1}\anglebr{\gamma,v}\big)
    \\
    \deno{\Gamma \vdash
      \pbmap{\lambda x.\pbterms}{\simterms} :\oneform\sigma_1
    } \gamma & =
    \lambda xv.
      \deno{\simterms}\gamma
      (\deno{\pbterms}\anglebr{\gamma,x}) \\
    & \qquad\qquad
      (D[\curry{\deno{\pbterms}}\gamma]\anglebr{v,x})
    \\
    \deno{\Gamma \vdash \dual{\PCF{\colvec{r_1\\\vdots\\r_n}}}:\dual{\PCFReal^n} } \gamma
    & =
    \colvec{v_1\\\vdots\\v_n} \mapsto \sum_{i=1}^n r_i\,v_i
  \end{align*}

  \section{Extended Differential Lambda-Calculus}
  \label{appendix:extended diff lam cal}

  Differential substitution of the extended differential $\lambda$-terms are defined as follows.
  \begin{align*}
    \dsub{\pi_i(s)}{T}{x}
    \equiv\ &
      \pi_i\big(\dsubup{s}{T}{x} \big)
    \\
    \dsub{\anglebr{s_1,s_2}}{T}{x}
    \equiv\ &
      \anglebr{\dsubup{s_1}{T}{x}, \dsubup{s_2}{T}{x} }
    \\
    \dsubup{\PCF{r}}{T}{x}
    \equiv\ &
      0
    \\
    \dsub{(\PCF{f}(s))}{T}{x}
    \equiv\ &
      \Big(\diff{\PCF{f}}{\big(\dsubup{s}{T}{x} \big)}\Big)\,s
    \\
    \dsub{(\diff{\PCF{f}}{s})}{T}{x}
    \equiv\ &
      \diff{\PCF{f}}{\big(\dsubup{s}{T}{x} \big)}
  \end{align*}

  Consider the term $\PCF{f}(s)$.
  There are no linear occurrences of $x$ in $\PCF{f}$.
  Hence, we ignore $\PCF{f}$ and
  perform differential substitution to $s$ directly and obtain
  $\Big(\diff{\PCF{f}}{\big(\dsubup{s}{T}{x} \big)}\Big)\,s$.

  We can interpret the extended differential $\lambda$-calculus with
  a differential $\lambda$-category,
  which is the categorical semantics of differential $\lambda$-calculus.
  Hence, what is left to show is the interpretations of
  the extended terms.
  \begin{align*}
    \deno{\pi_i(s)}
      & = \pi_i \circ \deno{s} \\
    \deno{\anglebr{s_1,s_2}}
      & = \anglebr{\deno{s_1},\deno{s_2}} \\
    \deno{\PCF{r}}
      & = \lambda \gamma.r \\
    \deno{\PCF{f}(s)}
      & = f \circ \deno{s} \\
    \deno{\diff{\PCF{f}}{s}}
      & = \lambda \gamma x.D[f]\anglebr{\deno{s}\gamma,x}
  \end{align*}

  \subsection*{Translation to Differential Lambda Calculus}

  \begin{align*}
    \trans{0} & := 0
    &
    \trans{\pi_i({\simterms})} & := \pi_i(\trans{\simterms}) \\
    \trans{(\simterms + \pbterms)} & := \trans{\simterms} + \trans{\pbterms}
    &
    \trans{(\anglebr{\simterms_1,\simterms_2})}
      & := \anglebr{\trans{(\simterms_1)},\trans{(\simterms_2)}} \\
    \trans{y} & := y
    &
    \trans{\PCF{r}} & := \PCF{r} \\
    \trans{(\lambda y.\simterms)} & := \lambda y.\trans{\simterms}
    &
    \trans{(\PCF{f}(\pbterms))} & := \PCF{f}(\trans{\pbterms}) \\
    \trans{(\simterms\,\pbterms)} & := \trans{\simterms}\,\trans{\pbterms}
    &
    \trans{(\jac{f}{\simterms})}
      & := \diff{\PCF{f}}{\trans{(\simterms)}}
    \\
    \omit\rlap{$
      \trans{\dual{\PCF{\colvec{r_1\\\vdots\\r_n}}}}
      := \lambda v.\sum_{i=1}^n \PCF{f_i}(\pi_i(v))
    $} \\
    \omit\rlap{$
      \trans{\big(\dualmap{(\lambda y.\simterms_1)}{\simterms_2}\big)}
      := \lambda v.\trans{(\simterms_2)}\big((\lambda y.\trans{(\simterms_1)})v\big)
    $} \\
    \omit\rlap{$
      \trans{(\pbmap{\lambda y.\pbterms}{\simterms})}
      := \lambda xv.
            \trans{\simterms}
            \big((\lambda y.\trans{\pbterms})x\big)
            \big((\diff{(\lambda y.\trans{\pbterms})}{v})x\big)
    $}
  \end{align*}
  where $f := r_i\times -$.

  \section{Proofs}
  \label{appendix:proofs}

  \begin{proposition}
    \label{prop: diff of constant function}
    The derivative of any constant morphism $f$ in
    a differential $\lambda$-category is $0$, \ie
    $D[f] = 0$.
  \end{proposition}

  \begin{proof}
    A constant morphism $f:A \to B$
    that maps all of $A$ to $b \in B$ can be written as
    $f = (\lambda z.b) \circ 0$ where $0:A \to B$ and $\lambda z.b:B \to B$.
    So by [CD1,2,5] we have
    $D[f] = D[(\lambda z.b) \circ 0]
    = D[\lambda z.b] \circ \anglebr{D[0],0 \circ \pi_2}
    = D[\lambda z.b] \circ \anglebr{0,0\circ \pi_2}
    = 0$.
  \end{proof}

  \begin{lemma}
    \label{lemma: cons is a differential lambda-cat}
    $\ConS$ is a differential $\lambda$-category with
    the differential operator \\
    $
      D[f]\anglebr{v,x} := \jacob(f)(x)(v) =
      \lim_{t \to 0} (f(x+tv)-f(x))/t .
    $
  \end{lemma}

  \begin{proof}
    \cite{FrolicherK88,MichorK97} have shown that $\ConS$ is Cartesian closed,
    and \cite{DBLP:journals/corr/abs-1006-3140} have shown that $\ConS$ is a
    Cartesian differential category.
    What is left to show is that $\lambda(-)$ preserves the additive structure
    and $D[-]$ satisfies the (D-curry) rule, \ie
    $
      D[\lambda(f)]
      =
      \lambda \big(D[f] \circ \anglebr{\pi_1 \times 0, \pi_2\times \Id} \big)
    $.

    We first show that $\lambda(-)$ is additive, \ie
    $\lambda(f+g)\ =\ \lambda(f)+\lambda(g)$ and $\lambda(0)\ =\ 0$.
    Note that for $f,g,0: A\times B \to C$ and $a \in A$, $b \in B$,
    $
    \lambda(f+g)(a)(b)
    \ =\ (f+g)\anglebr{a,b}
    \ =\ f\anglebr{a,b} + g\anglebr{a,b}
    \ =\ \lambda(f)(a)(b) + \lambda(g)(a)(b)
    $
    and
    $
    \lambda(0)(a)(b)
    \ =\ 0\anglebr{a,b}
    \ =\ 0
    \ =\ 0(a)(b).
    $

    Now we show that $D[-]$ satisfies the (D-curry) rule.
    Let $f: A \times B \to C$, $v,x \in A$ and $b \in B$.
    \begin{align*}
      D[\lambda(f)]\,\anglebr{v,x}\,b
      & =
      \Big(
        \lim_{t \to 0} \frac{\lambda(f)(x+vt)- \lambda(f)(x) }{t}
      \Big)\,b \\
      & = \lim_{t \to 0} \frac{f \anglebr{x+vt,b} - f \anglebr{x,b} }{t} \\
      & = \lim_{t \to 0}
        \frac{f (\anglebr{x,b} + t\,\anglebr{v,0}) - f \anglebr{x,b} }{t} \\
      & = D[f]\anglebr{\anglebr{v,0},\anglebr{x,b}} \\
      & = \big(D[f] \circ \anglebr{\pi_1 \times 0, \pi_2\times \Id} \big)
        \anglebr{\anglebr{v,x},b} \\
      & = \lambda \big(D[f] \circ \anglebr{\pi_1 \times 0, \pi_2\times \Id} \big)
        \,\anglebr{v,x}\,b
    \end{align*}
  \end{proof}



  \begin{proposition}
    \label{prop:ConS is separated}
    Let $E$ be a convenient vector space and $x,y \in E$ be distinct elements in $E$.
    Then, there exists a bornological linear map $l:E \to \Real$ that separates $x$ and $y$,
    \ie $l(x) \not= l(y)$.
  \end{proposition}

  \begin{proof}
    This follows from the fact that convenient vector space is separated.

    $x \not= y$ implies that $x - y \not= 0$.
    Hence by separation,
    there is a bornological linear map $l:E\to \Real$ such that
    $l(x-y) \not= 0$.
    Notice that $l$ is linear, so we have $l(x)-l(y) \not= 0$
    which implies $l(x) \not= l(y)$.
  \end{proof}

  \linearity*
  \begin{proof}
    Induction on the structure of $\pbterms$ on the following two statements.
    \begin{enumerate}[{I}H.1]
      \item
        If $\Gamma_1 \cup \set{x:\sigma_1} \vdash \pbterms:\tau$
        and $x \in \lin{\pbterms}$, then
        for any $\gamma_1 \in \deno{\Gamma_1}$,
        $\curry{\deno{\pbterms}}\gamma_1$ is linear, \ie
        $D[\curry{\deno{\pbterms}}\gamma_1]
        = (\curry{\deno{\pbterms}}\gamma_1) \circ \pi_1$.
      \item
        If $\Gamma_2 \vdash \pbterms:\dual{\sigma}$, then
        for any $\gamma_2 \in \deno{\Gamma_2}$,
        $\deno{\pbterms}\gamma$ is linear, \ie
        $D[\deno{\pbterms}\gamma] = (\deno{\pbterms_2}\gamma) \circ \pi_1$.
    \end{enumerate}

    \begin{itemize}
      \item[(var)]
        Say $\pbterms \equiv x$.

        (1)
        If $\Gamma_1 \cup \set{x:\sigma_1} \vdash x:\sigma_1$
        and $x \in \lin{x}$, then
        $D[\curry{\deno{x}}\gamma_1]
        = D[\Id]
        = \pi_1
        = \Id \circ \pi_1
        = (\curry{\deno{x}}\gamma_1) \circ \pi_1$.

        (2)
        If $\Gamma_2 \vdash x:\dual{\sigma}$, then
        $\Gamma_2 = \Gamma_3 \cup \set{x:\dual{\sigma}}$ so
        for any $\anglebr{\gamma_3,z} \in \deno{\Gamma_2}$,
        $z$ is linear and
        $D[\deno{x}\anglebr{\gamma_3,z}]
        = D[z]
        = z \circ \pi_1
        = (\deno{\pbterms_2}\anglebr{\gamma_3,z}) \circ \pi_1$.

      \item[(dual)]
        Say $\pbterms \equiv \dualmap{(\lambda x.\simterms_1)}{\simterms_2}$.

        (1)
        Let $\Gamma_1 \cup \set{x:\sigma_1} \vdash
        \dualmap{(\lambda x.\simterms_1)}{\simterms_2}:\tau$
        and $x \in \lin{\dualmap{(\lambda x.\simterms_1)}{\simterms_2}}
        := \big(\lin{\simterms_1}\setminus\freevar{\simterms_2}\big) \cup
        \big(\lin{\simterms_2}\setminus\freevar{\simterms_1}\big)$, then
        for any $\gamma_1 \in \deno{\Gamma_1}$ and
        since $\deno{\simterms_2}\anglebr{\gamma_1,x}$ is of a dual type,
        by IH.2,
        \begin{align*}
          &
          D[\curry{\deno{\dualmap{(\lambda x.\simterms_1)}{\simterms_2}}}\gamma_1]
          \anglebr{v,x}
          \\
          & =
          \lambda z.
          \big(D[\deno{\simterms_2}\anglebr{\gamma,-}]\anglebr{v,x}\big)
          g(\anglebr{x,z}) \\
          & \qquad \qquad+
          D[\deno{\simterms_2}\anglebr{\gamma,x}]
          \anglebr{
            D[g(\anglebr{-,z})]\anglebr{v,x} ,
            g(\anglebr{x,z})
          } \\
          & =
          \lambda z.
          \big(D[\deno{\simterms_2}\anglebr{\gamma,-}]\anglebr{v,x}\big)
          g(\anglebr{x,z}) \\
          & \qquad \qquad+
          \deno{\simterms_2}\anglebr{\gamma,x}
          (D[g(\anglebr{-,z})]\anglebr{v,x})
        \end{align*}
        where $g:\anglebr{x,z} \mapsto \deno{\simterms_1}\anglebr{\gamma_1,x,z}$.
        Note that $x$ can only be in
        either $\lin{\simterms_1}\setminus\freevar{\simterms_2}$
        or $\lin{\simterms_2}\setminus\freevar{\simterms_1}$
        but not both.
        Say $x \in \lin{\simterms_1}\setminus\freevar{\simterms_2}$, then
        by Proposition \ref{prop: diff of constant function} and IH.1,
        \begin{align*}
          &
          D[\curry{\deno{\dualmap{(\lambda x.\simterms_1)}{\simterms_2}}}\gamma_1]
          \anglebr{v,x}
          \\
          & =
          \lambda z.
          \big(D[\deno{\simterms_2}\anglebr{\gamma,-}]\anglebr{v,x}\big)
          (\deno{\simterms_1}\anglebr{\gamma_1,x,z}) \\
          & \qquad \qquad+
          \deno{\simterms_2}\anglebr{\gamma,x}
          (D[\deno{\simterms_1}\anglebr{\gamma_1,-,z}]\anglebr{v,x}) \\
          & =
          \lambda z.
          \deno{\simterms_2}\anglebr{\gamma,x}
          (D[\deno{\simterms_1}\anglebr{\gamma_1,-,z}]\anglebr{v,x}) \\
          & =
          \lambda z.
          \deno{\simterms_2}\anglebr{\gamma,v}
          (\deno{\simterms_1}\anglebr{\gamma_1,v,z}) \\
          & =
          \deno{\dualmap{(\lambda x.\simterms_1)}{\simterms_2}}
          \anglebr{\gamma_1,v}
          \\
          & =
          \big(\curry{\deno{
            \dualmap{(\lambda x.\simterms_1)}{\simterms_2}
          }}\gamma_1) \circ \pi_1\big)\anglebr{v,x}
        \end{align*}

        (2)
        Let $\Gamma_2 \vdash
        \dualmap{(\lambda x.\simterms_1)}{\simterms_2}:\dual{\sigma}$
        and $\gamma_2 \in \deno{\Gamma_2}$.
        Then, by IH.1 and IH.2,
        \begin{align*}
          & D[\deno{\dualmap{(\lambda x.\simterms_1)}{\simterms_2}}\gamma_2] \\
          & = D[(\deno{\simterms_2}\gamma_2) \circ
              \big(\curry{\deno{\simterms_1}}\gamma_2 \big)] \\
          & = D[\deno{\simterms_2}\gamma_2] \circ
            \anglebr{
              D[\curry{\deno{\simterms_1}}\gamma_2],
              (\curry{\deno{\simterms_1}}\gamma_2) \circ \pi_2
            } \\
          & = (\deno{\simterms_2}\gamma_2) \circ
            (\curry{\deno{\simterms_1}}\gamma_2) \circ \pi_1
            \\
          & = (\deno{\dualmap{(\lambda x.\simterms_1)}{\simterms_2}}\gamma_2)
          \circ \pi_1
        \end{align*}
        All other cases are straight forward inductive proofs.
    \end{itemize}
  \end{proof}

  \betasublemma*
  \begin{proof}
    The only interesting cases are dual and pullback maps.
    \begin{itemize}
      \item[(dual)]
        $\big(\dualmap{(\lambda x.\simterms_1)}{\simterms_2}\big)[\pbterms'/y]
        \equiv
        \dualmap{(\lambda x.\simterms_1[\pbterms'/y])}{\simterms_2[\pbterms'/y]}$
        \begin{align*}
          &
          \deno{\big(\dualmap{(\lambda x.\simterms_1)}{\simterms_2}\big)[\pbterms'/y]}\gamma
          \\
          & =
          \deno{\dualmap{(\lambda x.\simterms_1[\pbterms'/y])}{\simterms_2[\pbterms'/y]}} \gamma \\
          & =
          \deno{\simterms_2[\pbterms'/y]}\gamma \circ
          \curry{\deno{\simterms_1[\pbterms'/y]}}\gamma \\
          & =
          \lambda x.
          \deno{\simterms_2} \anglebr{\gamma,\deno{\pbterms'}\gamma}
          (\deno{\simterms_1}\anglebr{\gamma,\deno{\pbterms'}\gamma,x}) \tag{IH} \\
          & =
          \deno{\dualmap{(\lambda x.\simterms_1)}{\simterms_2}}
          \anglebr{\gamma,\deno{\pbterms'}\gamma}
        \end{align*}

      \item[(pb)]
        $
          \big(\pbmap{\lambda x.\pbterms}{\simterms}\big)[\pbterms'/y]
          \equiv
          \pbmap{\lambda x.\pbterms[\pbterms'/y]}{\simterms[\pbterms'/y]}
        $
        \begin{align*}
          & \deno{\big(\pbmap{\lambda x.\pbterms}{\simterms}\big)[\pbterms'/y]}\gamma \\
          & =
          \deno{\pbmap{\lambda x.\pbterms[\pbterms'/y]}{\simterms[\pbterms'/y]}}
          \gamma \\
          & = \lambda xv.
          \big(\deno{\simterms}\anglebr{\gamma,\deno{\pbterms'}\gamma} \big)
          \big(\deno{\pbterms}\anglebr{\gamma,x,\deno{\pbterms'}\gamma} \big)
          \\
          & \qquad\qquad\qquad\big(
            D[\deno{\pbterms}\anglebr{\gamma,-,\deno{\pbterms'}\gamma}]
            \anglebr{v,x}
          \big) \tag{IH} \\
          & =
          \deno{\pbmap{\lambda x.\pbterms}{\simterms}}
          \anglebr{\gamma,\deno{\pbterms'}\gamma}
        \end{align*}
    \end{itemize}
  \end{proof}

  \adequacy*
  \begin{proof}
    \begin{enumerate}
      \item Easy induction on $\red_A$.

      \item Case analysis on reductions of pullback terms.
        Let $\gamma \in \deno{\Gamma}$.

      \begin{itemize}
        \item[(\hyperlink{eq:1}{1}-\hyperlink{eq:4}{4})]
          $\deno{(\lambda x.\simterms)\,{\values}}
            = \deno{\simterms[\values/x]}$,
          $\pi_i{(\anglebr{\values_1,\values_2})}
            = \deno{\values_i}$,
          $\PCF{f}(\PCF{r})
            = \deno{\PCF{f(r)}}$ and
          $\deno{\PCF{\jacob(f)(\vec{r})}(\vec{r'})}
            = D[f]\anglebr{\vec{r'},\vec{r}}$
          are easily verified using the Substitution Lemma
          \ref{lemma:betasublemma}.

        \item[(\hyperlink{eq:5}{5})]
          Let $\jacob(f)(\vec{r}) = [a_{ij}]_{i=1,\dots,m, j=1,\dots,n}$
          and $r' = [r'_i]_{i=1,\dots,m}$.
          \begin{align*}
            &
            \deno{\dualmap{(\lambda v.\PCF{\jacob(f)(\vec{r})}(v))}
            {\dual{\vec{r'}}}}\gamma \\
            & =
            \dual{(\jacob(f)(\vec{r}))}(\lambda v.\sum_{i=1}^m r'_i v_i)
            =
            \lambda v.\sum_{i=1}^m r'_i \sum_{j=1}^n a_{ij} v_{j} \\
            & =
            \lambda v.\sum_{j=1}^n \sum_{i=1}^m (r'_i\cdot a_{ij}) v_{j}
            =
            \lambda v.\sum_{j=1}^n ((\jacob(f)(\vec{r}))^\top \times r)_{j} v_{j}
            \\
            & =
            \deno{\dual{((\jacob(f)(\vec{r}))^\top \times r)}}\gamma
          \end{align*}

        \item[(\hyperlink{eq:6}{6})]
          Say $\Gamma \cup \set{v_2:\sigma_2} \vdash \values_2 :\tau$.
          Let $\Gamma \cup \set{v_1:\sigma_1, v_2:\sigma_2} \vdash \values_2':\tau$
          where $v_1$ is not a free variable in $\values_2$.
          \begin{align*}
            & \deno{
              \dualmap{(\lambda v_1.\values_1)}{\Big(
                \dualmap{(\lambda v_2.\values_2)}{\values_3}
              \Big)}
            }\gamma \\
            & =
            \big(\curry{\deno{\values_1}}\gamma \big)^*
            \Big(
              \big(\curry{\deno{\values_2}}\gamma \big)^*
              (\deno{\values_3}\gamma)
            \Big)
            \\
            & =
            \Big(
              \big(\curry{\deno{\values_2}}\gamma \big) \circ
              \big(\curry{\deno{\values_1}}\gamma \big)
            \Big)^*
            (\deno{\values_3}\gamma)
            \\
            & =
            \Big(
              v_1 \mapsto
              \deno{\values_2}
              \anglebr{
                \gamma,
                \deno{\values_1}\anglebr{\gamma,v_1}
              }
            \Big)^*
            (\deno{\values_3}\gamma)
            \\
            & =
            \Big(
              v_1 \mapsto
              \deno{\values_2'}
              \anglebr{
                \anglebr{\gamma,v_1},
                \deno{\values_1}\anglebr{\gamma,v_1}
              }
            \Big)^*
            (\deno{\values_3}\gamma)
            \\
            & =
            \Big(
              v_1 \mapsto
              \big(
              \deno{\values_2'}
              \circ
              \anglebr{\Id,\deno{\values_1}}
              \big)\anglebr{\gamma,v_1}
            \Big)^*
            (\deno{\values_3}\gamma)
            \\
            & =
            \Big(
              v_1 \mapsto
              \big(
              \deno{\values_2'[\values_1/v_2] }
              \big)\anglebr{\gamma,v_1}
            \Big)^*
            (\deno{\values_3}\gamma)
            \\
            & =
            \Big(
              \curry{\deno{\values_2'[\values_1/v_2] }}\gamma
            \Big)^*
            (\deno{\values_3}\gamma)
            \\
            & =
            \deno{
              \dualmap{(\lambda v_1. \values_2'[\values_1/v_2])}{\values_3}
            }
          \end{align*}

        \item[(\hyperlink{eq:7}{7})]
          Using the Substitution Lemma \ref{lemma:betasublemma},
          $\deno{\pbmap{(\lambda y.\Let{x=\elemterms}{x})}{\omega}}
            =
          \deno{\pbmap{(\lambda y.\elemterms)}{\omega}}$
          follows immediately from
          $
            \deno{\lambda y.\Let{x=\elemterms}{x}}
            =
            \curry{\deno{\Let{x=\elemterms}{x}}}
            =
            \curry{\deno{\elemterms}}
            =
            \deno{\lambda y.\elemterms}.
          $

        \item[(\hyperlink{eq:8}{8})]
          Consider
          $
            \pbmap{(\lambda y.\Let{x=\elemterms}{\letterms})}{\omega}
            \red
            \pbmap
              {(\lambda y. \anglebr{y,\elemterms})}
              {\big(\pbmap{(\lambda z.\widehat{\letterms})}{\omega}\big)}
          $
          where
          $\Gamma \cup \set{z:\sigma_1 \times \sigma_2} \vdash \widehat{\letterms}
          \ \equiv\ \letterms[\pi_1(z)/y][\pi_2(z)/x]:\tau$.
          \begin{align*}
            & \deno{\pbmap{(\lambda y.\Let{x=\elemterms}{\letterms})}{\omega}}\gamma \\
            & =
            \oneform\Big(
              \curry{\deno{\Let{x=\elemterms}{\letterms}}} \gamma
            \Big)\,(\deno{\omega}\gamma) \\
            & =
            \oneform\Big(
              \curry{\deno{\letterms} \circ \anglebr{\Id,\deno{\elemterms}}} \gamma
            \Big)\,(\deno{\omega}\gamma) \\
            & =
            \oneform\Big(
              s_1 \mapsto \deno{\letterms}
                \anglebr{\anglebr{\gamma,s_1},\deno{\elemterms}\anglebr{\gamma,s_1} }
            \Big)\,(\deno{\omega}\gamma) \\
            & =
            \oneform\Big(
              s_1 \mapsto \deno{\widehat{\letterms}}
                \anglebr{\gamma, \anglebr{s_1,\deno{\elemterms}\anglebr{\gamma,s_1} }}
            \Big)\,(\deno{\omega}\gamma) \\
            & =
            \oneform\Big(
              \big(\curry{\deno{\widehat{\letterms}}}\gamma \big) \circ
              \anglebr{\Id_{\deno{\sigma_1}} , \curry{\deno{\elemterms}}\gamma }
            \Big)\,(\deno{\omega}\gamma) \\
            & =
            \oneform\big(\anglebr{\Id_{\deno{\sigma_1}} , \curry{\deno{\elemterms}}\gamma }\big)
            \Big(
              \oneform\big(\curry{\deno{\widehat{\letterms}}}\gamma \big)\,(\deno{\omega}\gamma)
            \Big) \\
            & =
            \oneform\big(\anglebr{\Id_{\deno{\sigma_1}} , \curry{\deno{\elemterms}}\gamma }\big)
            \Big(
              \deno{\pbmap{\lambda z.\widehat{\letterms}}{\omega}}\gamma
            \Big) \\
            & =
            \oneform\big(\curry{\deno{\anglebr{y,\elemterms}}}\gamma \big)
            \Big(
              \deno{\pbmap{\lambda z.\widehat{\letterms}}{\omega}}\gamma
            \Big) \\
            & =
            \deno{
              \pbmap
                {\lambda y. \anglebr{y,\elemterms}}
                {\big(\pbmap{\lambda z.\widehat{\letterms}}{\omega}\big)}
            }
          \end{align*}
          \cm{Still need to change}


        \item[(\hyperlink{eq:9}{9})]
          Say $y$ is not free in $\elemterms$ and
          $\pbat{\big(\pbmap{(\lambda y.\elemterms)}{\omega}\big)}
          {\values}
          \red 0$. Then,
          \begin{align*}
            & \deno{\pbat{(\pbmap{\lambda y.\elemterms}{\omega})}{\values}}\gamma \\
            & =
            \big(
              \lambda xv.\deno{\omega} \gamma
              (\deno{\elemterms}\anglebr{\gamma,x} )
              (D[\curry{\deno{\elemterms}}\gamma]\anglebr{v,x} )
            \big)\,(\deno{\values}\gamma) \\
            & =
            \big(
              \lambda xv.\deno{\omega} \gamma
              (\deno{\elemterms}\anglebr{\gamma,x} )
              0
            \big)\,(\deno{\values}\gamma) \\
            & =
            (
              \lambda xv.0
            )\,(\deno{\values}\gamma)
            =
            \lambda v.0
            =
            \deno{0} \gamma
          \end{align*}
          since $\curry{\deno{\elemterms}}\gamma$ is a constant function
          and the derivative of any constant function is $0$
          by Proposition \ref{prop: diff of constant function}.

        \item[(\hyperlink{eq:10a}{10})]
          We present the proof for (\hyperlink{eq:10b}{10b})
          $
          \pbat{\big(\pbmap{\lambda y.\proj{y}{i}+\proj{y}{j}}{}\\
          \omega\big)}{\values}
          \red
          \dualmap
            {(\lambda v.\proj{v}{i}+\proj{v}{j})}
            {
              \pbat{\omega}{\big({\values}_{\pi i} + {\values}_{\pi j}\big)}
            }
          $
          which leads to (\hyperlink{eq:10.1}{10.1}).
          \begin{align*}
            &
            \deno
              {\pbat{(\pbmap{\lambda y.\proj{y}{i}+\proj{y}{j}}{\omega})}
              {\values}}\gamma \\
            & =
            \big(
              \lambda xv.\deno{\omega} \gamma
              (\deno{\proj{y}{i}+\proj{y}{j}}\anglebr{\gamma,x} )
              (D[{\pi_i}+{\pi_j}]\anglebr{v,x} )
            \big)\,(\deno{\values}\gamma) \\
            & =
            \big(
              \lambda xv.\deno{\omega} \gamma
              ({x}_{\pi i} + {x}_{\pi j})
              (\proj{v}{i} + \proj{v}{j})
            \big)\,(\deno{\values}\gamma) \\
            & =
            \lambda v.\deno{\omega} \gamma
              (({(\deno{\values}\gamma)}_{\pi i} + ({(\deno{\values}\gamma)}_{\pi j})
              (\proj{v}{i} + \proj{v}{j}) \\
            & =
            \lambda v.
              \deno{\pbat{\omega}{({\values}_{\pi i} + {\values}_{\pi j})}}
              \gamma
              (\proj{v}{i} + \proj{v}{j}) \\
            & =
            \deno{\dualmap
              {(\lambda v.\proj{v}{i} + \proj{v}{j})}
              {\pbat{\omega}{({\values}_{\pi i} + {\values}_{\pi j})}}}\gamma
          \end{align*}

        \item[(\hyperlink{eq:11}{11})]
          $
          \pbat{\big(\pbmap{\lambda y.y}{\omega}\big)}{\values}
          \ \red\
          \dualmap{(\lambda v.v)}{\pbat{\omega}{\values}}
          $
          \begin{align*}
            &
            \deno{\pbat{(\pbmap{\lambda y.y}{\omega})}{\values}}\gamma \\
            & =
            \big(
              \lambda xv.\deno{\omega} \gamma
              (\deno{y}\anglebr{\gamma,x} )
              (D[\Id]\anglebr{v,x} )
            \big)(\deno{\values}\gamma) \\
            & =
            \big(
              \lambda xv.\deno{\omega} \gamma x v
            \big)(\deno{\values}\gamma) \\
            & =
            \lambda v.\deno{\omega} \gamma (\deno{\values}\gamma) v
            =
            \lambda v.\deno{\pbat{\omega}{\values}}\gamma v
            =
            \deno{\dualmap{(\lambda v.v)}{\pbat{\omega}{\values}}}\gamma
          \end{align*}

        \item[(\hyperlink{eq:12}{12})]
          $
          \pbat{(\pbmap{\lambda y.\proj{y}{i}}{\omega})}{\values}
          \ \red\
          \dualmap{(\lambda v.\proj{v}{i})}{\pbat{\omega}{{\values}_{\pi i}}}
          $
          \begin{align*}
            & \deno{\pbat{(\pbmap{\lambda y.\proj{y}{i}}{\omega})}{\values}}\gamma
            \\
            & =
            \big(
              \lambda xv.\deno{\omega} \gamma
              (\deno{\proj{y}{i}}\anglebr{\gamma,x} )
              (D[\pi_i]\anglebr{v,x})
            \big)(\deno{\values}\gamma) \\
            & =
            \big(
              \lambda xv.\deno{\omega} \gamma {x}_{\pi i} \proj{v}{i}
            \big)(\deno{\values}\gamma) \\
            & =
            \lambda v.\deno{\omega} \gamma (\deno{{\values}_{\pi i}} \gamma) \proj{v}{i}
            =
            \deno{\dualmap{(\lambda v.\proj{v}{i})}{\pbat{\omega}{{\values}_{\pi i}}}}\gamma
          \end{align*}

        \item[(\hyperlink{eq:13a}{13})]
          We prove for (\hyperlink{eq:13c}{13c}),
          $
          \pbat{(
            \pbmap{\lambda y. \anglebr{\proj{y}{i},\proj{y}{j}}}{\omega}
          )}{\values}
          \ \red\
          \dualmap
            {(\lambda v.\anglebr{\proj{v}{i},\proj{v}{j}})}
            {\pbat{\omega}{\anglebr{{\values}_{\pi i},{\values}_{\pi j}}}}
          $
          which leads to
          (\hyperlink{eq:13a}{13a}) and (\hyperlink{eq:13b}{13b}).
          \begin{align*}
            & \deno{
              \pbat{(
                \pbmap{\lambda y. \anglebr{\proj{y}{i},\proj{y}{j}}}{\omega}
              )}{\values}
            }\gamma \\
            & =
            \big(
              \lambda xv.\deno{\omega} \gamma
              (\deno{\anglebr{\proj{y}{i},\proj{y}{j}}}\anglebr{\gamma,x} )
              (D[\anglebr{\pi_i,\pi_j}]\anglebr{v,x})
            \big)(\deno{\values}\gamma) \\
            & =
            \big(
              \lambda xv.\deno{\omega} \gamma
              \anglebr{{x}_{\pi i}, {x}_{\pi j}}
              \anglebr{\proj{v}{i}, \proj{v}{j}}
            \big)(\deno{\values}\gamma) \\
            & =
            \lambda v.\deno{\omega} \gamma
              \anglebr{{(\deno{\values}\gamma)}_{\pi i}, {(\deno{\values}\gamma)}_{\pi j}}
              \anglebr{\proj{v}{i}, \proj{v}{j}} \\
            & =
            \lambda v.
              \deno{
                \pbat
                  {\omega}
                  {\anglebr{{\values}_{\pi i}, {\values}_{\pi j}}}
              }
              \gamma
              \anglebr{\proj{v}{i}, \proj{v}{j}} \\
            & =
            \deno{\dualmap
              {(\lambda v.\anglebr{\proj{v}{i}, \proj{v}{j}})}
              {\pbat
                {\omega}
                {\anglebr{{\values}_{\pi i}, {\values}_{\pi j}}}
              }}\gamma
          \end{align*}

        \item[(\hyperlink{eq:14}{14})]
          $
            \pbat{\big(
              \pbmap{\lambda y. \jac{f}{\proj{y}{i}}}{\omega}
            \big)}{\values}
              \Pred
              \dualmap{(\lambda v.
                \jac{f}{\proj{v}{i}}
              )}{\big(\pbat{\omega}{(\jac{f}{\proj{\values}{i}})}\big)}
          $
          By [CD3,4,5,6],
          \begin{align*}
            & D[\lambda yz.D[f]\anglebr{\proj{y}{i},z}] \\
            & =
            D[\curry{D[f]}\circ \pi_i] \\
            & =
            D[\curry{D[f]}] \circ (\pi_i\times\pi_i) \\
            & =
            \curry{D[D[f]] \circ \anglebr{\pi_1\times 0, \pi_2\times \Id}}
            \circ (\pi_i\times\pi_i) \\
            & =
            \curry{D[f] \circ (\pi_1\times \Id)} \circ (\pi_i\times\pi_i)
          \end{align*}
          Hence
          \begin{align*}
            & \deno{
              \pbat{\big(
              \pbmap{\lambda y. \jac{f}{\proj{y}{i}}}{\omega}
            \big)}{\values}
            }\gamma \\
            & =
              \lambda v.\deno{\omega} \gamma
              \big(\lambda x.D[f] \anglebr{
                  \deno{\proj{\values}{j}}\gamma, x
              }\big)
            \\
            & \qquad
              \big(
                D[\lambda yz.D[f] \anglebr{\proj{y}{i},z}]
                \anglebr{v,\deno{\values}\gamma}
              \big) \\
            & =
              \lambda v.\deno{\omega} \gamma
              \big(\lambda x.D[f] \anglebr{
                  \deno{\proj{\values}{j}}\gamma, x
              }\big)
            \\
            & \qquad
              \Big(
                \big(
                \curry{D[f] \circ (\pi_1\times \Id)} \circ (\pi_i\times\pi_i)
                \big)
                \anglebr{v,\deno{\values}\gamma}
              \Big) \\
            & =
              \lambda v.\deno{\omega} \gamma
              \big(
                \deno{(\jac{f}{\proj{\values}{i}})}\gamma
              \big)
              (\lambda x.D[f]\anglebr{\proj{v}{i},x}) \\
            & =
              \deno{
                \dualmap{(\lambda v.
                  \jac{f}{\proj{v}{i}}
                )}{\big(\pbat{\omega}{(\jac{f}{\proj{\values}{i}})}\big)}
              }\gamma
          \end{align*}

        \item[(\hyperlink{eq:15}{15})]
          $
            \pbat{(\pbmap{\lambda y. \PCF{f}(\proj{y}{i})}{\omega})}{\values}
            \red
            \dualmap
              {(\lambda v.(\jac{f}{\proj{v}{i}})\proj{\values}{i})}
              {\big(\pbat{\omega}{(\PCF{f}{(\proj{\values}{i})})}\big)}
          $
          \begin{align*}
            & \deno{
              \pbat{(\pbmap{\lambda y. \PCF{f}(\proj{y}{i})}{\omega})}{\values}
            }\gamma \\
            & =
            \big(
              \lambda xv.\deno{\omega} \gamma
              \big(f x_{\pi i}\big)
              \big(D[f] \anglebr{\proj{v}{i},x_{\pi i}}\big)
            \big)(\deno{\values}\gamma) \\
            & =
            \lambda v.
              \deno{\omega} \gamma
              \big(f(\deno{\proj{\values}{i}}\gamma)\big)
              \big(D[f] \anglebr{\proj{v}{i},{\deno{\proj{\values}{i}}\gamma}}\big)
            \\
            & =
            \lambda v.
              \deno{\omega} \gamma
              \big(f(\deno{\proj{\values}{i}}\gamma)\big)
              \big(\deno{
                (\jac{f}{\proj{v}{i}})\proj{\values}{i}
              }\anglebr{\gamma,v}\big)
            \\
            & =
            \deno{
              \dualmap
                {(\lambda v.(\jac{f}{\proj{v}{i}})\proj{\values}{i})}
                {\big(\pbat{\omega}{(\PCF{f}{(\proj{\values}{i})})}\big)}
            } \gamma
          \end{align*}

        \item[(\hyperlink{eq:16a}{16})]
          We prove for the most complicated case
          (\hyperlink{eq:16c}{16c})
          which leads to
          (\hyperlink{eq:16a}{16a}) and (\hyperlink{eq:16b}{16b}).

          By IH,
          $
            \deno{\pbat{\big(\pbmap{\lambda y.\letterms}{\omega}\big)}{\values}}
            =
            \deno{\dualmap{(\lambda v.\simterms)}{\pbat{\omega}{\values'}}}
          $ implies
          for any 1-form $\phi$, $\gamma$ and $x,v$,
          \begin{align*}
            &\phi\,
            (\deno{\letterms}\anglebr{\gamma,\deno{\values}\gamma,x})\,
            (D[\deno{\letterms}\anglebr{\gamma,-,x}]
            \anglebr{v,\deno{\values}\gamma})
            \\
            &=
            \phi\,
            (\deno{\values}\gamma)\,
            (\deno{\simterms}\anglebr{\gamma,x,v}).
          \end{align*}
          By Hahn-Banach Theorem, we have
          $
            D[\deno{\letterms}\anglebr{\gamma,-,x}]
            \anglebr{v,\deno{\values}\gamma}
            =
            \deno{\simterms}\anglebr{\gamma,x,v}.
          $

          First, note that since $\proj{\values}{i}$ is of the dual type,
          hence by Lemma \ref{lemma:linearity of syntax} (2),
          $D[\deno{\proj{\values}{i}}\gamma]
          = (\deno{\proj{\values}{i}}\gamma) \circ \pi_1$.
          \begin{align*}
            &
            D[\curry{\deno{
              \dualmap{(\lambda z.\letterms)}{\proj{y}{i}}
            }}\gamma] \anglebr{v,\deno{\values}\gamma} \\
            & =
            D[\lambda y.\lambda z.
              \proj{y}{i} (\deno{\letterms}\anglebr{\gamma,y,z})
            ] \anglebr{v,\deno{\values}\gamma} \\
            & =
            D[\curry{\catev \circ
              \anglebr{\pi_i \circ \pi_1, g}
            }] \anglebr{v,\deno{\values}\gamma} \\
            & =
            \lambda z.D[\catev \circ
              \anglebr{\pi_i \circ \pi_1, g}
            ] \anglebr{\anglebr{v,0},\anglebr{\deno{\values}\gamma,z}} \\
            & =
            \lambda z.
            \big(
              \catev \circ \anglebr{D[\pi_i \circ \pi_1], g \circ \pi_2} +
              D[\uncurry{\pi_i \circ \pi_1}] \\
            & \circ
              \anglebr{\anglebr{0,D[g]},\anglebr{\pi_2,g \circ \pi_2}}
            \big)
            \anglebr{\anglebr{v,0},\anglebr{\deno{\values}\gamma,z}} \\
            & =
            \lambda z.
              \proj{v}{i} (g\anglebr{\deno{\values}\gamma,z}) +
              D[\uncurry{\pi_i\circ \pi_1}] \\
            & \anglebr{
                \anglebr{0,
                  D[g]\anglebr{\anglebr{v,0},\anglebr{\deno{\values}\gamma,z}}
                },
                \anglebr{
                  \anglebr{\deno{\values}\gamma,z},
                  g\anglebr{\deno{\values}\gamma,z}
                }
              } \\
            & =
            \lambda z.
              \proj{v}{i} (g\anglebr{\deno{\values}\gamma,z}) +
              D[\uncurry{\pi_i}] \\
            & \anglebr{
                \anglebr{0,
                  D[g]\anglebr{\anglebr{v,0},\anglebr{\deno{\values}\gamma,z}}
                },
                \anglebr{
                  \deno{\values}\gamma,
                  g\anglebr{\deno{\values}\gamma,z}
                }
              } \\
            & =
            \lambda z.
              \proj{v}{i} (g\anglebr{\deno{\values}\gamma,z}) +
              D[\uncurry{\pi_i}\anglebr{\deno{\values}\gamma,-}] \\
            & \anglebr{
                D[g]\anglebr{\anglebr{v,0},\anglebr{\deno{\values}\gamma,z}},
                g\anglebr{\deno{\values}\gamma,z}
              } \\
            & =
            \lambda z.
              \proj{v}{i} (g\anglebr{\deno{\values}\gamma,z}) \\
            & +
              \big(
                D[\deno{\proj{\values}{i}}\gamma]
                \circ \anglebr{D[g],g\circ \pi_2}
              \big)
              \anglebr{\anglebr{v,0},\anglebr{\deno{\values}\gamma,z}}
            \\
            & =
            \lambda z.
              \proj{v}{i} (g\anglebr{\deno{\values}\gamma,z}) \\
            & +
              \big(
                \proj{v}{i} \circ \pi_1
                \circ \anglebr{D[g],g\circ \pi_2}
              \big)
              \anglebr{\anglebr{v,0},\anglebr{\deno{\values}\gamma,z}}
            \\
            & =
            \lambda z.
              \proj{v}{i} (g\anglebr{\deno{\values}\gamma,z})
             +
            \deno{\proj{\values}{i}}\gamma)
              \big(
                D[g\anglebr{-,z}]
                \anglebr{v,\deno{\values}\gamma}
              \big)
            \\
            & =
            \lambda z.
              \proj{v}{i}
              (\deno{\letterms}\anglebr{\gamma,\deno{\values}\gamma,z}) +
            \deno{\proj{\values}{i}}\gamma
              \big(
                D[\deno{\letterms}\anglebr{\gamma,-,z}]
                \anglebr{v,\deno{\values}\gamma}
              \big)
            \\
            & =
            \lambda z.
              \proj{v}{i}
              (\deno{\letterms}\anglebr{\gamma,\deno{\values}\gamma,z}) +
            \deno{\proj{\values}{i}}\gamma
              \big(
                \deno{\simterms}\anglebr{\gamma,x,v}
              \big)
            \\
            & =
            \deno{
              \dualmap{(\lambda z.\letterms[\values/y])}{\proj{v}{i}} +
              \dualmap{(\lambda z.\simterms)}{\proj{\values}{i}}
            }\anglebr{\gamma,v}.
          \end{align*}
          where $g : \anglebr{y,z} \mapsto \deno{\letterms}\anglebr{\gamma,y,z}$.

          Now we have
          \begin{align*}
            & \deno{
              \pbat{\big(\pbmap{\lambda y.
                \dualmap{(\lambda z.\letterms)}{\proj{y}{i}}
              }{\omega}\big)}{\values}
            } \gamma \\
            & =
            \lambda v.
              \deno{\omega}\gamma
              (
                \deno{\dualmap{(\lambda z.\letterms)}{\proj{y}{i}}}\anglebr{\gamma,\deno{\values}\gamma}
              )\\
            & \qquad\qquad\qquad\qquad\quad
              \big(
                D[\curry{\deno{\dualmap{\lambda z.\letterms}{\proj{y}{i}}}}\gamma]
                \anglebr{v,\deno{\values}\gamma}
              \big)
            \\
            & =
            \lambda v.
              \deno{\omega}\gamma
              (
                \deno{\dualmap{(\lambda z.\letterms[\values/y])}{\proj{\values}{i}}}\gamma
              ) \\
            & \qquad\qquad\quad
              \big(
                \deno{
                  \dualmap{(\lambda z.\letterms[\values/y])}{\proj{v}{i}} +
                  \dualmap{(\lambda z.\simterms)}{\proj{\values}{i}}
                }\anglebr{\gamma,v}
              \big) \\
            & =
            \llbracket
              \dualmap{(
                \lambda v.
                  \dualmap{(\lambda z.\simterms)}{\proj{\values}{i}} +
                  \dualmap{(\lambda z.\letterms[\values/y])}{\proj{v}{i}}
              )}{} \\
            & \qquad\qquad\qquad\qquad\qquad\qquad
              \pbat{\omega}{(
                \dualmap{(\lambda z.\letterms[\values/y])}{{\values}_{\pi i}}
              )}
            \rrbracket\gamma
          \end{align*}

        \item[(\hyperlink{eq:17}{17})]
          $
            \pbat{\big(\pbmap{
              \lambda y. \pbmap{\lambda x.\letterms}{\proj{y}{i}}
            }{\omega}\big)}{\values}
            \ \red\
            \pbat{\big(\pbmap{
              \lambda y. \lambda a.
              \dualmap{(\lambda v.\simterms)}{\pbat{z}{\letterms[a/x]}}
            }{\omega}\big)}{\values}
          $
          if
          $
            \pbat{\big(\pbmap{\lambda x.\letterms}{z})}{a}
            \ \redplus\
            \dualmap{(\lambda v.\simterms)}{\pbat{z}{\letterms[a/x]}}
          $ for fresh variable $a$.

          By IH,
          $
            \deno{\pbat{\big(\pbmap{\lambda x.\letterms}{z})}{a}}
            =
            \deno{\dualmap{(\lambda v.\simterms)}{\pbat{z}{\letterms[a/x]}}}
          $
          implies
          for any $\phi$, $\gamma$ $y, a, v$,
          \begin{align*}
            & \phi\,
            (\deno{\letterms}\anglebr{\gamma,a,y})\,
            (D[\deno{\letterms}\anglebr{\gamma,-,y}] \anglebr{v,a}) \\
            & =
            \phi\,
            (\deno{\letterms}\anglebr{\gamma,a,y})\,
            (\deno{\simterms}\anglebr{\gamma,y,a,v}).
          \end{align*}
          By Hahn-Banach Theorem,
          $
            D[\deno{\letterms}\anglebr{\gamma,-,y}] \anglebr{v,a}
            = \deno{\simterms}\anglebr{\gamma,y,a,v}.
          $
          \begin{align*}
            & \deno{\pbmap{\lambda x.\letterms}{z}} \anglebr{\gamma,y} \\
            & = \lambda a v.
              \deno{z}\anglebr{\gamma,y}\,
              (\deno{\letterms}\anglebr{\gamma,v,a})\,
              (D[\deno{\letterms}\anglebr{\gamma,-,y}] \anglebr{v,a}) \\
            & = \lambda a v.
              \deno{z}\anglebr{\gamma,y}\,
              (\deno{\letterms[a/x]}\anglebr{\gamma,v})\,
              (\deno{\simterms}\anglebr{\gamma,y,a,v}) \\
            & = \lambda a.
              \deno{
                \dualmap{(\lambda v.\simterms)}{
                  \pbat{z}{\letterms[a/x]}
                }
              } \anglebr{\gamma,y,a} \\
            & =
              \deno{
                \lambda a.
                \dualmap{(\lambda v.\simterms)}{
                  \pbat{z}{\letterms[a/x]}
                }
              } \anglebr{\gamma,y}
          \end{align*}
          Hence we have
          \begin{align*}
            & \deno{
              \pbat{\big(\pbmap{
                \lambda y. \pbmap{\lambda x.\letterms}{z}
              }{\omega}\big)}{\values}
            } \gamma \\
            & = \lambda v.
              \deno{\omega}\gamma
              \big(
                \deno{\pbmap{\lambda x.\letterms}{z}} \anglebr{\gamma,\deno{\values}\gamma}
              \big) \\
            & \qquad\qquad\qquad\qquad
              \big(
                D[\curry{\deno{\pbmap{\lambda x.\letterms}{z}}}\gamma]
                \anglebr{v,\deno{\values}\gamma}
              \big) \\
            & =
            \lambda v.
              \deno{\omega}\gamma
              \big(
                \deno{
                  \lambda a.
                  \dualmap{(\lambda v.\simterms)}{
                    \pbat{z}{\letterms[a/x]}
                  }
                }
                \anglebr{\gamma,\deno{\values}\gamma}
              \big)\\
            & \qquad\qquad\qquad\qquad
              \big(
                D[\curry{\deno{\pbmap{\lambda x.\letterms}{z}}}\gamma]
                \anglebr{v,\deno{\values}\gamma}
              \big) \\
            & =
            \deno{
              \pbat{\big(\pbmap{
                \lambda y. \lambda a.
                \dualmap{(\lambda v.\simterms)}{\pbat{z}{\letterms[a/x]}}
              }{\omega}\big)}{\values}
            } \gamma
          \end{align*}

        \item[(\hyperlink{eq:18}{18})]
          If
          $
            \pbat{\big(\pbmap{\lambda y.\letterms}{\omega}\big)}{\values}
            \ \redplus\
            \dualmap{(\lambda v.\simterms)}{\pbat{\omega}{\values}}
          $ and $x \not\in \freevar{\values}$,
          then
          $
            \pbat{\big(\pbmap{\lambda y.\lambda x.\letterms}{\values}\big)}{\values}
            \ \red\
            \dualmap{(
              \lambda v.\lambda x.\simterms
            )}{
              \pbat{\values}{\lambda x.\letterms[\values/y]}
            }.
          $

          Recall the (D-curry) rule,
          $D[\curry{f}] = \curry{D[f]\circ\anglebr{\pi_1\times 0,\pi_2\times\Id}}.$
          By IH, we have
          $
            \deno{\pbat{\big(\pbmap{\lambda y.\letterms}{\omega}\big)}{\values}}
            =
            \deno{\dualmap{(\lambda v.\simterms)}{
              \pbat{\omega}{(\letterms[\values/y])}
            }},
          $
          which means for any 1-form $\phi$, $\gamma$ and $x,v$,
          \begin{align*}
            &\phi\,
            (\deno{\letterms}\anglebr{\gamma,\deno{\values}\gamma,x})\,
            (
              D[\deno{\letterms}\anglebr{\gamma,-,x}]
              \anglebr{v,\deno{\values}\gamma}
            ) \\
            & =
            \phi\,
            (\deno{\letterms}\anglebr{\gamma,\deno{\values}\gamma,x})\,
            (\deno{\simterms}\anglebr{\gamma,x,v}).
          \end{align*}
          By Hahn-Banach Theorem,
          $
            D[\deno{\letterms}\anglebr{\gamma,-,x}]
            \anglebr{v,\deno{\values}\gamma}
            =
            \deno{\simterms}\anglebr{\gamma,x,v}.
          $
          Now
          \begin{align*}
            & D[\curry{\deno{\lambda x.\letterms}}\gamma]
            \anglebr{v,\deno{\values}\gamma}
            \\
            & =
            D[\curry{\deno{\letterms}}\anglebr{\gamma,-}]
            \anglebr{v,\deno{\values}\gamma}
            \\
            & =
            D[\curry{f}]
            \anglebr{v,\deno{\values}\gamma}
            \\
            & =
            \curry{
              D[f]\circ\anglebr{\pi_1\times 0,\pi_2 \times \Id}
            }
            \anglebr{v,\deno{\values}\gamma}
            \\
            & =
            \lambda x.
            (D[f]\circ\anglebr{\pi_1\times 0,\pi_2 \times \Id})
            \anglebr{\anglebr{v,\deno{\values}\gamma},x}
            \\
            & =
            \lambda x.
            D[f\anglebr{-,x}]
            \anglebr{v,\deno{\values}\gamma}
            \\
            & =
            \lambda x.
            D[\deno{\letterms}\anglebr{\gamma,-,x}]
            \anglebr{v,\deno{\values}\gamma} \\
            & =
            \lambda x.
              \deno{\simterms} \anglebr{\gamma,x,v}
          \end{align*}
          where $f := \uncurry{\curry{\deno{\letterms}}\anglebr{\gamma,-}}$.
          Hence, we have
          \begin{align*}
            & \deno{\pbat{(
              \pbmap{\lambda y.\lambda x.\letterms}{\values}
            )}{\values}}\gamma \\
            & =
            \big(
              \lambda xv.
              \deno{\values} \gamma
              (\deno{\lambda x.\letterms}\anglebr{x,\gamma})
              (D[\curry{\deno{\lambda x.\letterms}}\gamma]\anglebr{v,x} )
            \big)(\deno{\values}\gamma) \\
            & =
              \lambda v.
              \deno{\values} \gamma
              (\deno{\lambda x.\letterms}\anglebr{\deno{\values}\gamma,\gamma})
              (D[\curry{\deno{\lambda x.\letterms}}\gamma]\anglebr{v,\deno{\values}\gamma} )
            \\
            & =
              \lambda v.
              \deno{\values} \gamma
              ( \deno{\lambda x.\letterms[\values/y]}\gamma )
              ( \lambda x.\deno{\simterms} \anglebr{\gamma,x,v} )
            \\
            & =
              \lambda v.
              \deno{\pbat{\values}{(\lambda x.\letterms[\values/y])}}\gamma
              ( \deno{\lambda x.\simterms}\anglebr{\gamma,v} )
            \\
            & =
              \deno{
                \dualmap
                  {(\lambda v.\lambda x.\simterms)}
                  {\pbat{\values}{(\lambda x.\letterms[\values/y])}}
                }\gamma
          \end{align*}

        \item[(\hyperlink{eq:19a}{19})]
          We prove it for the complicated case (\hyperlink{eq:19c}{19c})
          and (\hyperlink{eq:19a}{19a}) and (\hyperlink{eq:19b}{19b}) follows.

          First note that by (D-eval) in \cite{DBLP:journals/mscs/Manzonetto12},
          we have
          $
            D[\catev \circ \anglebr{\pi_i,\pi_j}]\anglebr{v,x}
            =
            \pi_i(v)(\pi_j(x)) +
            D[\pi_i(x)]\anglebr{\pi_j(v),\pi_j(x)}.
          $
          By IH, and $\proj{\values}{i} \equiv \lambda z.\pbterms'$, we have
          $
            \deno{\pbat{\big(
              \pbmap{\lambda z.\pbterms'}{\omega}
            \big)}{\proj{\values}{j}}}
            =
            \deno{\dualmap{(\lambda v'.\simterms')}{
              \pbat{\omega}{(\pbterms'[\proj{\values}{j}/z])}}
            }
          $
          which means for any 1-form $\phi$, $\gamma$ and $v$,
          \begin{align*}
            & \phi\,
            (\deno{\pbterms'} \anglebr{\gamma,\deno{\proj{\values}{j}} \gamma})\,
            \big(
              D[\curry{\deno{\pbterms'}}\gamma]
              \anglebr{v,\deno{\proj{\values}{j}}\gamma}
            \big) \\
            & =
            \phi\,
            (\deno{\pbterms'} \anglebr{\gamma,\deno{\proj{\values}{j}} \gamma})\,
            (\deno{\simterms'}\anglebr{\gamma,v}).
          \end{align*}
          By Hahn-Banach Theorem,
          $
            D[\deno{\proj{\values}{i}}\gamma]
            \anglebr{\proj{v}{j},\deno{\proj{\values}{j}}\gamma}
            =
            D[\curry{\deno{\pbterms'}}\gamma]
            \anglebr{v,\deno{\proj{\values}{j}}\gamma}
            =
            \deno{\simterms'}\anglebr{\gamma,v}.
          $
          Hence we have
          \begin{align*}
            & \deno{
              \pbat{\big(
                \pbmap{\lambda y.\proj{y}{i} \proj{y}{j}}{\omega}
              \big)}{\values}} \gamma \\
            & =
              \lambda v.
              \deno{\omega} \gamma
              \big(\deno{\proj{y}{i} \proj{y}{j}}
                \anglebr{\gamma, \deno{\values}\gamma}
              \big)
              \big(D[\catev\circ\anglebr{\pi_i,\pi_j}]
                \anglebr{v,\deno{\values}\gamma)}
              \big)
            \\
            & =
              \lambda v.
              \deno{\omega} \gamma
              \big(
                \deno{\proj{\values}{i}\,\proj{\values}{j}}\gamma
              \big)
            \\
            & \qquad\qquad\qquad\quad
              \big(
                \proj{v}{i}(\deno{\proj{\values}{j}}\gamma) +
                D[\deno{\proj{\values}{i}}\gamma]
                \anglebr{\proj{v}{j},\deno{\proj{\values}{j}}\gamma}
              \big)
            \\
            & =
              \lambda v.
              \deno{\omega} \gamma
              \big(
                \deno{\proj{\values}{i}\,\proj{\values}{j}}\gamma
              \big)
              \big(
                \proj{v}{i}(\deno{\proj{\values}{j}}\gamma) +
                \deno{\simterms'}\anglebr{\gamma,\deno{\proj{\values}{j}}\gamma}
              \big)
            \\
            & =
              \deno{
                \dualmap{(\lambda v.
                  \proj{v}{i}\,{\values}_{\pi j} +
                  \simterms'[\proj{\values}{j}/v]
                )}{
                  \pbat{\omega}{({\proj{\values}{i}}\,{\proj{\values}{j}})}
                }
              }\gamma
          \end{align*}


        \item[(\hyperlink{eq:20a}{20a})]
          Say $y$ is a free variable in $\elemterms$,
          $
            \pbat
              {\big(\pbmap{(\lambda y.\anglebr{y,\elemterms})}{\omega}\big)}
              {\values}
            \ \red\
            \dualmap
              {(\lambda v.\anglebr{v, \simterms})}
              {\pbat{\omega}{\anglebr{\values,\elemterms[\values/y]}}}
          $
          if
          $
            \pbat{\big(\pbmap{\lambda y.\elemterms}{\omega}\big)}{\values}
            \ \red\
            \dualmap{(\lambda v.\simterms)}{\pbat{\omega}{(\elemterms[\values/y])}}.
          $
          By IH, we have
          $
            \deno{\pbat{\big(\pbmap{\lambda y.\elemterms}{\omega}\big)}{\values}}
            =
            \deno{\dualmap{(\lambda v.\simterms)}{\pbat{\omega}{(\elemterms[\values/y])}}},
          $
          which implies for any $\gamma \in \deno{\Gamma}$ and $v$,
          $
            \deno{\elemterms}\anglebr{\gamma,\deno{\values}\gamma}
            =
            \deno{\pbterms}\gamma
          $
          and
          $
            D[\deno{\elemterms}\anglebr{\gamma,-}]\anglebr{v,\deno{\values}\gamma}
            =
            \deno{\simterms}\anglebr{\gamma,v}.
          $
          Now,
          \begin{align*}
            & \deno{
              \pbat
                {\big(\pbmap{(\lambda y.\anglebr{y,\elemterms})}{\omega}\big)}
                {\values}
            }\gamma \\
            & =
              \lambda v.
              \deno{\omega} \gamma
              \big(\anglebr{
                \deno{\values}\gamma,
                \deno{\elemterms}\anglebr{\gamma, \deno{\values}\gamma}
              } \big) \\
            & \qquad\qquad\qquad
              \big(
                \anglebr{
                  D[\Id] \anglebr{v,\deno{\values}\gamma},
                  D[\deno{\elemterms}\anglebr{\gamma,-}] \anglebr{v,\deno{\values}\gamma}
                }
              \big)
              \\
            & =
              \lambda v.
              \deno{\omega} \gamma
              \big(\anglebr{\deno{\values}\gamma, \deno{\elemterms[\values/y]}\gamma } \big)
              \big(
                \anglebr{v, \deno{\simterms}\anglebr{\gamma,v}}
              \big)
              \\
            & =
              \lambda v.
              \deno{\pbat{\anglebr{\values,\elemterms[\values/y]}}{\omega}} \gamma
              \big(
                \deno{\lambda v.\anglebr{v,\simterms}} \gamma v
              \big)
              \\
            & =
              \deno{
                \dualmap
                  {(\lambda v.\anglebr{v, \simterms})}
                  {\pbat{\omega}{\anglebr{\values,\elemterms[\values/y]}}}
              }\gamma
          \end{align*}

        \item[(\hyperlink{eq:20b}{20b})]
          If $y \not \in \freevar{\elemterms}$, we have
          $$
            \pbat
              {\big(\pbmap{(\lambda y.\anglebr{y,\elemterms})}{\omega}\big)}
              {\values}
            \ \red\
            \dualmap
              {(\lambda v.\anglebr{v, 0})}
              {\pbat{\omega}{\anglebr{\values,\elemterms }}}
          $$
          and
          \begin{align*}
            & \deno{
              \pbat
                {\big(\pbmap{(\lambda y.\anglebr{y,\elemterms})}{\omega}\big)}
                {\values}
            }\gamma \\
            & =
              \lambda v.
              \deno{\omega} \gamma
              \big(\anglebr{
                \deno{\values}\gamma,
                \deno{\elemterms}\anglebr{\gamma, \deno{\values}\gamma}
              } \big) \\
            & \qquad\qquad\qquad
              \big(
                \anglebr{
                  D[\Id] \anglebr{v,\deno{\values}\gamma},
                  D[\deno{\elemterms}\anglebr{\gamma,-}] \anglebr{v,\deno{\values}\gamma}
                }
              \big)
              \\
            & =
              \lambda v.
              \deno{\omega} \gamma
              \big(\anglebr{
                \deno{\values}\gamma,
                \deno{\elemterms}\gamma
              } \big)
              \big(
                \anglebr{v,0}
              \big)
              \\
            & =
              \lambda v.
              \deno{\omega} \gamma
              \big(\deno{
                \anglebr{
                  \values,
                  \elemterms
                } \gamma
              }\big)
              \big(
                \deno{\lambda v.\anglebr{v,0}} \gamma v
              \big)
              \\
            & =
              \deno{
                \dualmap{(\lambda v.\anglebr{v,0})}{
                  \pbat{\omega}{\anglebr{\values,\elemterms}}
                }
              }\gamma
          \end{align*}
      \end{itemize}
    \end{enumerate}
  \end{proof}

  \TransRedModel*
  \begin{proof}
    \begin{enumerate}[1.]
      \item
        Easy induction on $\red$.
      \item
        We prove by induction on $\pbterms$.
        Most cases are trivial. Let $\gamma \in \deno{\Gamma}$.
        \begin{itemize}
          \item[(dual)]
            \begin{align*}
              \deno{
                \dualmap{(\lambda x.\simterms_1)}{\simterms_2}
              }\gamma
              & =
                \lambda v.
                \deno{\simterms_2}\gamma
                (\curry{\deno{\simterms_1}}\gamma v)
              \\
              & =
                \lambda v.
                \deno{\simterms_2} \anglebr{\gamma,v}
                \big(
                  \deno{\lambda x.\simterms_1} \anglebr{\gamma,v}
                  (\deno{v} \anglebr{\gamma,v})
                \big)
              \\
              & =
                \lambda v.
                \deno{\trans{\simterms_2}} \anglebr{\gamma,v}
                \big(
                  \deno{\trans{\lambda x.\simterms_1}} \anglebr{\gamma,v}
                  (\deno{\trans{v}} \anglebr{\gamma,v})
                \big)
              \\
              & =
                \lambda v.\deno{
                  \trans{\simterms_2}
                  \big( (\lambda x.\trans{\simterms_1})\,v\big)
                }\anglebr{\gamma,v}
              \\
              & =
                \deno{
                  \lambda v.
                  \trans{\simterms_2}
                  \big( (\lambda x.\trans{\simterms_1})\,v\big)
                }\gamma
            \end{align*}

          \item[(pb)]
            \begin{align*}
              & \deno{\pbmap{(\lambda y.\pbterms)}{\simterms}}\gamma
              \\
              & =
              \lambda xv.
                (\deno{\simterms}\gamma)
                ({\deno{\pbterms}}\anglebr{\gamma,x})
                (D[\curry{\deno{\pbterms}}\gamma]\anglebr{v,x})
              \\
              & =
              \lambda xv.
                (\deno{\simterms}\gamma)
                ({\deno{\pbterms}}\anglebr{\gamma,x})
                (D[\deno{\pbterms}]\anglebr{\anglebr{0,v},\anglebr{\gamma,x}})
                \\
              & =
              \lambda xv.
                (\deno{\simterms}\gamma)
                ({\deno{\pbterms}}\anglebr{\gamma,x})\\
              & \qquad\qquad\qquad\qquad
                \big(
                  D[\deno{\pbterms}]
                  \anglebr{
                    \anglebr{0, \deno{v}\anglebr{\gamma,x,v}},
                    \anglebr{\anglebr{\gamma,x,v},x}
                  }
                \big)
              \\
              & =
              \lambda xv.
                (\deno{\simterms}\gamma)
                \big(
                  \curry{\deno{\pbterms}}\gamma
                  (\deno{x}\anglebr{\gamma,x,v})
                \big)\\
              & \qquad\qquad\qquad\qquad
                \big(
                  \deno{\diff{(\lambda y.\pbterms)}{v}} \anglebr{\gamma,x,v}
                  (\deno{x}\anglebr{\gamma,x,v})
                \big)
              \\
              & =
              \lambda xv.
                (\deno{\simterms}\anglebr{\gamma,x,v})
                \big(
                  \curry{\deno{\pbterms}}\anglebr{\gamma,x,v}
                  (\deno{x}\anglebr{\gamma,x,v})
                \big) \\
              & \qquad\qquad\qquad\qquad
                \big(
                  \deno{\diff{(\lambda y.\pbterms)}{v}} \anglebr{\gamma,x,v}
                  (\deno{x}\anglebr{\gamma,x,v})
                \big)
              \\
              & =
              \lambda xv.
                \deno{
                  \trans{\simterms}
                  \big((\lambda y.\trans{D})\,x\big)
                  \big(
                    \big(\diff{(\lambda y.\trans{\pbterms})}{v}\big)\,x
                  \big)
                }\anglebr{\gamma,x,v}
              \\
              & =
              \deno{
                \lambda xv.
                \trans{\simterms}
                \big((\lambda y.\trans{D})\,x\big)
                \big(
                  \big(\diff{(\lambda y.\trans{\pbterms})}{v}\big)\,x
                \big)
              }\gamma
            \end{align*}
        \end{itemize}
    \end{enumerate}
  \end{proof}

  \StrongNorm*
  \begin{proof}
    If $\pbterms$ does not terminates,
    then we can form a reduction sequence in $\lang{D}$
    that does not terminates using
    Lemma \ref{lemma:translation respects reduction and model} (1) and
    confluent property of differential $\lambda$-calculus,
    proved in \cite{DBLP:journals/tcs/EhrhardR03}.
    Then, this contradicts
    the strong normalization property of
    differential $\lambda$-calculus.
  \end{proof}




\end{document}